\documentstyle[epsfig,floats,pre,aps]{revtex}
\input epsf
\input rotate
\begin{document}


\title{Scroll waves in isotropic excitable media :
linear instabilities, bifurcations and restabilized states}

\author{Herv\'e Henry and Vincent Hakim}

\address{Laboratoire de Physique Statistique\footnote{Associ\'e au CNRS et
aux Universit\'es Paris VI et VII.},
Ecole Normale Sup\'erieure,\\ 24 rue Lhomond, 75231 Paris Cedex 05, France}
\date{\today}
\maketitle
\begin{abstract}
Scroll waves are three-dimensional analogs of spiral waves. The linear stability
spectrum of untwisted and twisted scroll waves is computed for a 
two-variable reaction-diffusion model of an excitable medium. Different bands
of modes are seen to be unstable in different regions of parameter space. The
corresponding bifurcations and bifurcated states are characterized by
performing direct numerical simulations. In addition, computations of the 
adjoint linear stability operator eigenmodes are also performed 
and serve to obtain
a number of matrix elements characterizing the long-wavelength deformations of
scroll waves.
\end{abstract}
\section{Introduction}
Depolarization waves in cardiac muscle, oxidation waves in the
Belousov-Zhabotinsky (BZ) chemical medium or at the surface of 
certain metal catalyst and cAMP waves in colonies of slime molds
are different examples of wave propagation in excitable media\cite{Win0}.
 They can
be described in similar mathematical terms although the underlying processes
are of a very different nature. In  a two-dimensional (2D) or 
quasi-two dimensional situation,
the
propagation of
spiral waves has been observed in these three cases  as well as
in other excitable media \cite{Win0,Zyk}.
 Asides from their intrinsic scientific interest,
the potential role of 
these remarkable waves in cardiac arrhythmias and fibrillation \cite{heart}
has motivated detailed studies of their properties 
during the past two decades. In particular, the mechanisms of
different instabilities have been intensively
investigated as well as
their locations determined in parameter space of simple models
\cite{zykfg,winchaos} and 
of experiments \cite{belfles}.

The potential relevance to cardiac dysfunction of scroll waves,
the three dimensional analogs of spirals, 
has also
been emphasized \cite{win3d} but their dynamics is still
less thoroughly analyzed.
Visualization of the
BZ reaction in 3D gels \cite{pert,tombz}
has confirmed the existence of these complex waves. 
Numerical simulations have revealed that they are prone to
instabilities in several parameter regimes\cite{henze,panrud,bik3d,garfin}. 
In order to analyze more 
systematically the different possible instabilities, we report here the result
of computations of the full linear stability spectrum of a straight scroll wave in a simple
two-variable model of an excitable medium. This
enables us to follow the different
modes of deformation of a scroll wave and to investigate which type of modes
become unstable in different regions of parameter space. The modes of the 
adjoint operator are also determined in order to compute the value
of several coefficients given by  matrix
elements and to check proposed analytic relations. 
In addition,
direct  numerical simulations
are performed to investigate the nonlinear fate of the 
different instabilities
and to provide a detailed characterization of the restabilized bifurcated states
(when they exist).

In section \ref{sec:meth}, we define the studied two-variable reaction model and explain our numerical
methods. Some general properties of spiral waves are also
recalled. Then,
we first consider in section \ref{sec:untw} a straight 
untwisted scroll wave. It is the simplest (i.e. z-independent)
extension in the third (z-) 
dimension of
a two dimensional (xy) spiral wave. At the linear level,
two possible types of instabilities are found. Modes with positive real parts
can be observed on the translation bands, which correspond to
z-dependent translations of the 2d spiral in the different xy-planes, or
on the meander bands, which come from
z-deformations of the 2d spiral meander modes. At the non linear level, the
translation band instability gives rise to a scroll wave with a continuously
extending core \cite{panrud,bik3d} and does not lead to a restabilized non
linear state. We confirm that this type of instability  is directly 
related \cite{hk2} to the drift direction of a 2d spiral in an
external
field. In contrast, the meander band type of instability \cite{armit}
generally restabilizes in a distorted scroll wave and no simple relation
to 2d spiral drift is observed.

In section \ref{sec:tw}, we consider twisted scroll waves. A 3D steady wave
is built  by rotating (i.e. twisting)  the 2d spiral around its rotation center as one 
translates it along the z-direction. We find that twist exceeding a definite 
threshold can lead to the appearance of unstable modes in the translation 
brands
of the scroll wave. This ''sproing'' \cite{henze} instability
is seen to
take place a finite wavevector away from the scroll wave translation
symmetry mode. 
We  provide analytical arguments which show that this very
generally results from 
3D rotational invariance in an isotropic medium. 
Nonlinear 
development of this instability when a single unstable mode is present
results in a restabilized helical wave, as previously described \cite{henze}.
Properties of these nonlinear states are computed
and compared with
the linear characteristics (wavevector, frequency) of
the sproing instability. When several unstable modes are present in 
the simulation box, the scroll wave core filament takes a more complex 
shape 
which
is found to travel like a nonlinear wave of constant shape in the 
vertical direction. Three appendices provide the
details of our numerical algorithms, 
a derivation of
a general formula for spiral drift in an external field and useful
ribbon geometry
formulas. A fourth one explains the relation between the present calculations
and previously derived averaged equations for the motion of a weakly curved and twisted
scroll wave \cite{keen3d,bik3d}. 

The results of our linear stability analysis have previously been shortly
described in \cite{hhprl}.
\section{Methods and general results}
\label{sec:meth}
\subsection{Reaction-diffusion model}
\begin{figure}
\begin{center}
\includegraphics[height=4.2cm]{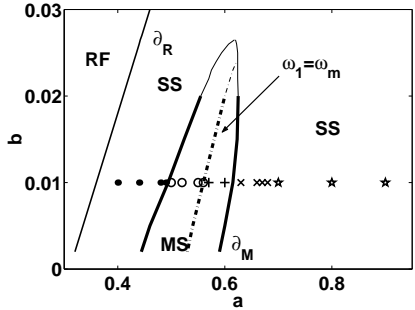}
\end{center}
\caption{
\label{phasd.fig} Spiral
 bifurcation diagram for Eq.~(\ref{eq1},\ref{eq2}) with $\epsilon=0.025$ and
the reaction terms of  
ref.~[19].
The bold line $\partial M$ is the meander
threshold instability line and separates steadily rotating spirals (SS) from
meandering spirals (MS)
 (above $b=0.02$ the thin line denotes our less accurate
determination of  $\partial M$).
The line $\partial R$ marks the boundary of spiral
wave existence on the left of $\partial R$ the wave tip retracts 
(see e.g.  
ref.~[14]
). The symbols along the line $b=0.01$ denotes 
the different
parameters at which scroll waves are studied in the present work. Stars
represent stable scroll waves, crosses ($\times$) and pluses($+$) 
meander-unstable scroll waves,  
circles ($o$), and  ($\bullet$) translation-band-unstable
scroll waves. 2D meandering spirals are represented by  ($+$) and
($o$).
}
\end{figure}

Two-variable reaction diffusion systems have been shown to describe
semi-quantitatively spiral waves dynamics and its ''generic'' features.
They have been used in various contexts since their original introduction
\cite{FitzH}
as a simplification of Hodgkin-Huxley dynamics. The analysis of such a simple
model appears in any case as a useful first step before going to a more
complicated description if required. We thus follow this classic path and
take for the
excitable medium dynamics 
\begin{eqnarray}
\partial_t u&=& \nabla^2u+f(u,v)/\epsilon\label{eq1}
\\
\partial_t v&=& g(u,v)
\label{eq2}
\end{eqnarray}
We only consider the singly diffusive case, the most relevant
to cardiac physiology.
For definiteness, we also  choose reaction terms 
$f(u,v)=u(1-u)[u-(v+b)/a],\ g(u,v)=u-v$ as proposed
in \cite{barkn}. This permits fast direct
simulations and tests of our numerics by comparison with previous results for 
spiral waves. The 2d-spiral bifurcation diagram for this model 
is shown in Fig.~\ref{phasd.fig} for variable
values of the parameters $a$ and $b$ at a fixed value of
$\epsilon=0.025$. The meander instability line ($\partial M$ with
the notation of \cite{winchaos}) is plotted
with its ''large core'' branch
at smaller values of $a$ than its ''small core''
branch.
The crossing line which separates meander trajectories with
outward petals from those with inward petals is also drawn as well as the
diverging core existence boundary of spiral waves ($\partial R$).

 In the 
following, the stability and dynamics 
of scroll waves are analyzed at several points along the line $b=0.01$,
as $a$ varies and crosses the different boundaries.
\subsection{Numerical strategy}
In order to obtain the steady scroll waves and compute their stability spectrum,
Eq.~(\ref{eq1},~\ref{eq2}) are written in cylindrical coordinates
with $u$ and
$v$  functions of $r, \phi=\theta-\omega t -\tau_w z, t$ and $z$,
\begin{eqnarray}
(\partial_t +2\tau_w\partial^2_{\phi z} -\partial^2_{zz})u&=&(\omega
\partial_\phi+
\tau^2_w\partial^2_{\phi\phi}+\nabla^2_{2D})u+f(u,v)/\epsilon\label{tdpolar1}\\
\partial_t v&=&\omega\partial_\phi v+g(u,v)
\label{tdpolar2}
\end{eqnarray}
\subsubsection{Steady states}
A steady scroll wave for a given imposed twist $\tau_w$ is 
a time independent solution of Eq.~(\ref{tdpolar1},~\ref{tdpolar2}) 
with $u(r,\phi,t,z)=u_0(r,\phi)$ and $v(r,\phi,t,z)=v_0(r,\phi)$ and rotation
frequency $\omega=\omega_1$,
\begin{eqnarray}
(\nabla^2_{2D}+\omega_1\partial_\phi +
\tau_w^2 \partial^2_{\phi\phi}) u_0 +f(u_0,v_0)/
\epsilon
&=& 0
\label{2dst1}
\\
\omega_1\partial_\phi v_0 +g(u_0,v_0)&=&0
\label{2dst2}
\end{eqnarray}
This nonlinear fixed point problem for the function $u_0, v_0$ and nonlinear
eigenvalue $\omega_1$ is solved by using Newton's method after discretization
of Eq.~(\ref{2dst1},\ref{2dst2}), as detailed in appendix \ref{appss}. 
It should be noted that Eq.~(\ref{2dst1},~\ref{2dst2}) are purely 
two-dimensional due to the scroll wave translation symmetry along the z axis
in the introduced coordinates.

\subsubsection{Linear stability}
\label{linstab.sec}
Once a steady scroll wave is obtained, one can linearize Eq.~(\ref{tdpolar1},
\ref{tdpolar2}) around it. Invariance of Eq.~(\ref{tdpolar1},~\ref{tdpolar2})
and of the steady state by translation along the $z$-direction (in the introduced twisted
rotating frame) leads to decompose a general perturbation depending
on the three spatial coordinates on its Fourier components
along the $z$-axis. We thus consider perturbations under the form
$u=u_0+
\exp[\sigma(k_z) t -i k_z z] u_1(r,\phi),
v=v_0+ \exp[\sigma(k_z) t -i k_z z]
v_1(r,\phi)$. The linear equations obeyed by $u_1, v_1$ and the (complex)
growth rate $\sigma(k_z)$ read,
\begin{eqnarray}
\sigma u_1 &=& (-k_z^2+2 i\tau_w k_z\partial_{\phi}) u_1+
(\omega_1\partial_{\phi}+\tau_w^2 \partial_{\phi\phi}^2+ \nabla^2_{2D}
)u_1+[\partial_u f(u_0,v_0)u_1+\partial_v f(u_0,v_0)v_1]/\epsilon
\label{eqlin1}
\\
\sigma v_1 &=& \omega_1\partial_{\phi}v_1 +
\partial_u g(u_0,v_0)u_1+\partial_v g(u_0,v_0)v_1
\label{eqlin2}
\end{eqnarray}
or symbolically,
\begin{equation}
\sigma(k_z)  \left( \begin{array}{c} u_1\\v_1\end{array}\right) =
{\mathcal{L}}_{k_z}
        \left( \begin{array}{c} u_1\\v_1\end{array}\right)\label{formelle}
\end{equation}

So in a second step, the ($\sim 10$) eigenvalues of largest real parts of
${\mathcal{L}}_{k_z}$ are precisely determined (for given $k_z$ and $\tau_w$)
using an iterative algorithm \cite{Gold87} detailed
in appendix \ref{appls}.
The whole numerical procedure
is quite similar to the spiral linear stability analysis of 
ref.~\cite{barkl} and extends it to 3D.
It should be noted that the twist rate $\tau_w$ can be prescribed at will
so the procedure is not confined to weak twist (of course, for too large 
a twist, a steady scroll may no longer exist and the Newton steady state
finding procedure fails to converge).

 Two points are worth emphasizing:\\
-taking  a Fourier transform has eliminated
the $z$-direction 
so
Eq.~(\ref{eqlin1},~\ref{eqlin2}) are purely two-dimensional
 (but $k_z$-dependent) as the steady state equations (\ref{2dst1},~\ref{2dst2}),
\\
-correlatively, each mode of the 2D-spiral is replaced by a band of modes
indexed by the wavevector $k_z$.\\

Some general properties of the spectrum can be noted at this stage.

For zero twist, Eq.~(\ref{eqlin1},~\ref{eqlin2}) depend only on $k_z^2$ so
the spectrum bands are even functions of $k_z$. Moreover, 
${\mathcal{L}}_{k_z}$ is a real operator and its complex eigenvalues come
in complex conjugate (c.c.) pairs. So, bands of complex modes also come
in complex conjugate pairs.

For non-zero twist ($\tau_w\ne 0$), these symmetries non-longer hold. It only
remains true that ${\mathcal{L}}_{k_z}={\mathcal{L}^*}_{-k_z}$. So
bands of complex modes can be grouped in pairs $\sigma_1(k_z),\sigma_2(k_z)$
with $\sigma_2(k_z)=\sigma_1^*(-k_z)$
\subsubsection{Direct numerical simulations and instantaneous 
filament definition}
In order to determine the bifurcation type and the fate of each
instabilities, 
we performed direct numerical simulations of Eq.~(\ref{eq1},\ref{eq2}) as
explained in appendix \ref{appdns}. 
In two dimensions, it is usual 
to define the spiral tip as the intersection point 
of two (somewhat arbitrary) particular level lines $u=u_{tip},v=v_{tip}$.
The spiral tip trajectory is then a convenient way to visualize the spiral
dynamics and its core instabilities. Similarly in 3D, we choose here to
define the 
instantaneous filament as the intersection line of the
two particular level surfaces $u=u_{tip}=0.5$ and $v=v_{tip}=0.75(0.5 a-b)$. 
It can be thought 
of as the line of spiral tips in the different $xy$-plane.

\subsection{Special eigenmodes}
\label{speceigen.sec}
In the 2D case,
there are five
dominant modes of spiral dynamics in the simple case described by
Eq.~(\ref{eq1},~\ref{eq2}) 
\cite{alt}
:\\
-one neutral mode with $\sigma=0$, the rotation mode, 
which comes from the rotational invariance of 
Eq.~(\ref{tdpolar1},~\ref{tdpolar2}).\\
-two complex conjugate purely imaginary modes
with $\sigma=\pm i \omega_1$, the translation modes,
 coming from the translation invariance of 
the starting equations (\ref{eq1},
\ref{eq2})\\
- two complex conjugate modes, the meander modes,
 corresponding to the oscillatory meander 
instability the real part of which crosses zero on the meander instability line.

In the following, the five bands of modes originating from 
these special modes are found to play the most important role 
in scroll wave dynamics
in the sense that they have the largest real parts and that each
instability of a scroll wave can be ascribed to one of these
bands (i.e.  the modes on
a part of that particular band acquire a positive real part).

Thus, before proceeding it is worth recalling the expression of these symmetry 
eigenmodes
for spirals \cite{barkl}
 and their straightforward generalization for scroll waves.

The rotation mode is the simplest. Differentiation of 
Eq.~(\ref{2dst1},~\ref{2dst2}) directly shows that $(\partial_{\phi}u_0,
\partial_{\phi}v_0)$, the rotation mode, is a solution of Eq.~(\ref{eqlin1},
\ref{eqlin2}) for $k_z=0$ and $\sigma=0$.

The eigenmodes associated to invariance under $x,y$-translations of the spiral
rotation center are less straightforwardly obtained because the spiral is
a time independent solution in a rotating frame (with $(x',y')$ coordinates).
A steady spiral rotating around the origin is given by 
\begin{equation}
U_0[x',y']=U_0[\cos(\omega_1 t) x+\sin(\omega_1 t) y,-\sin(\omega_1 t) x+
\cos(\omega_1 t) y]
\end{equation}
The corresponding spiral rotating around the point $(x_0,y_0)$ is then
$$
U_0[\cos(\omega_1 t) (x-x_0)+\sin(\omega_1 t) (y-y_0),-\sin(\omega_1 t) (x-x_0)
+
\cos(\omega_1 t) (y-y_0)].$$
Using the $(x',y')$ rotating coordinates, this translated spiral reads,
\begin{eqnarray}
U_0[x'-x_0 \cos(\omega_1 t)-y_0 \sin(\omega_1 t),
y'+x_0\sin(\omega_1 t)-y_0\cos(\omega_1 t)]=\nonumber\\
U_0[x'-(\frac{x_0-i y_0}{2} \exp(i\omega_1 t)+c.c.), 
y'- (i\frac{x_0-i y_0}{2} \exp(i\omega_1 t)+c.c.)]
\label{transspi}
\end{eqnarray}
Expansion of Eq.~(\ref{transspi}) for small $x_0,y_0$
gives the expression of the
translation modes\begin{equation}
\left( \begin{array}{c} u_t\\v_t\end{array}\right) =
\left( \begin{array}{c} (\partial_{x'}+i \partial_{y'})u_0\\
(\partial_{x'}+i \partial_{y'})v_0\end{array}\right) =
\exp(i\phi)
\left( \begin{array}{c}
\partial_r u_0+i\partial_{\phi} u_0/r\\
\partial_r v_0+i\partial_{\phi} v_0/r\end{array}\right) 
\label{tm} 
\end{equation}
with eigenvalue $i\omega_1$, and the complex conjugate eigenvector
with eigenvalue $-i\omega_1$.

This computation immediately generalizes to untwisted scroll waves. This
is also the case for twisted scroll waves but one should recall that 
Eq.~(\ref{2dst1},~\ref{2dst2}) are written in a referential that rotates in time
but also as one moves along the z-axis. This modifies the exponential factor in
Eq.~(\ref{transspi}) which 
becomes $\exp(i\omega_1 t+\tau_w z)$ to include the $z$-rotation. So, for
a twisted scroll wave the translation mode $(u_t,v_t)$ is an eigenvector
of ${\cal L}_{k_z=-\tau_w}$ with eigenvalue $i \omega_1$. The other (complex
conjugate) translation eigenvector $(u_t^*,v_t^*)$ is associated to the
eigenvalue $-i\omega_1$ of the now {\em different} linear operator
${\cal L}_{k_z=\tau_w}$.

A direct algebraic proof of these facts can be given. If ones defines
the two operators
\begin{eqnarray}
T &\equiv& \exp(i\phi)(\partial_r+i/r\partial_{\phi})\nonumber\\
M &\equiv& \omega_1\partial_{\phi}+\tau_w^2\partial^2_{\phi\phi}+
\nabla^2_{2D}
\end{eqnarray}
a direct computation gives the commutators
\begin{eqnarray}
[T,M]&=&-i\omega_1 T -\tau_w^2 (1+2i\partial_{\phi})T
\nonumber\\
\left[T,\partial_{\phi}\right]&=&-i T
\label{com}
\end{eqnarray}
With these notations, the fixed point equations reads
$Mu_0+f(u_0,v_0)=0,\ 
\omega_1\partial_{\phi}v_0+g(u_0,v_0)=0$.
Action of $T$ on these two equations gives $TMu_0+\partial_u f \,\ Tu_0+
\partial_v f\, Tv_0=0,\,\omega_1 T\partial_{\phi}v_0+\partial_u g\, Tu_0+
\partial_v g\, Tv_0=0$. Commutations of $T$ and $M$ (in the first one) and
$T$ and $\partial_{\phi}$ (in the second one) using
Eq.~\ref{com} directly show that $(u_t,v_t)=(Tu_0,Tv_0)$ satisfies
Eq.~(\ref{eqlin1}, \ref{eqlin2}) with $\sigma=+i\omega_1$ and $k_z=-\tau_w$.

\subsection{Left eigenvectors and scalar product}

The linear stability computation can be extended to determine the left 
eigenvectors of ${\cal L}_{k_z}$ 
 We have found it worth in particular
to compute the left eigenvectors of  the 2D spiral stability operator
${\cal L}$
(i.e. ${\cal L}_{k_z=0}$ for $\tau_w=0$) 
corresponding to the translation
and rotation modes since they often appear in perturbation
calculation \cite{keen3d,bik3d} (for examples, see section \ref{stabdrift.sec}
 and appendix\ref{keenav.sec}).

We simply define the scalar product between a left 
$(\tilde{u},\tilde{v})$ and  right $(u_r,v_r)$ two-component function
by integration over the whole 2D space
as $\langle\tilde{u},u_r\rangle+\langle\tilde{v},v_r\rangle$ where
\begin{equation}
\langle f,g\rangle\equiv\int\!\!\!\int\!\! dr d\phi r f(r,\phi)\ g(r,\phi)
\label{scaprod}
\end{equation}

The left eigenmodes of ${\cal L}$ thus obey,
\begin{eqnarray}
\sigma \tilde{u}_1 &=&
(-\omega_1\partial_{\phi}+ \nabla^2_{2D}
)\tilde{u}_1+
\partial_u f(u_0,v_0)\tilde{u}_1/\epsilon+\partial_u g(u_0,v_0)\tilde{v}_1
\label{eqlin1adj}
\\
\sigma \tilde{v}_1 &=&-\omega_1\partial_{\phi}\tilde{v}_1 +
\partial_v f(u_0,v_0)\tilde{u}_1/\epsilon+\partial_v g(u_0,v_0)\tilde{v}_1
\label{eqlin2adj}
\end{eqnarray}
\label{lefteigen}

\begin{figure}[h]
\begin{center}
\begin{tabular}{ccc}
(a)&&(b)\\
\includegraphics[width=4.cm]{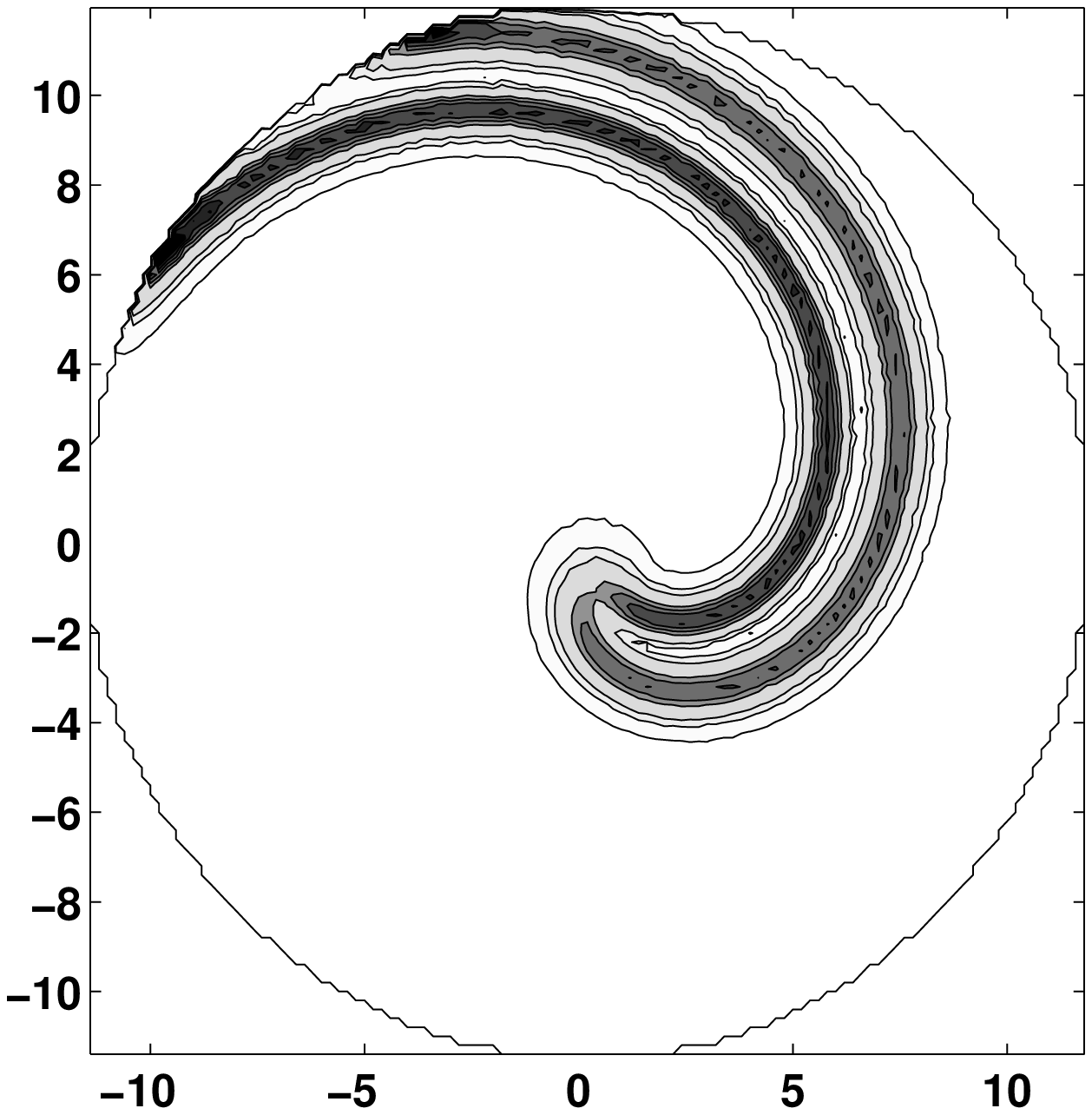}
&&
\includegraphics[width=4.cm]{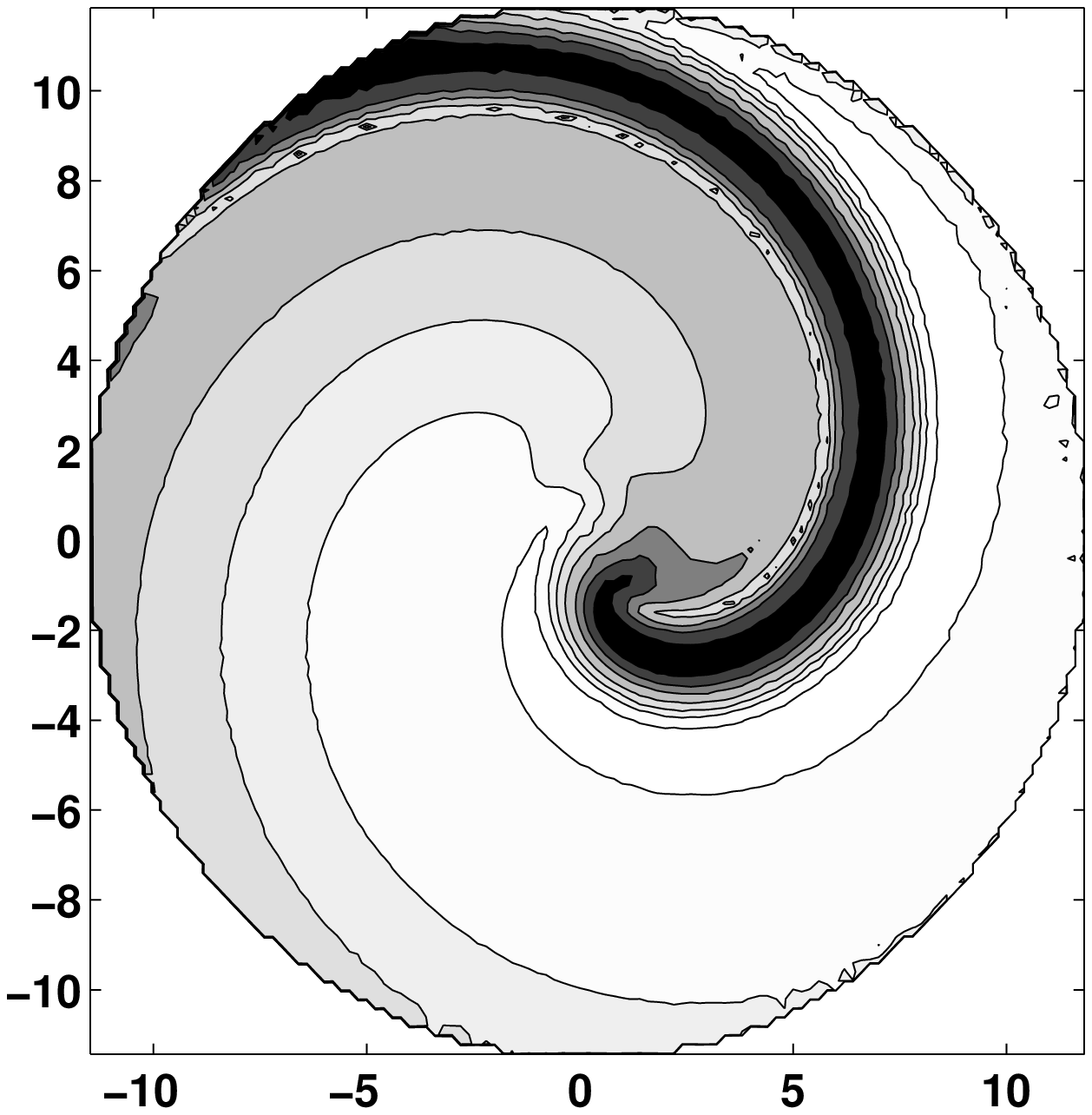}\\
(c)&&(d)\\
\includegraphics[width=4.cm]{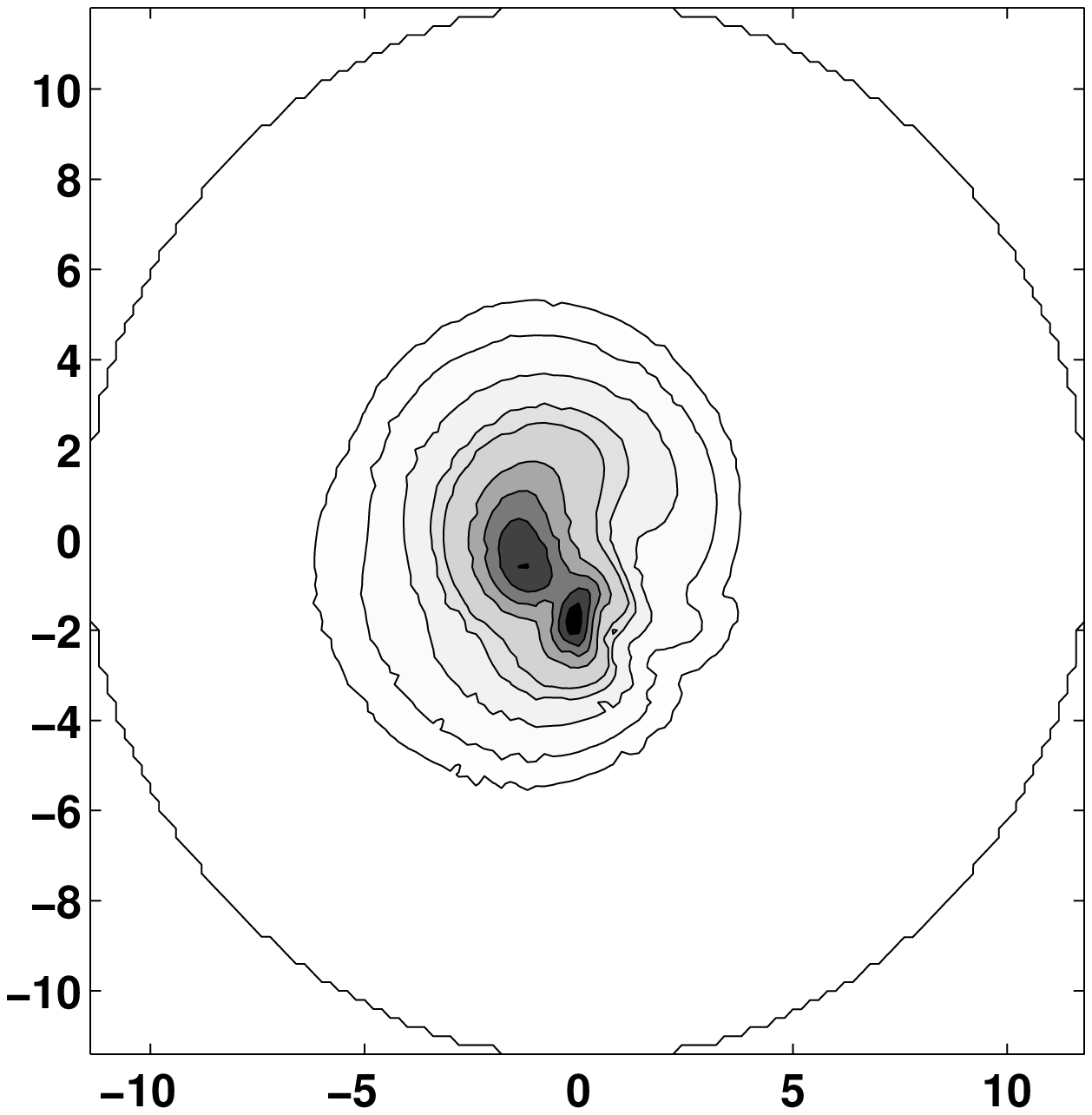}
&&
\includegraphics[width=4.cm]{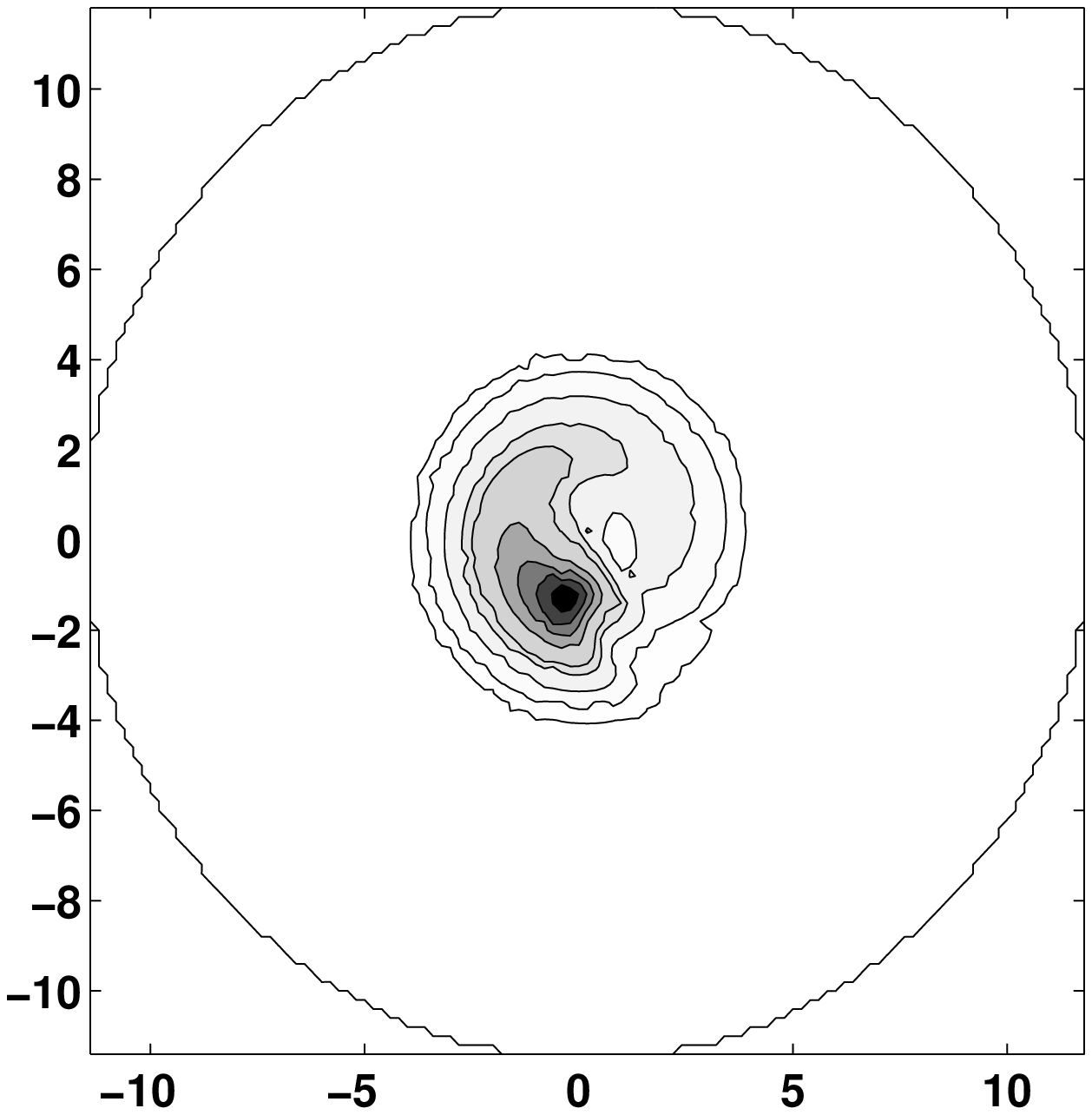}
\end{tabular}

\caption{\label{contour-modespropres} Contour plots of the translation
eigenmode, (a) $u$-component modulus, (b) $v$- component modulus, and
of the corresponding left eigenvector , (c) $u$-component modulus,
(d) $v$- component modulus. 
The maximum value of  the fields 
is set independently for $u$ and $v$ 
equal to $1.$  and the contours are plotted for  
(a)$u=0.01,\
 0.05,\ 0.1,\ 0.3,\ 0.4,\
0.5,\ 0.6 \mbox{ and } 0.7$, (b)$v=0.01,\ 0.05,\ 0.1,\ 0.2,\
0.4,\  0.6  \mbox{ and }  0.8$, (c) and (d)
$u,v=0.0001,\ 0.001,\
0.01,\  0.05,\  0.1,\ 0.3,\ 0.5,\ 0.7\mbox{ and } 0.9$. 
The
parameter values are $a=0.44$, $b=0.01$ and $\epsilon=0.025$. The
pulsation of the steady rotating spiral is $\omega=1.1612$. The
circles represent the limit of the simulation box.
\label{eigent.fig}}
\end{center}
\end{figure}

\begin{figure}
\begin{center}
\begin{tabular}{ccc}
(a)&&(b)\\
\includegraphics[width=4.cm]{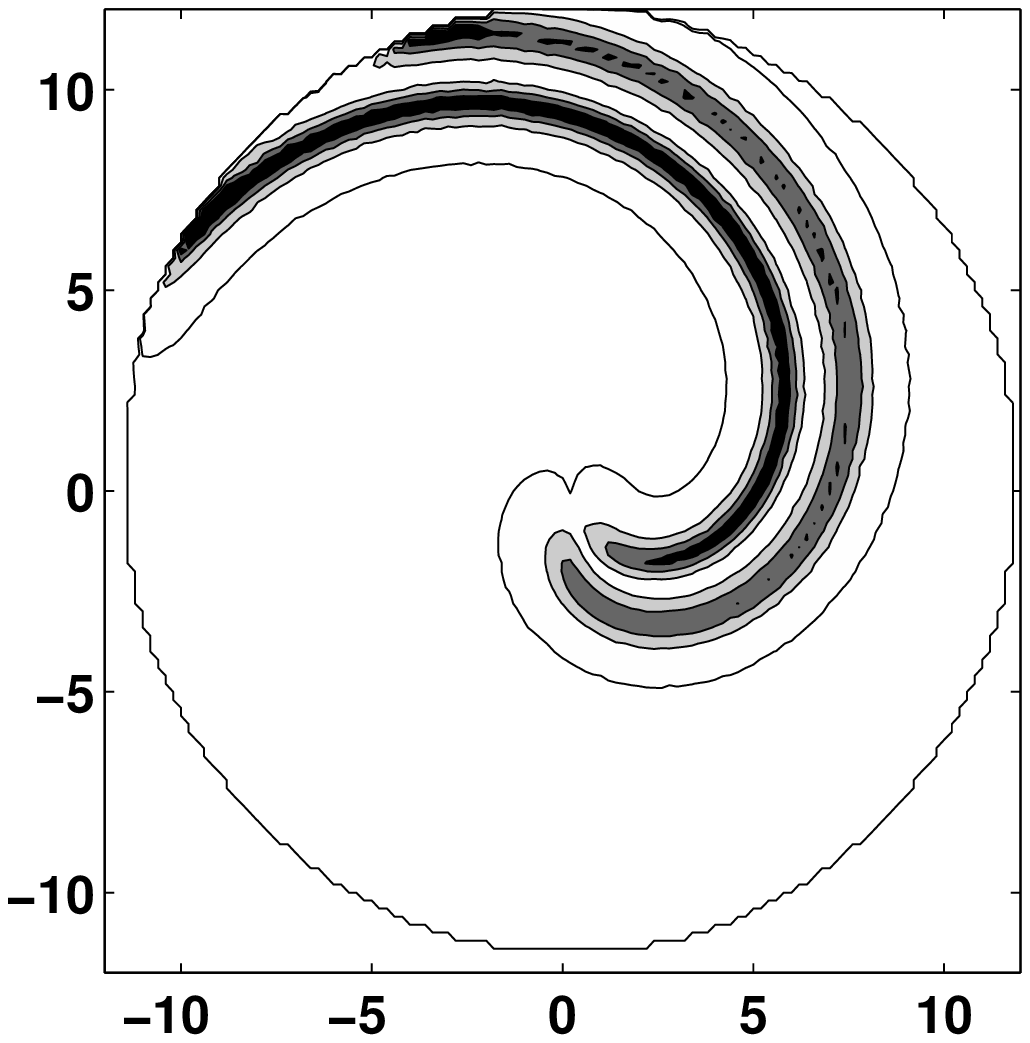}
&&
\includegraphics[width=4.cm]{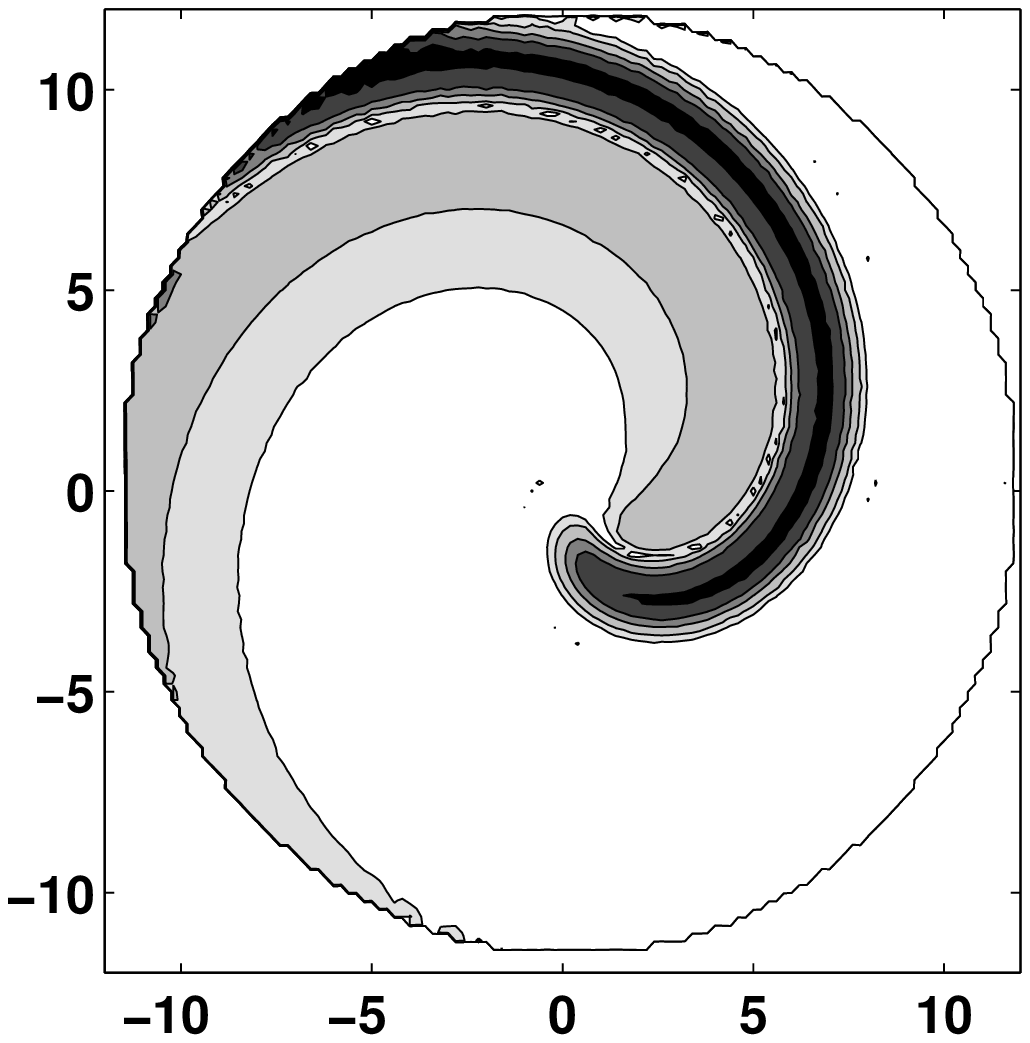}\\
(c)&&(d)\\
\includegraphics[width=4.cm]{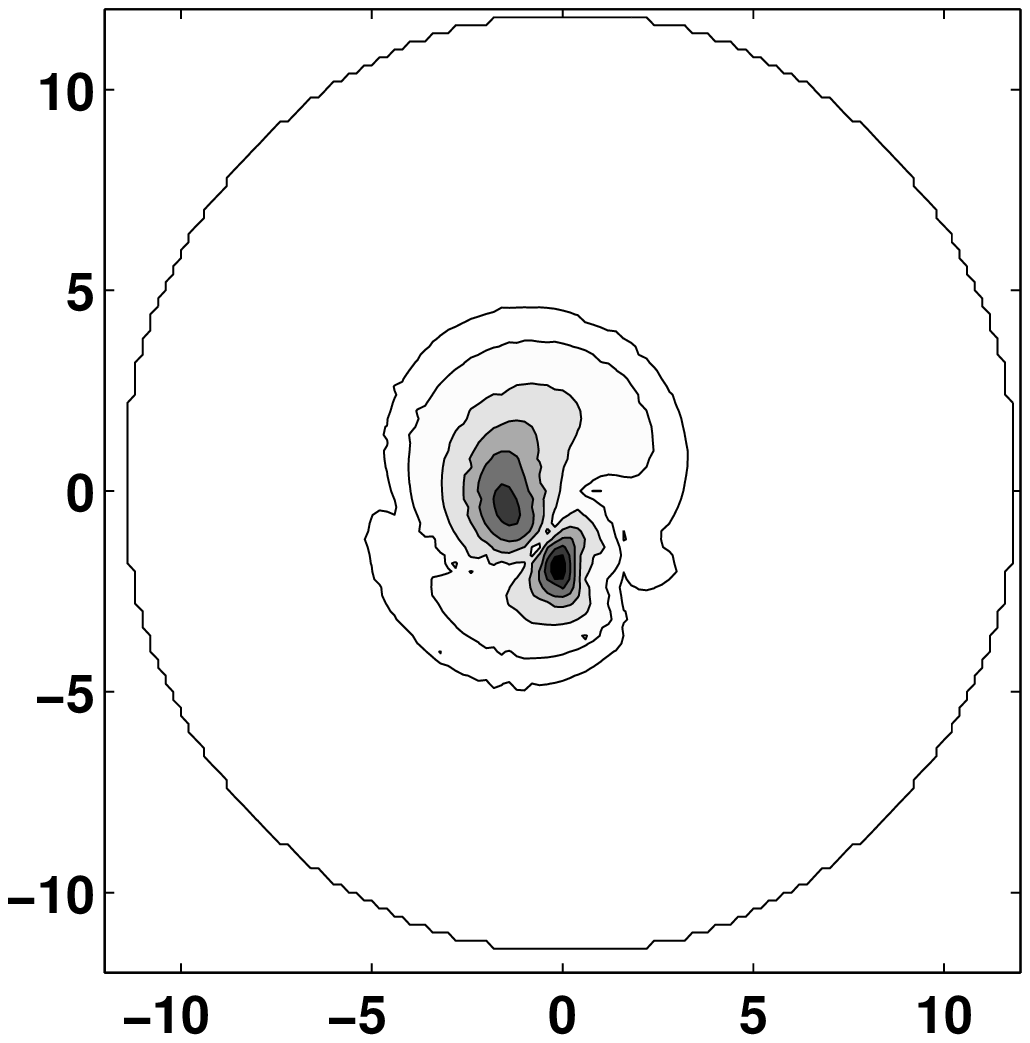}
&&
\includegraphics[width=4.cm]{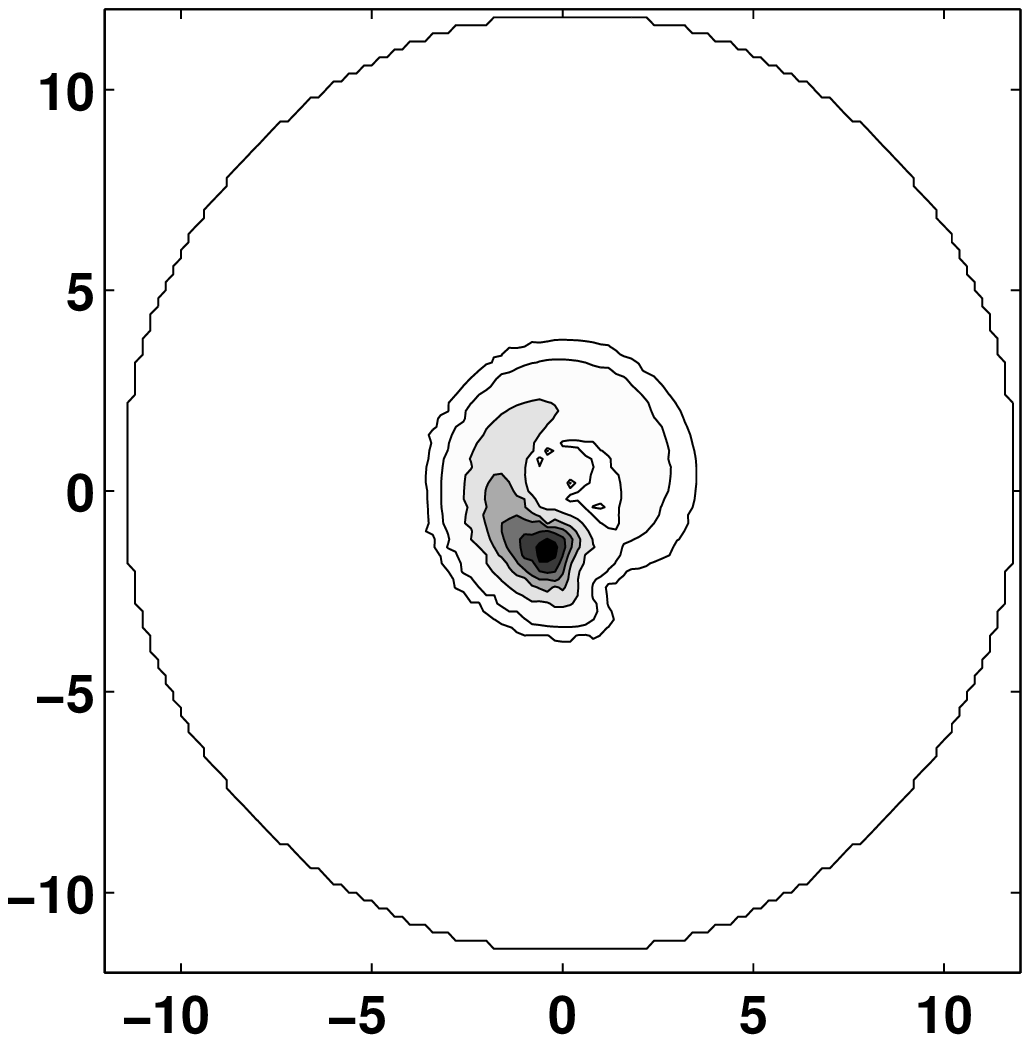}
\end{tabular}
\caption{\label{contour-modespropres-rot}
Contour plots of the rotation      
eigenmode, (a) $u$-component modulus, (b) $v$- component modulus, and  
of the corresponding left eigenvector , (c) $u$-component modulus,
(d) $v$- component modulus.
The maximum value of the fields is set equal to
$1.$ and the contours are plotted for (a) $u=$  0.001, 0.1,  0.3,  0.5 and
0. 7 , (b) $v=$ 0.001,  0.1,  0.2,  0.4, 0.6 and 0.8, (c) and (d) 
$u,\ v =$ 0.,  0.001, 0.01, 0.1, 0.3,  0.5, 0.7 and 0.9. 
Same parameters as in Fig.~\ref{eigent.fig}. 
\label{eigenr.fig}}
\end{center}
\end{figure}

The result of one such computation for the left translation eigenmode 
$(\tilde{u}_t,\tilde{v}_t)$ (the solution $(\tilde{u}_1,\tilde{v}_1)$ 
for $\sigma=i\omega_1$)
 is shown
in Fig.~\ref{eigent.fig}. In contrast to
the translation mode $(u_t,v_t)$, the left eigenmode 
$(\tilde{u}_t, \tilde{v}_t)$  quickly decreases away from the spiral core
as argued in \cite{bik3d} and explicitly obtained in the free-boundary limit
\cite{hk2,mak} (but opposite to what is supposed in \cite{keen3d}).
This also holds for the left rotation mode 
$(\tilde{u}_{\phi},\tilde{v}_{\phi})$ (the solution $(\tilde{u}_1,\tilde{v}_1)$
for $\sigma=0$) as shown in Fig.~\ref{eigenr.fig}.
As a consequence, the scalar product (\ref{scaprod})
between these left functions 
$(\tilde{u},\tilde{v})$ and any right function $(u_r,v_r)$ (even slowly
increasing) is well defined\footnote{The fast decay of the left eigenmodes
makes the space integration converge without any additional factor. Adding
one such extra factor as suggested in \cite{keen3d} would actually make all the
scalar product vanish.}. We
do not find it useful to include a time integration in the scalar product
as in \cite{keen3d,bik3d}. 

\section{Untwisted filaments}
\label{sec:untw}
\begin{figure}
\begin{center}
\begin{tabular}{ccc}
(a)&&(b)\\[.3cm]
\includegraphics[height=4.cm,width=4.cm]{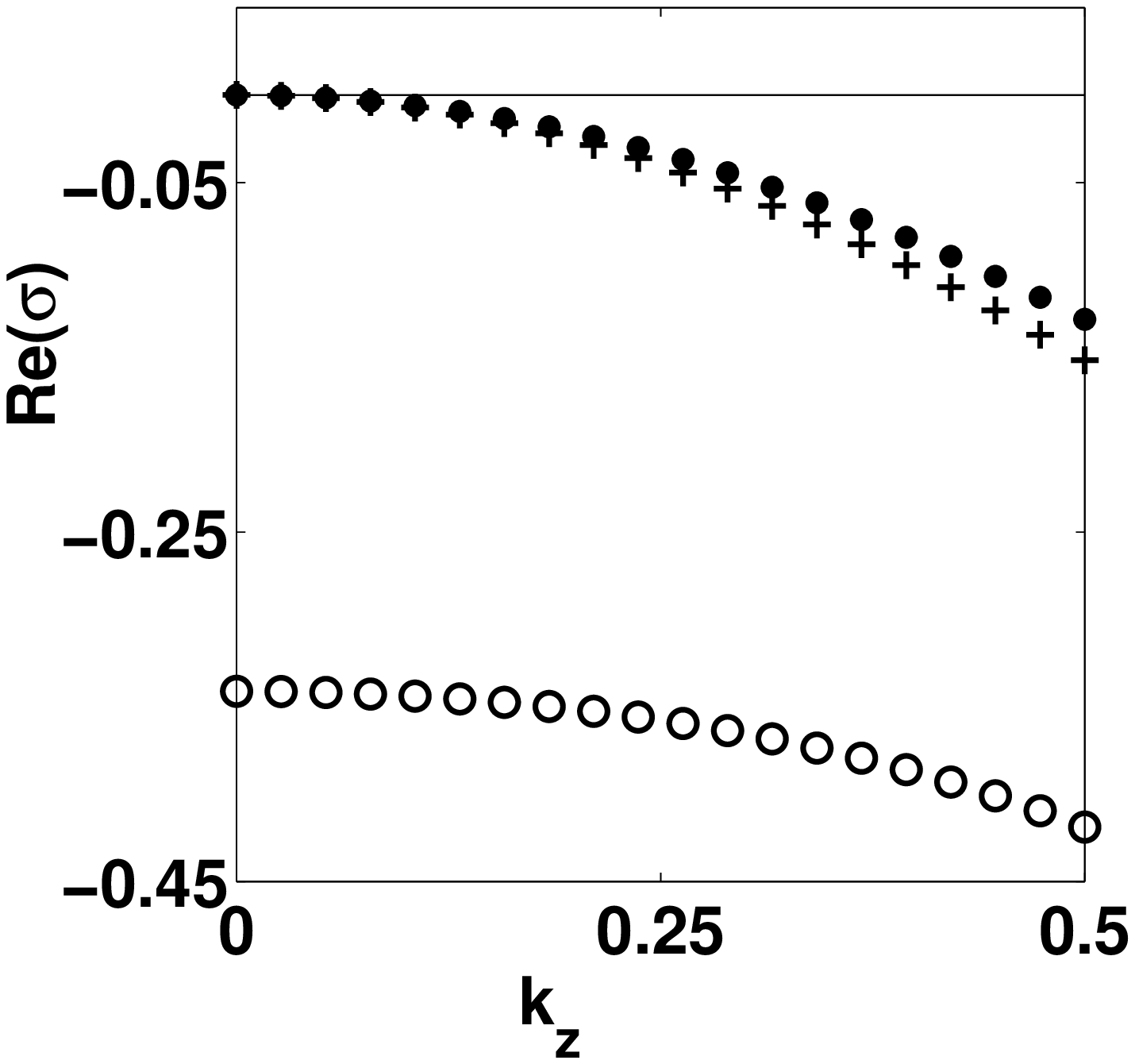}
&&
\includegraphics[height=4.cm,width=4.cm]{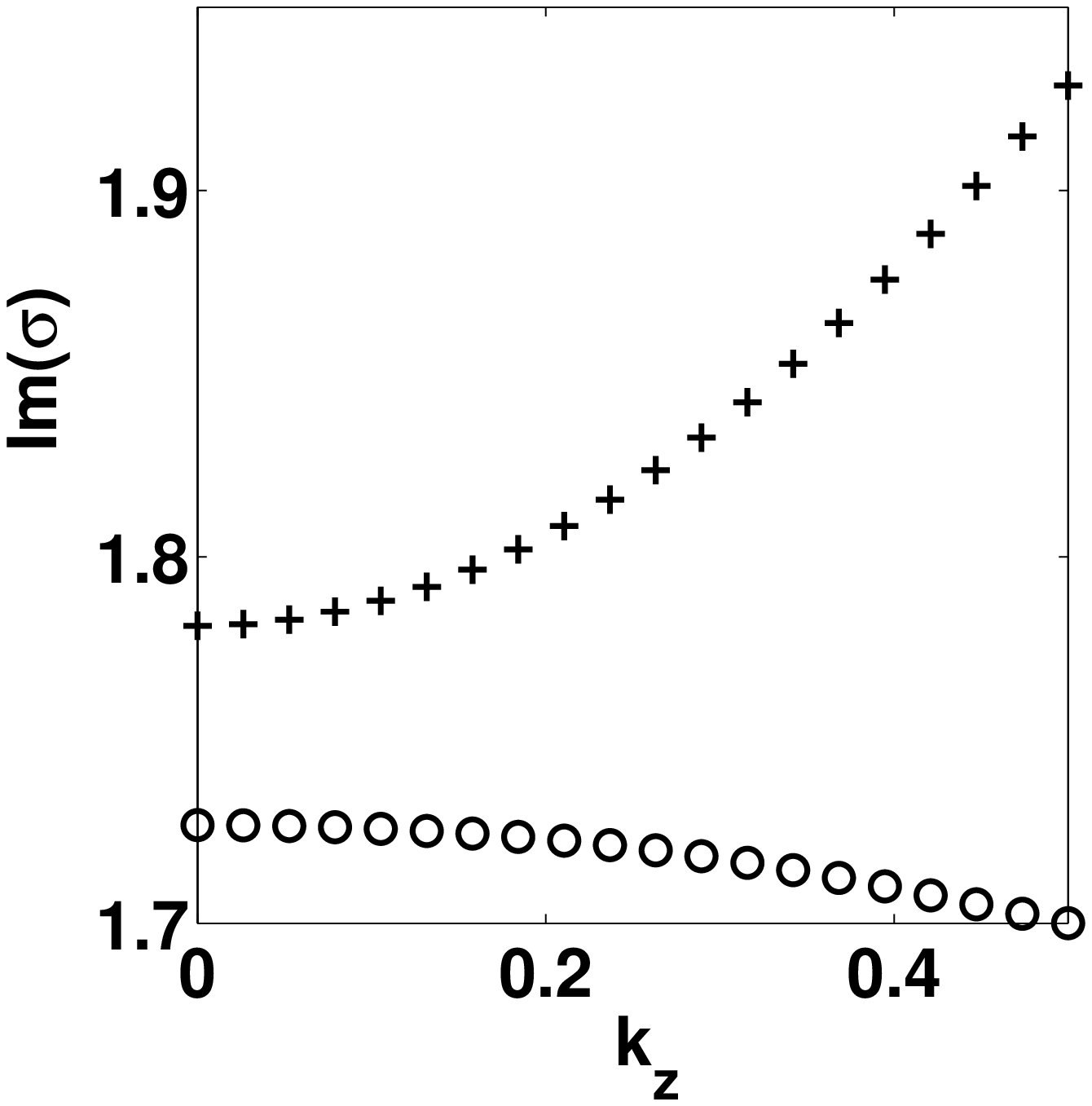}
\end{tabular}
\caption{(a) Real and (b) imaginary parts of the growth rate $\sigma(k_z)$ as a 
function of
the wavenumber $k_z$ 
for the translation ($+$), rotation ($\bullet$) and
meander ($\circ$) bands. The parameter values are $a=0.9,\ b=0.01$ 
 and $\epsilon=0.025$. Fig.~\ref{lsuts.fig}a shows that
the meander mode at $k_z=0$ is stable and that
the growth rate decreases on the meander band with $k_z$. The
translation mode is also restabilized for finite values of $k_z$. The 
translation and meander bands
are well approximated by respectively
 $\sigma_t(k_z)=i 1.769
+(-0.65+i 0.61)k_z^2$ and $\sigma_m(k_z)=-0.3411+i 1.720
+(-0.25+i 0.01)k_z^2$. The value of $\sigma_t(0)$ is in good agreement with
the independently determined pulsation of the steady scroll wave
$\omega_1=1.769$
\label{lsuts.fig}
}
\end{center}
\end{figure}

We begin with the simplest case of scroll waves with no twist ($\tau_w=0$).
In this case, the steady scroll equations (\ref{2dst1},~\ref{2dst2})
 are clearly identical
to those of a 2d spiral. We take as an example the parameter value $a=0.9,
b=0.01$ (and $\epsilon=0.025$).
A steady spiral/scroll wave is found for a rotation frequency $\omega_1=1.769$.
The linear spectrum of modes around this steady scroll is plotted in 
Fig.~\ref{lsuts.fig} (only the upper quadrant upper $k_z>0, Im(\sigma)>0$ is 
shown
since the other quadrants can be deduced by parity and complex conjugation)
 The five translation, rotation and meander modes
of the spiral wave stand at $k_z =0$. The steady spiral is stable as shown
by the negative real parts of the meander modes. As stated above, the spectrum
around the scroll wave is organized in several bands of modes which
 originates from the spiral modes at $k_z=0$. Only the five less stable bands are shown in Fig.~\ref{lsuts.fig}. At these parameter values, extension to
the third dimension does not bring any instability (at least at the
linear level) since as seen on
Fig.~\ref{lsuts.fig} the real part of $\sigma(k_z)$ becomes more negative on
 each band as
$k_z$ increases.
For other parameter values, a straight scroll wave can however be unstable while
2D-spiral are stable. This can happen in two different ways. Depending on
position in parameter space, either the translation or the meander bands
become unstable for $k_z\neq 0$. We examine these two cases in turn in the
following two subsections.
\subsection{Translation band instability}
\label{tbi.sec}
\begin{figure}[h]
\begin{center}
\begin{tabular}{ccc}
(a)&& (b)\\[.1cm]
\includegraphics[width=4.cm]{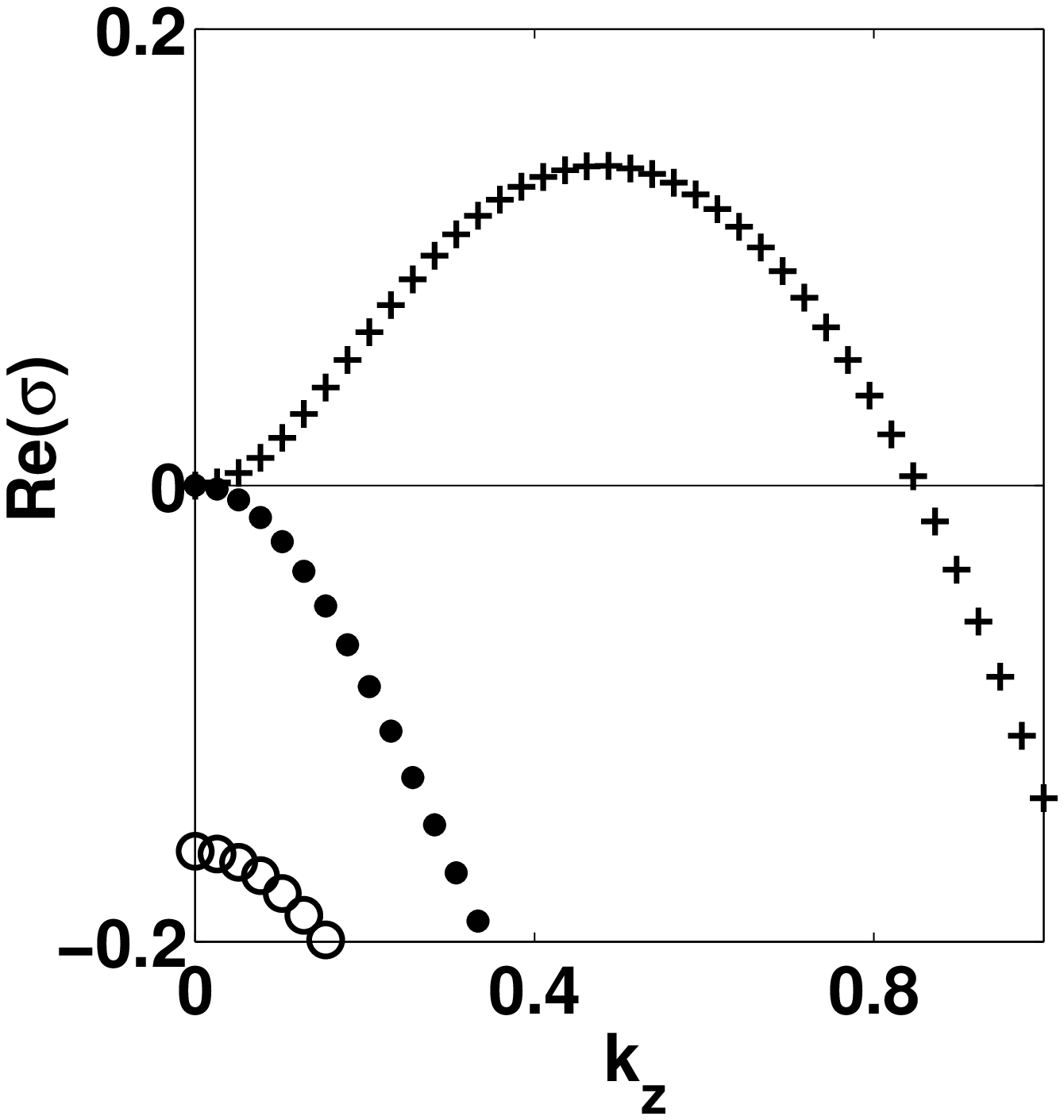}
&&
\includegraphics[width=4.cm]{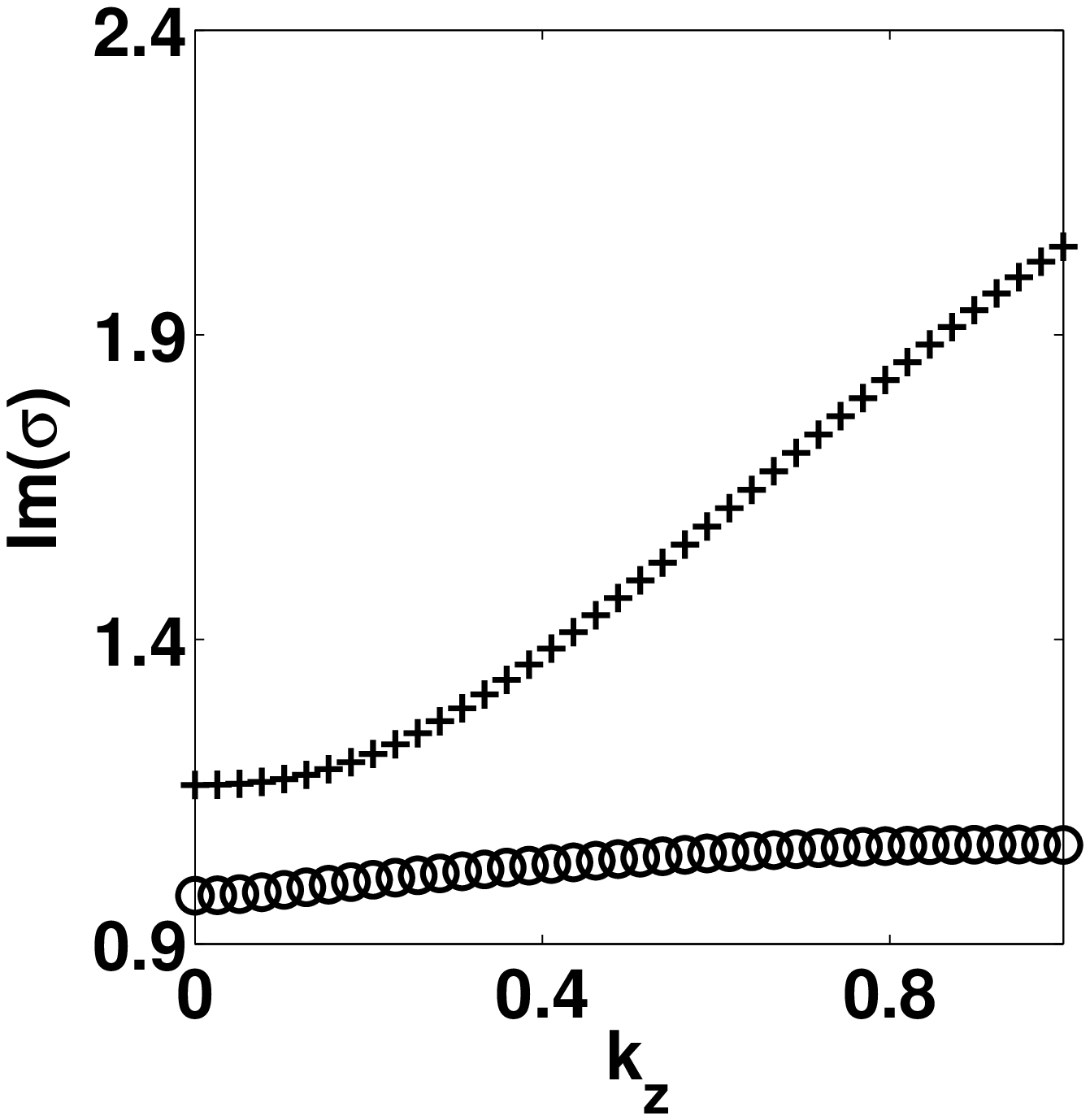}
\end{tabular}
\caption{(a) Real and (b) imaginary part 
of the growth rate $\sigma(k_z)$ as a function of
the wavenumber for the parameter value $a=0.44$, $b=0.01$,
$\epsilon=0.025$. The meander mode, ($\circ$) is stable and
the growth
rate decreases with the wave number on the meander band.
 The modes of the rotation band ($\bullet$), are also 
stable. The mode of  the translation band ($+$), are 
unstable for finite values of
$k_z$.
\label{lstruns.fig}}
\end{center}
\end{figure}

The translation bands can have unstable modes for $k_z\neq 0$ while the
2D spirals are stable. An example of this phenomenon is shown on 
Fig.~\ref{lstruns.fig} for $a=0.44,\ b=0.01$ and $\epsilon=0.025$. A qualitatively similar spectrum
is obtained for all points in Fig.~\ref{phasd.fig}
denoted by filled dots ($\bullet$)
near
the large core spiral existence boundary $\partial R$.
As seen on Fig.~\ref{lstruns.fig}, the instability takes place for
small $k_z$ as soon as $k_z$ is non-zero. It corresponds to the "negative
line tension" instability of ref.~\cite{bik3d}. We show below that the curvature
of the translation modes at $k_z=0$ is given by the spiral drift coefficients
in an external field. So, this translation band instability is directly
related to the fact that 2D spiral drift opposite to an applied external field
in this parameter region.
\subsubsection{Long wavelength stability 
and 2d spiral drift in an external field}
\label{stabdrift.sec}
As recalled and shown in detail in appendix \ref{drift.app}, a small applied
external field $\boldmath{E}$ induces a drift of the spiral rotation center
at a velocity $\boldmath{v}$ such that
\begin{equation}
\boldmath{v}=\alpha_{\parallel}\boldmath{E}+\alpha_{\perp}
\boldmath{\omega_1\times E}
\label{2ddrift}
\end{equation}
where $\boldmath{\omega_1}$
is the spiral rotation vector.
It has previously been noted \cite{hk2} that a weak scroll wave curvature
acts as an external field and therefore that a straight 
scroll wave is unstable if $\alpha_{\parallel} < 0$ since a small curvature
tends to grow. More precisely, 
the small $k_z$ behavior of the two translation bands
is given by
\begin{equation} 
\sigma_{\pm}(k_z)=\pm i \omega_1 +(-\alpha_{\parallel}
\pm i \alpha_{\perp})\ k_z^2 +O( k_z^4)
\label{drifttrans}
\end{equation}
Eq.~(\ref{drifttrans}) is simply derived by a first-order perturbative
calculation as follows. The
linear eigenvalue problem (\ref{eqlin1},~\ref{eqlin2}) reads
\begin{equation}
\sigma(k_z)  \left( \begin{array}{c} u_1\\v_1\end{array}\right) =
-k_z^2 \left( \begin{array}{c} u_1\\0\end{array}\right) +
{\mathcal{L}}_{k_z=0}
        \left( \begin{array}{c} u_1\\v_1\end{array}\right).
\label{pertkz}
\end{equation}
For small $k_z$, the modes of the translation bands can be obtained
by perturbation around the known translation modes at $k_z=0$ (Eq.~\ref{tm}).
 For definiteness, we  consider
the upper band (which start at $\sigma(k_z=0)=+i\omega_1$) and write
$\sigma(k_z)=i\omega_1 +\delta\sigma$ where $\delta\sigma\ll 1$ is the
sought perturbative correction,
\begin{equation}
\left( \begin{array}{c} u_1\\v_1\end{array}\right)=
\left( \begin{array}{c} u_t\\v_t\end{array}\right)+
\left( \begin{array}{c} \delta u_1\\\delta v_1\end{array}\right)
\label{notpert}
\end{equation}
Substitution in Eq.~(\ref{pertkz}) gives
\begin{equation}
\delta\sigma \left( \begin{array}{c} u_t\\v_t\end{array}\right)+
i\omega_1 \left( \begin{array}{c} \delta u_1\\\delta v_1\end{array}\right)=
-k_z^2 \left( \begin{array}{c} u_t\\0\end{array}\right)+
{\mathcal{L}}_{k_z=0} \left( \begin{array}{c} \delta u_1\\\delta v_1
\end{array}\right)
\label{pertkz2}
\end{equation}
The first-order expression of $\delta\sigma$ is obtained in a usual way by
taking the scalar product of Eq.~(\ref{pertkz2}) with the left eigenvector
$(\tilde{u}_t,\tilde{v}_t)$ of
${\mathcal{L}}_{k_z=0}$ for the eigenvalue $i\omega_1$ (section\ref{lefteigen})
\begin{equation}
\delta\sigma=-k_z^2 \frac{\langle\tilde{u}_t,u_t\rangle}{
\langle\tilde{u}_t,u_t\rangle+
\langle\tilde{v}_t,v_t\rangle}
\label{fop}
\end{equation}
Eq.~ (\ref{fop}) is equivalent to the announced formula (\ref{drifttrans}) since
the matrix coefficients on its r.h.s also gives the spiral
drift
coefficients, as shown in appendix \ref{drift.app} (see Eq.~(\ref{driftcoef})).

In Table \ref{curv.tab}, the spiral drift coefficients $\alpha_{\parallel}$ and $\alpha_{\perp}$ are compared to the results of independent 
computations of
the curvature
of $\sigma(k_z)$ at $k_z=0$ for
the translation bands, from diagonalisations of ${\mathcal{L}}_{k_z}$ at different 
values of $a$.
The good agreement between these results is a check both of
the analytic formula (\ref{drifttrans}) and 
of  our numerics.
\begin{table}
\begin{center}
\begin{tabular}{|c|c|c|c|c|}
\hline
$a$ & $\omega_1$ & $\sigma_t''(k_z=0)/2$ &
$\sigma_m''(k_z=0)/2$ & $\alpha_\parallel -i \alpha_\perp$\\
\hline 
$0.44$ &  $1.16$ &  $1.9  +   0.82 i$  &  $-1.6 +
0.78i$ & $-1.97  -0.84i$  \\ 
\hline 
$0.48$ &  $1.38$ &  $3.2  +   0.44i$   &  $-3.7  +   
1.10 i$ & $-3.0  -0.49i$ \\
\hline  
$0.67$ & $1.76$ & $-2.14 +0.85i$ & $1.61 +0.25i$ & $2.2  -0.9i$   \\
\hline
$0.7$  &  $1.78$ & $-1.63 +   0.83 i$  &  $ 1.04  +
0.21 i$ & $1.62  -0.83i$\\
\hline
$0.8$  & $1.81$ &  $-0.87  + 0.70i$    &   $0.08   
-0.06i$   & $0.854  -0.71i$   
\\
\hline
$0.9$  &  $1.77$  & $-0.65  +   0.61i$ &   $-0.25  -0.089i$ & $0.66 -0.61i$\\
\hline
\end{tabular}
\caption{ 
The scroll wave pulsation $\omega_1$, half the second derivative 
$\sigma_t''(k_z=0)/2 $ of  the translation band at $k_z=0$
(with $\sigma_t(0)=i\omega_1$), half the second derivative
$\sigma_m''(k_z=0)/2 $ of the meander band at $k_z=0$ (with
$Im(\sigma_m(0))>0$) 
 and the drift coefficients of the
2D-spiral in an electric field $\alpha_\parallel-i\alpha_\perp$ for
$b=0.01$, $\epsilon=0.025$ and
different values of $a$.
\label{curv.tab}
}
\end{center}
\end{table}
To recapitulate, the translation band instability is found to be 
a long wavelength
instability (i.e. the band of unstable wavelength starts at $k_z=0$) which
is present in the whole domain of parameters where a 2D-spiral drifts opposite
(given our sign convention in Eq.~\ref{eqE1}) to the applied field.
\subsubsection{Nonlinear evolution of the instability}
\begin{figure}
\begin{center}
\includegraphics[width=4.cm,height=4.cm]{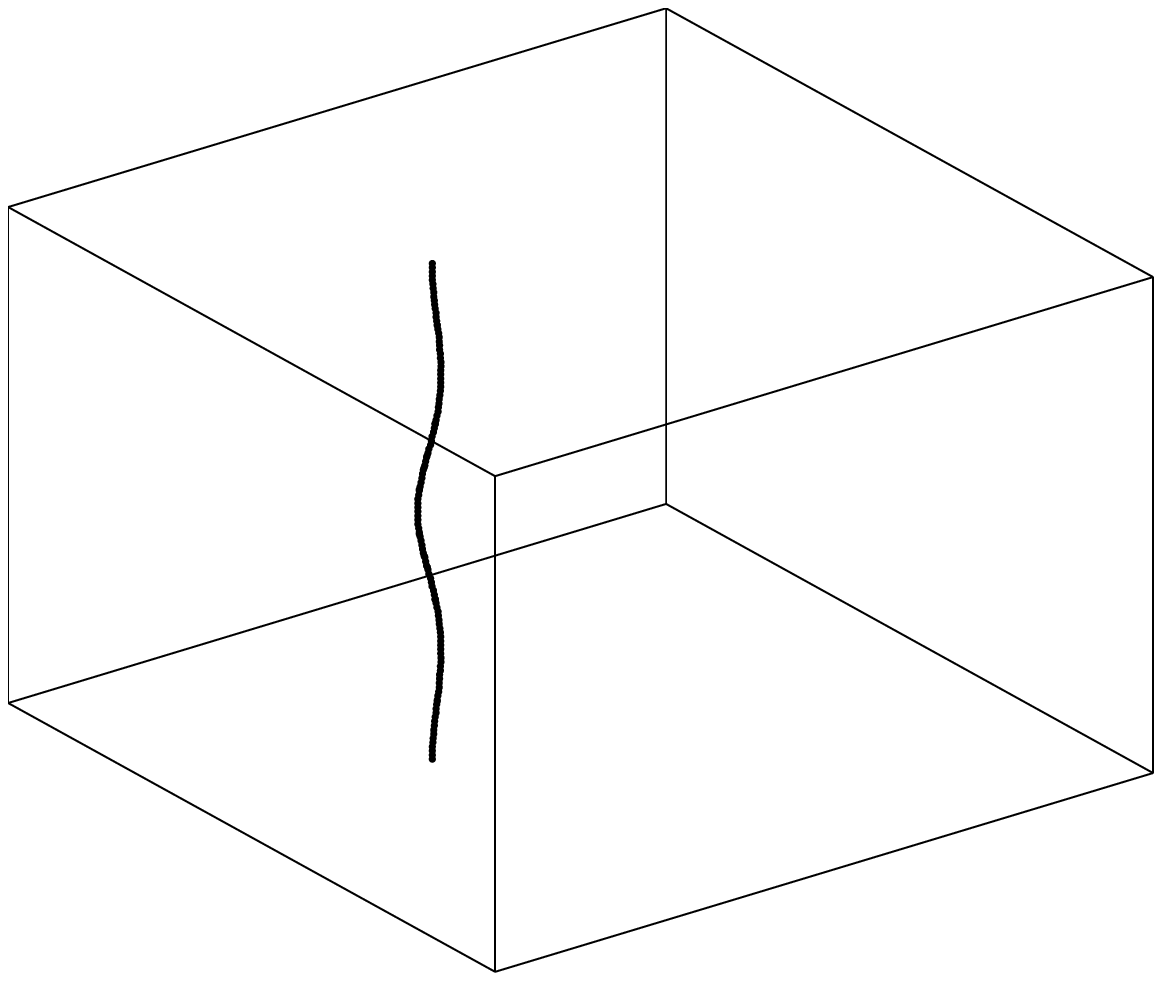}\hspace{.5cm} 
\includegraphics[width=4.cm,height=4.cm]{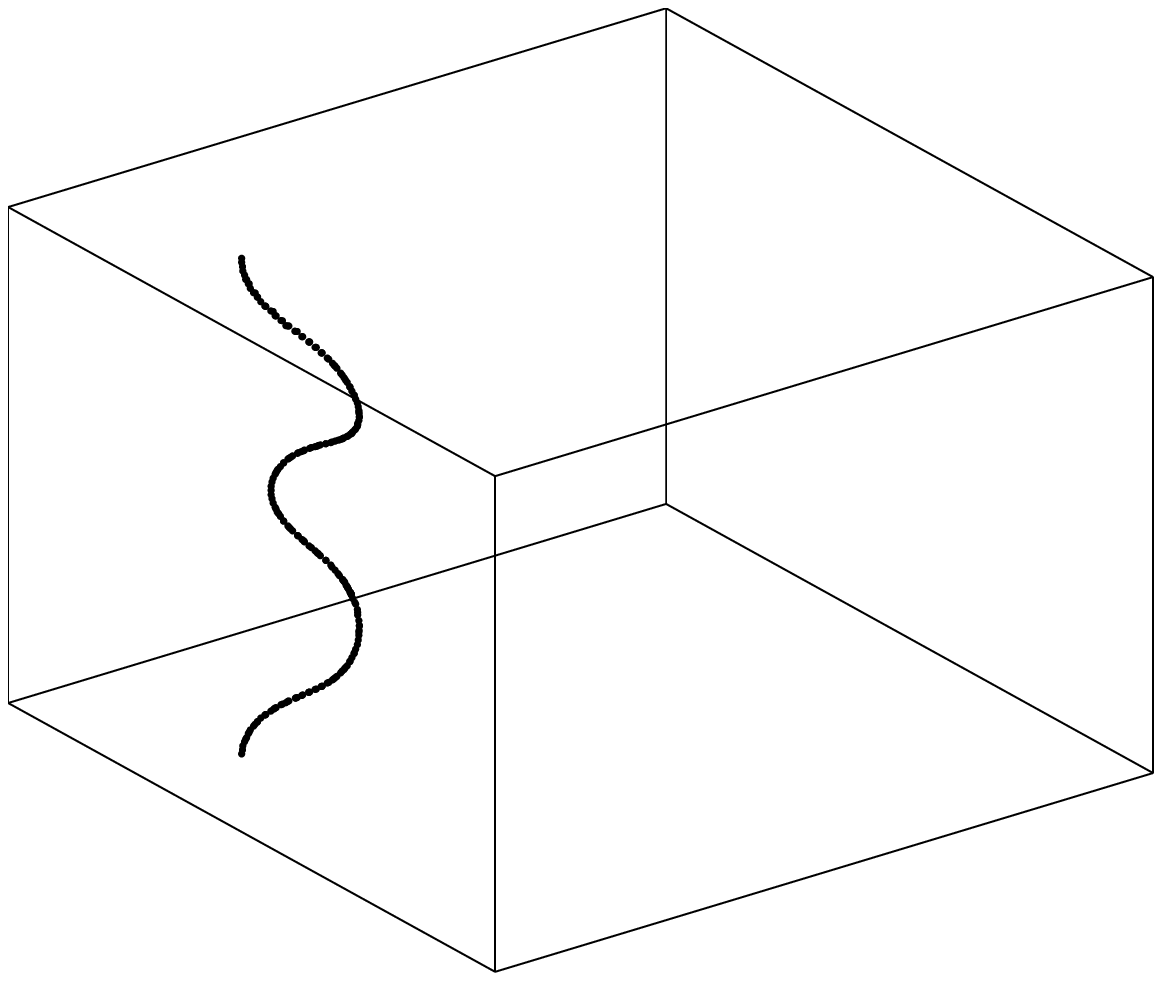}
\\
\includegraphics[width=4.cm,height=4.cm]{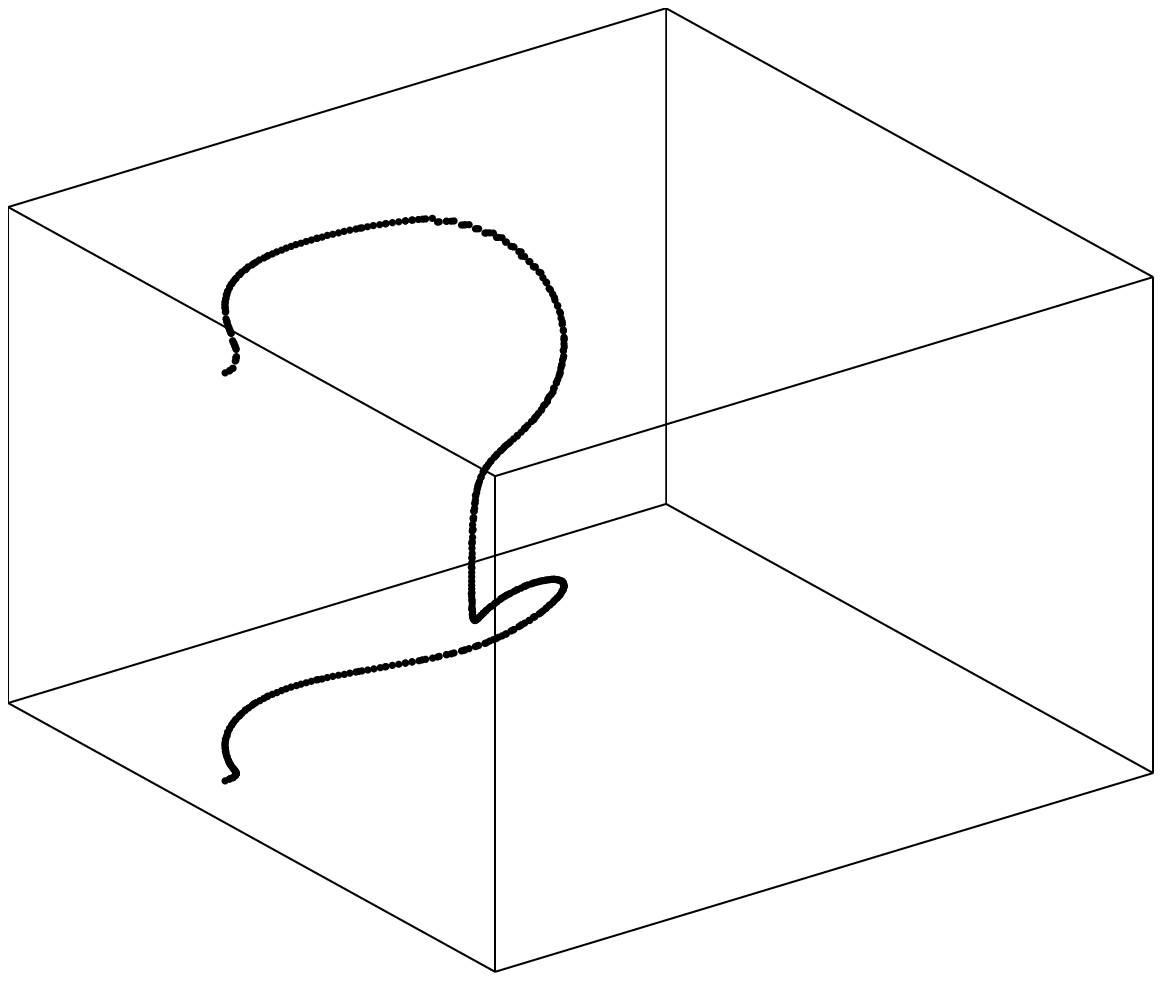}\hspace{.5cm}  
\includegraphics[width=4.cm,height=4.cm]{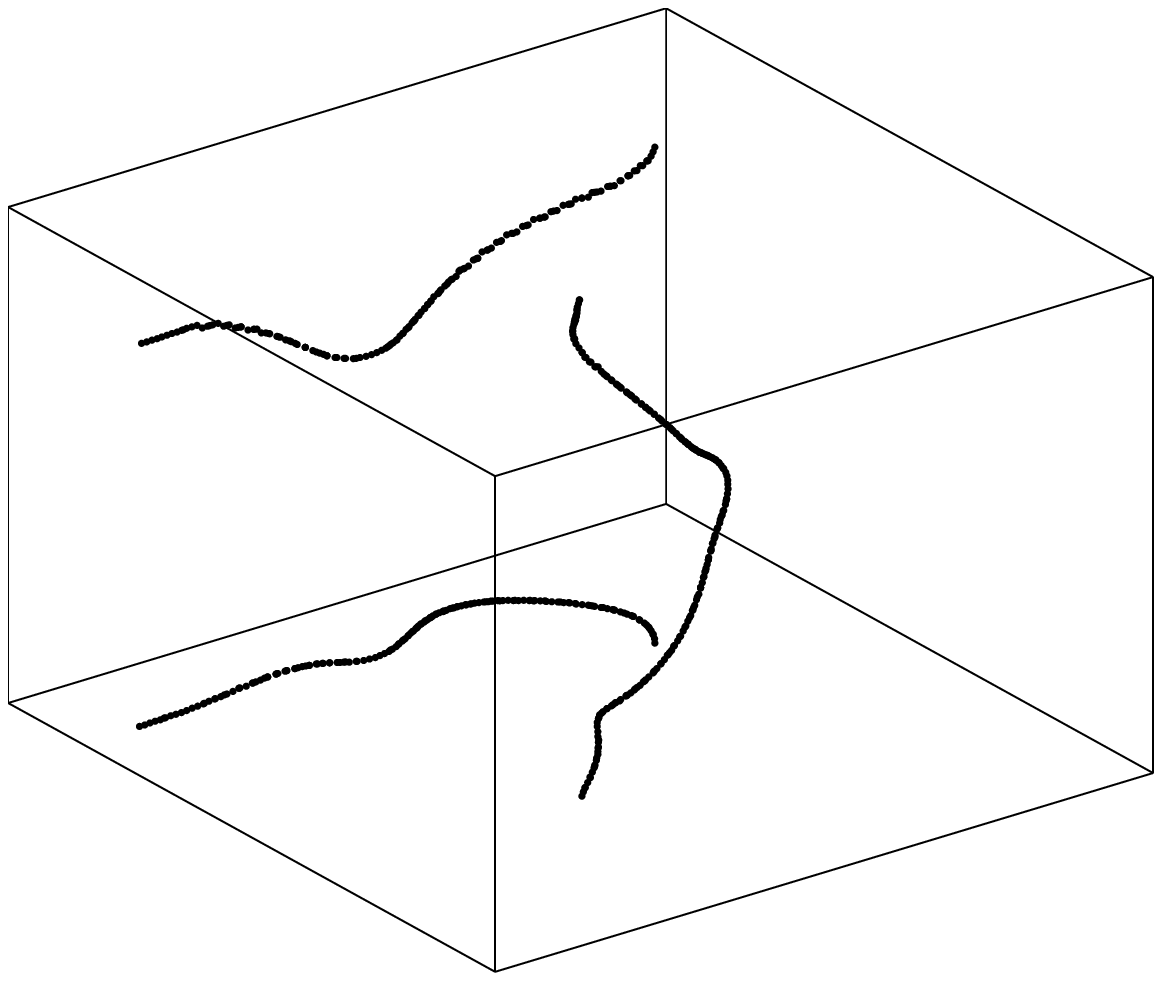}
\end{center}
\caption{Instantaneous filament evolution starting from
a slightly perturbed straight scroll for
equally spaced times  ($t=25.,\ t=50.,\ t=75.\mbox{ and } t=100.$)
during a simulation in a simulation box of size $(128\times 128
\times 120)$ with a space discretization step $dx=0.2$
 using periodic boundary conditions. The parameters are
$a=0.44,\ b=0.01\mbox{ and } \epsilon=0.025$ and correspond to the
linear spectrum shown in Fig.~\ref{lstruns.fig}.   
\label{coubperiononlin}}
\end{figure}

The nonlinear fate of this translation band instability was studied by
direct dynamical simulations of Eq.~(\ref{eq1},~\ref{eq2}). In the parameter
regime
of Fig.~\ref{lstruns.fig} when modes of the translation bands are unstable
for finite values of $k_z$, an initially straight scroll wave was observed to
be unstable provided that the simulation box was large enough to accommodate an
unstable mode. In agreement with previous observations \cite{bik3d},
the filament was observed to increasingly depart from its straight initial
configuration and its length was observed to grow in the simulation box.
When the filament eventually
collided with the boundaries of the simulation box, it split into
two filaments.
This repeated again and no
restabilisation was observed. A typical evolution is shown in 
Fig.~\ref{coubperiononlin}.
The minimum simulation box size which allowed the instability development
closely agreed with the results of
the linear stability analysis. For instance, in the case $a=0.44,\
b=0.01,\mbox{ and }\epsilon=0.025$, the maximal $k_z$ of the unstable band
is obtained to be $k_z=0.84$ via the linear stability analysis
whereas the direct numerical
simulations show a minimal simulation box height corresponding
to $k_z=0.81$
\footnote{For the observed largest unstable
$k_z$, numerical simulations were not carried long enough
to observe the full nonlinear development of the instability.
However, it was observed with a box
height increase of a single space step.}.

The choice of boundary conditions on the simulation
box top and bottom faces influences the minimum box height for the instability
development (but, apart
from that, was not observed to qualitatively modify the instability
nonlinear development). This critical size was found to be twice bigger
for periodic periodic boundary condition than for no-flux boundary
conditions which can accommodate linear modes of wavelength twice as long
as the box height.

\subsection{The third dimension-induced meander instability}
In the region where a 2D-spiral drifts toward an applied field (i.e.
$\alpha_{\parallel}>0$), the modes of an untwisted scroll wave translation
bands are stable. As pointed in \cite{armit}, an untwisted
scroll wave can nonetheless
be unstable in a parameter region where a 2D-spiral is stable. This happens
when the meander bands are destabilized by deformation in the $z$-direction
as we study below.
\subsubsection{Linear analysis}
\begin{figure}[h]
\begin{center}
\begin{tabular}{ccc}
(a)&&(b)\\[.1cm]
\includegraphics[height=4.cm,height=4.1cm]{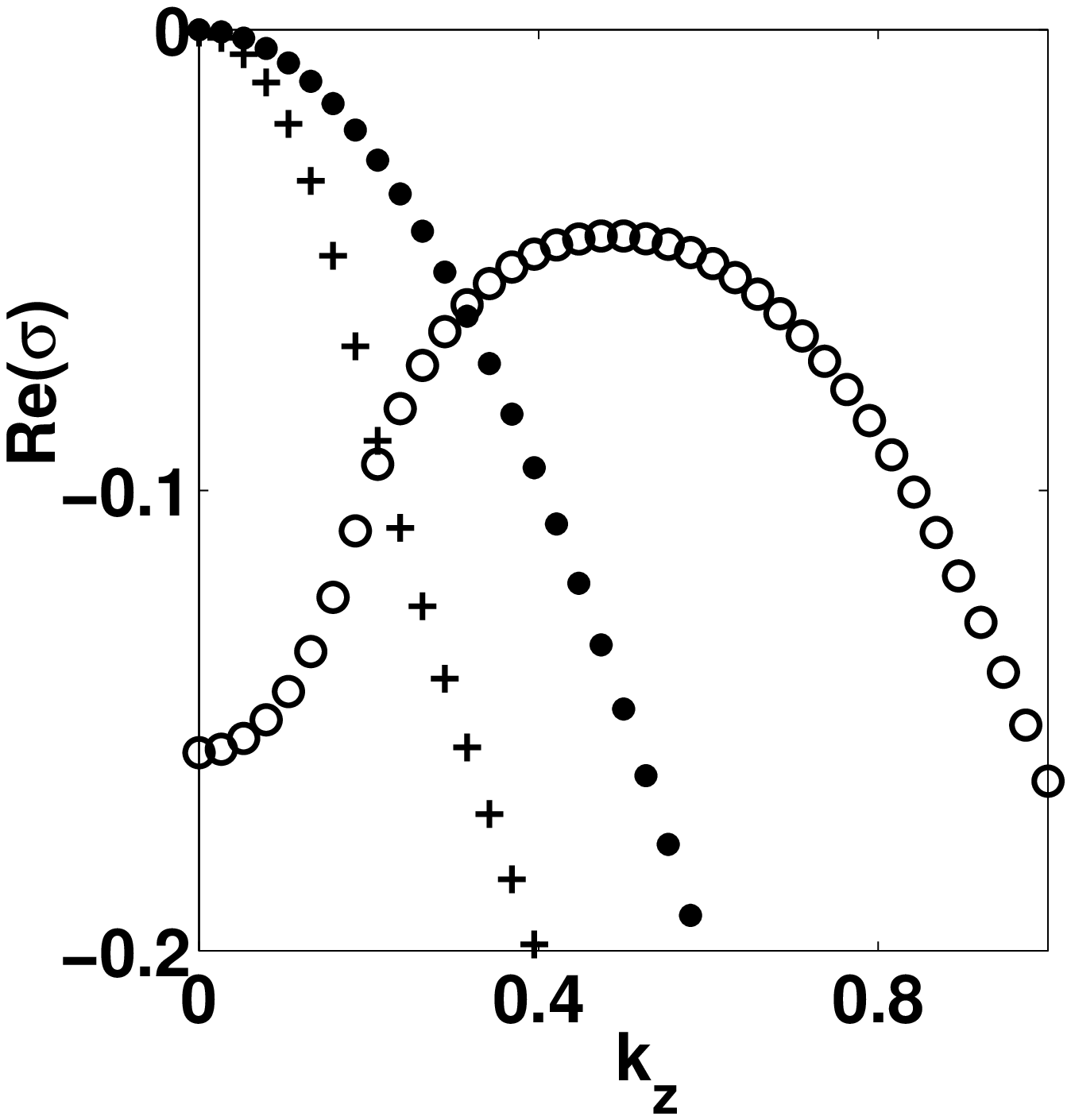}&&
\includegraphics[height=4.cm,height=4.cm]{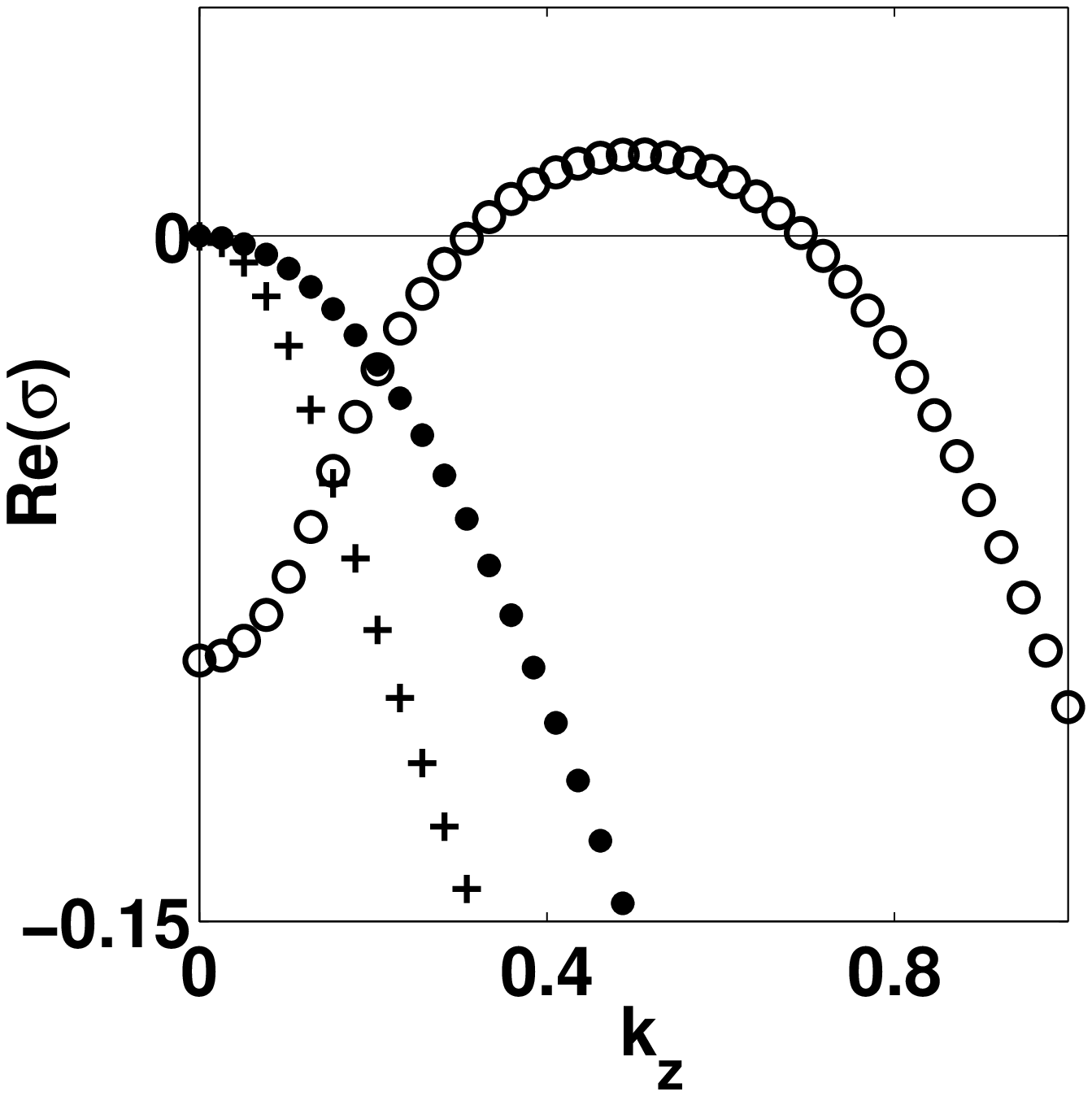}
\end{tabular}
\caption{The growth rate $Re[\sigma(k_z)]$ as a function of
the wavenumber for the parameter value (a) $a=0.7$ and (b) $a=0.67$(b) 
The other parameter are $b=0.01$ and
$\epsilon=0.025$. The meander mode, ($\circ$), is stable for $k_z=0$ 
both  in cases (a) and (b). The growth
rate increases on the meander band with $k_z$ . 
In case (a), it always remains negative and there is no instability.
In case (b), it becomes positive for $k_z$ higher than 0.30  and lower
than 0.69 showing the finite-$k_z$ instability of the steady scroll wave. 
The rotation ($\bullet$) and
 translation bands ($+$) are stable. In the phase diagram
Fig.\ref{phasd.fig} the points with a spectrum similar to the (b) case are
represented by crosses $(\times)$. 
\label{lsmeanuns.fig}}
\end{center}

\end{figure}

\begin{figure}[h]
\begin{center}

\begin{tabular}{ccc}
(a)& & (b)
\\[.1cm]
\includegraphics[height=4.2cm,width=4.2cm]{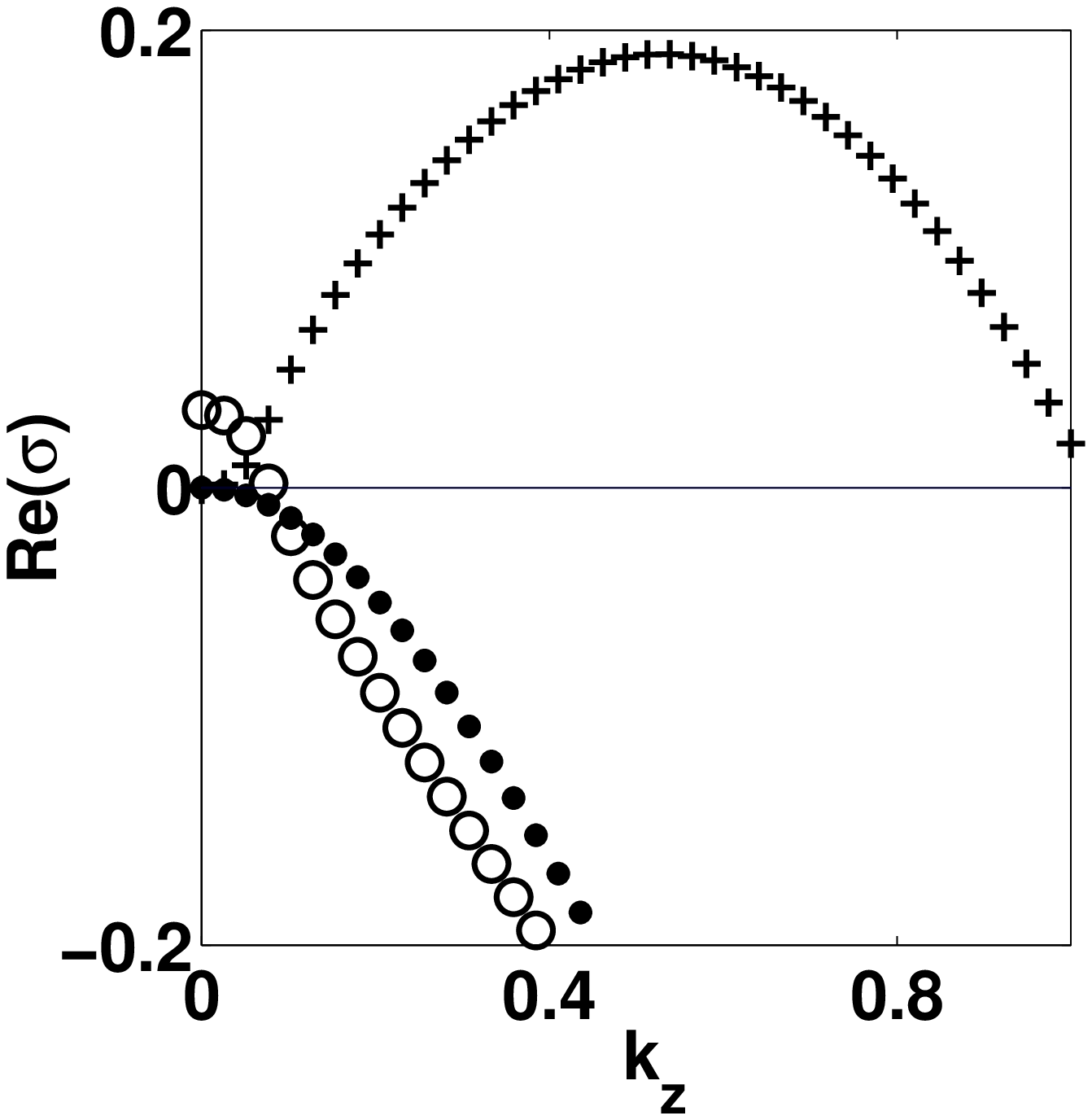}
& &
\includegraphics[height=4.2cm,width=4.2cm]{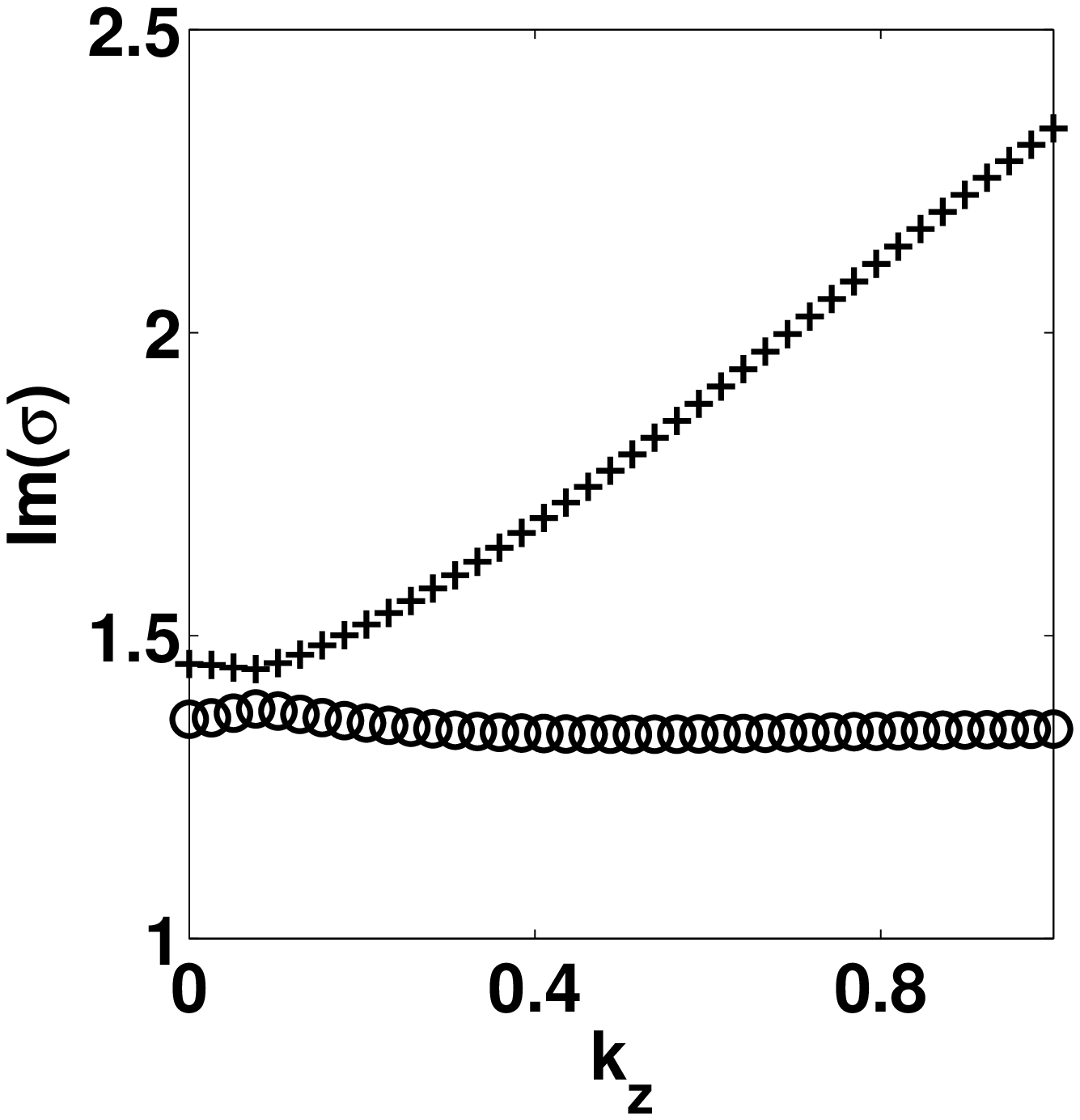}
\\
 (c)& & (d)
\\[.2cm]
\includegraphics[height=4.2cm,width=4.2cm]{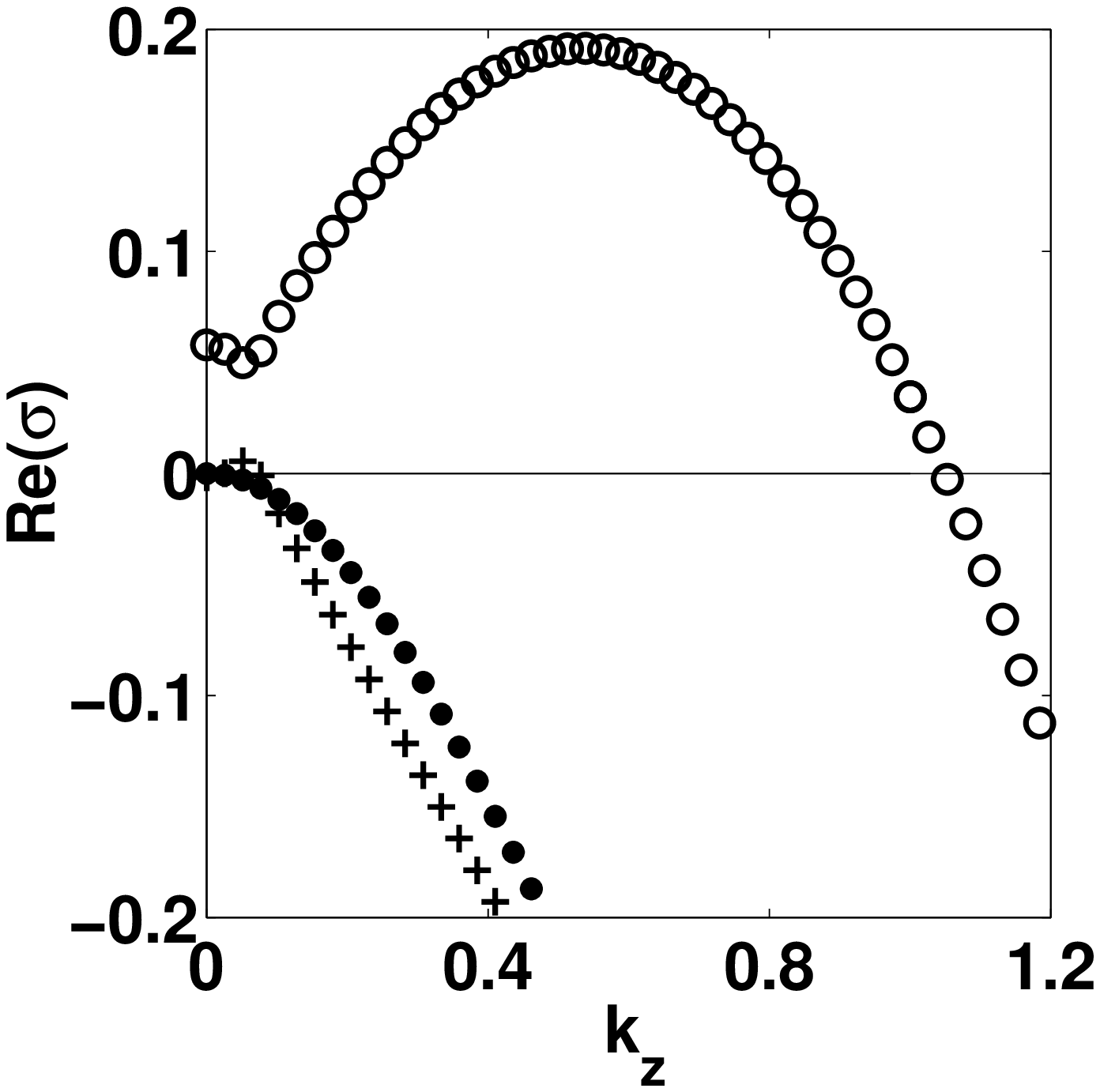}
& &
\includegraphics[height=4.2cm,width=4.2cm]{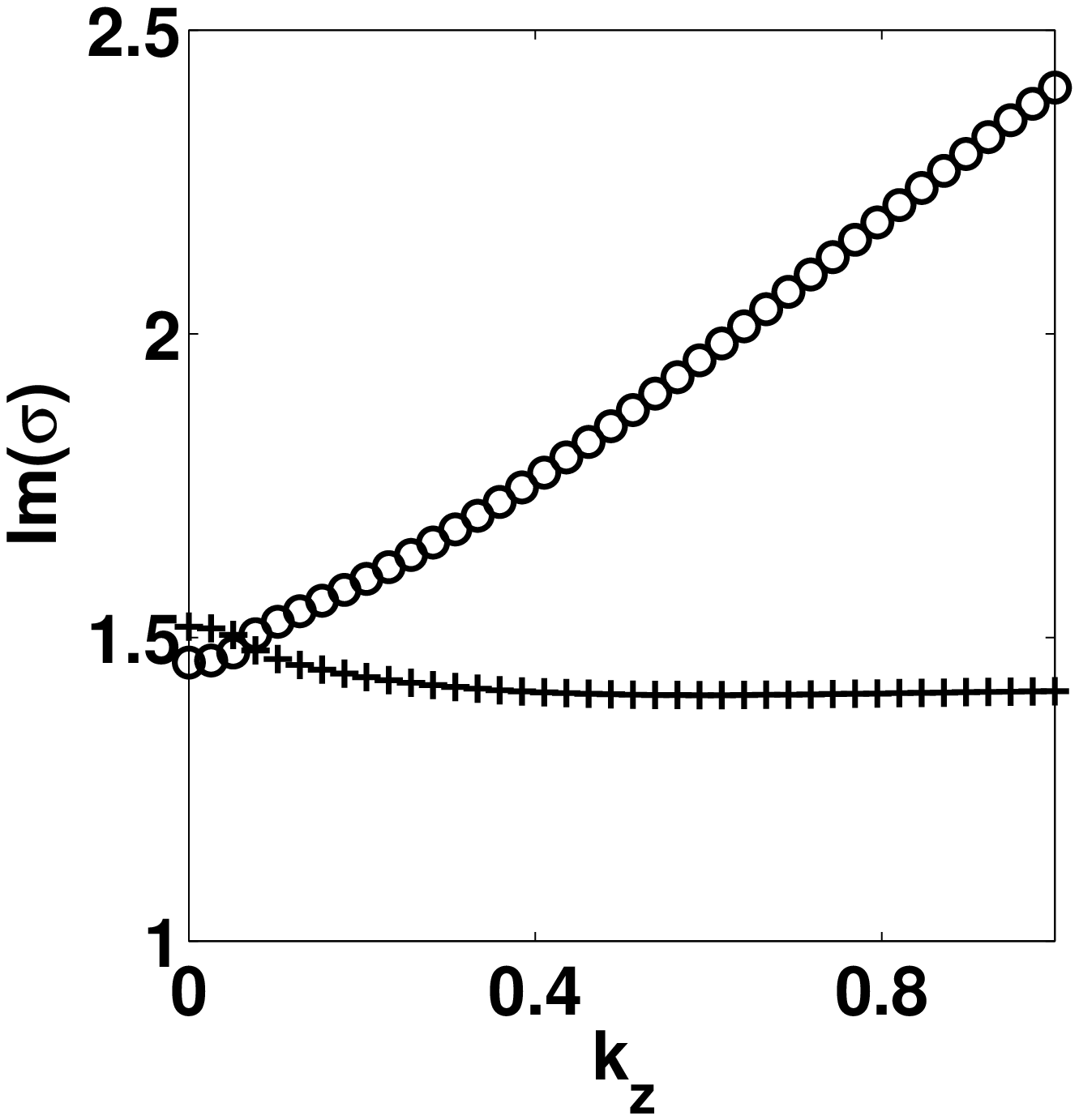}
\end{tabular}

\caption{Real and  imaginary part 
of the growth rate $\sigma(k_z)$ as a function of
the wavenumber for $b=0.01$,
$\epsilon=0.025$. (a),\,(b): $a=0.5$; (c),\,(d): 
$a=0.52$.
In both cases, the real part decreases on the meander band and increases on
the translation band with $k_z$ for small values of $k_z$. In the $a=0.52$ case,
 hybridization of the
eigenvectors of the two bands exchange  these trends
at slightly 
higher values of $k_z$ i.e. the real part of the meander band increases
whereas the real part of the translation band decreases. 
In the phase diagram of Fig. \ref{phasd.fig}, the points
where the linear spectrum is similar to one of these cases (with both the
meander and translation bands unstable at small $k_z$) are represented by
represented by circles (o).
\label{lsmeantruns.fig}}
\end{center}
\end{figure}

This induction of the meander instability by three-dimensional effects is
shown in Fig.~\ref{lsmeanuns.fig}. For the parameters of 
Fig.~\ref{lsmeanuns.fig}a, all modes have  negative real parts and the
scroll wave is stable. However, one sees that the real part of the modes
on the 
meander band  starts by increasing as $k_z$ increases from zero.
For the parameters of Fig.~\ref{lsmeanuns.fig}b which stand closer
to the 2D meander boundary, a finite band of modes with $k_z\neq 0$ has
acquired
a positive real part while the real part of 2D spiral meander mode at $k_z=0$ is still 
negative . Thus, for these parameter values close to the ''small core''
 side
of the 2D meander instability boundary, a 3D scroll wave is unstable to meander
while the steadily rotating 2D spiral is still stable as pointed out in
\cite{armit}.
On the ''large core'' side of the meander 
instability boundary, three-dimensional
modulations have on the contrary a stabilizing effect 
on the meander instability and the 
2D and 3D meander instability thresholds coincide as shown on 
Fig~\ref{lsmeantruns.fig} (note however  that the translation band instability
of section \ref{tbi.sec} renders the scroll wave unstable for these parameters).
The fact that the curvature of the translation and meander bands are
of opposite sign in Fig.~\ref{lstruns.fig} and Fig.~\ref{lsmeanuns.fig} 
may lead to think that the finite $k_z$ behavior
of the meander bands is also related to 2D spiral drift (as indeed
proposed in \cite{armit}). However, a quantitative computation shows
no simple relation between  the
meander bands curvature at $k_z=0$ and the spiral drift coefficient as
reported in table \ref{curv.tab}. Moreover, even at the qualitative level,
there is no general validity to the
opposite sign rule between the translation and meander bands curvature
at $k_z=0$, as shown by the data of Fig.~\ref{lsuts.fig}.

It is of some interest to see how the spectrum of Fig.~\ref{lstruns.fig}
is transformed into the spectrum of  Fig.~\ref{lsmeanuns.fig} as one traverses
the 2D meander unstable region. This happens through hybridization between
the translation and meander bands as illustrated in Fig.~\ref{lsmeantruns.fig}.

\subsubsection{Restabilized bifurcated states}
We performed direct numerical simulations to examine the nonlinear
development of the meander bands instability at $k_z\neq 0$
and to characterize
the restabilized nonlinear states. For boxes of medium size in the 
$z$-direction as used in our numerical simulations, the restabilized
nonlinear states strongly depend on the top and bottom boundary
conditions. Two different ones were implemented
(no flux boundary conditions were
always implemented on the side boundaries). We used either periodic boundary
conditions which permit the development of a single pair of 
unstable modes and are
simpler to analyze or no-flux boundary conditions since they are clearly
more relevant in an experimental context. We describe the results in turn.

{\em Periodic boundary conditions}.

Direct numerical simulations were performed in the parameter regime of
Fig.~ \ref{lsmeanuns.fig}.
Three types of initial conditions were used which all led to a
stationary  restabilized scroll wave after a transient regime.

For the first one, the perturbation of the steady state was
chosen to contain the unstable wavelengths $\pm k_z$.
The initial fields were chosen
in the form $u(x,y,z)=u_{2D}(x,y)(1+\alpha \cos(k_z
z)), v(x,y,z)=v_{2D}(x,y)(1+\alpha \sin(k_z
z))$ with
the magnitude $\alpha$
of the 3D perturbation  being of order $10^{-2}$ ($u_{2D}$
and $v_{2D}$ correspond to a steady spiral wave which is stable in 2D in
this regime). This perturbation transforms the straight unperturbed 
instantaneous filament into an ''helix'' of small elliptical 
cross section
and pitch $2\pi/k_z$.
When the wavelength $k_z$ corresponded to an
unstable mode, this helix was observed to grow and to reach a helical
restabilized state of periodicity $k_z$.

A slightly more complicated time development was observed with an initial
perturbation of the straight scroll wave in the form 
$u(x,y,z)=u_{2D}(x,y)(1+\alpha \cos(k_z
z)), v(x,y,z)=v_{2D}(x,y)(1+\alpha \cos(k_z z))$. This gives the 
instantaneous filament $(x_f(z),y_f(z))$
a planar ''zig-zag'' shape of small amplitude, 
$x(z)=x_1 \cos(k_z z),\ y(z)=y_1 \cos(k_z z)$. The zig-zag perturbation
was first observed to grow before turning into an helical
restabilized state of periodicity $k_z$. 
This type of time development is expected on general ground
in system where left and right progressive wave compete
(see e.g. \cite{sfbegrohu}).

Finally, competition between several different unstable modes (up to
$4$  different
$k_z$) was examined with initial conditions such as
$u(x,y,z)= u_{2D}(x,y) [1+\alpha\exp (-(z-z_o)^2/L_c)],
v(x,y,z)=v_{2D}(x,y)[1+\beta\exp (-(z-z_o)^2/L_c)] $. 
(typical values are $\|(\alpha,\beta)\|_2=0.01$ and $L_c$ about a few
tens of space steps). 
In that case, it was generally observed
that 
the most unstable wavelength compatible with the box height was selected.
At a qualitative level, the transient regime was found to
mainly depend of the planar or non
planar character of the initial perturbed filament~: When it was
planar (case $\beta=\alpha$) a zig-zag filament first grew
before taking  an helical shape. When it
was non planar, $ \beta \ne \alpha$ then an helical filament grew directly.

\begin{figure}[!h]
\begin{center}

\begin{tabular}{ccc}
(a)& & (b)\\
\includegraphics[height=5.cm,width=3.cm]{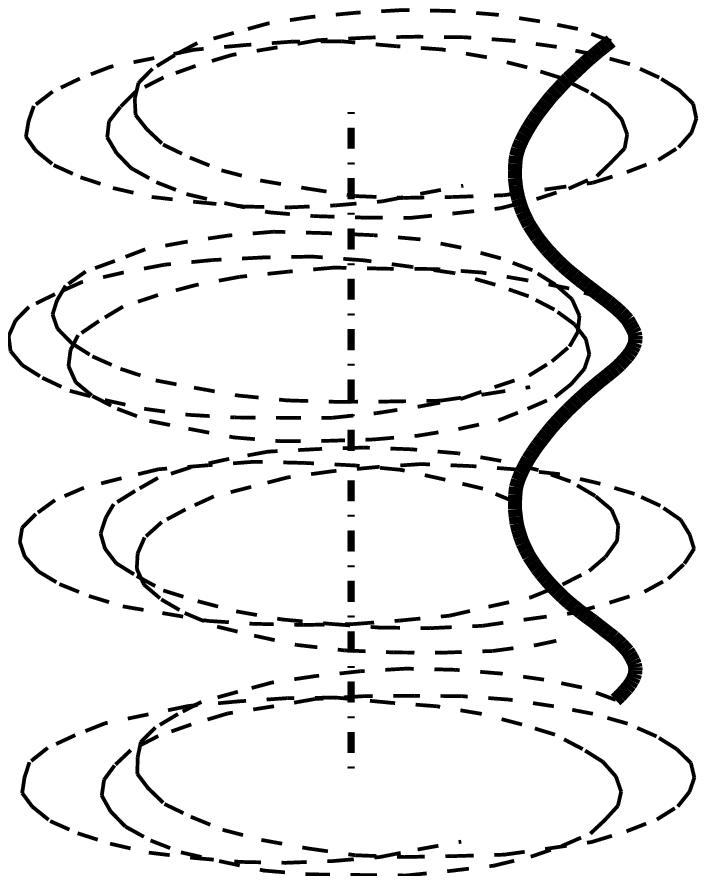}&
&
\includegraphics[height=5.cm,width=3.cm]{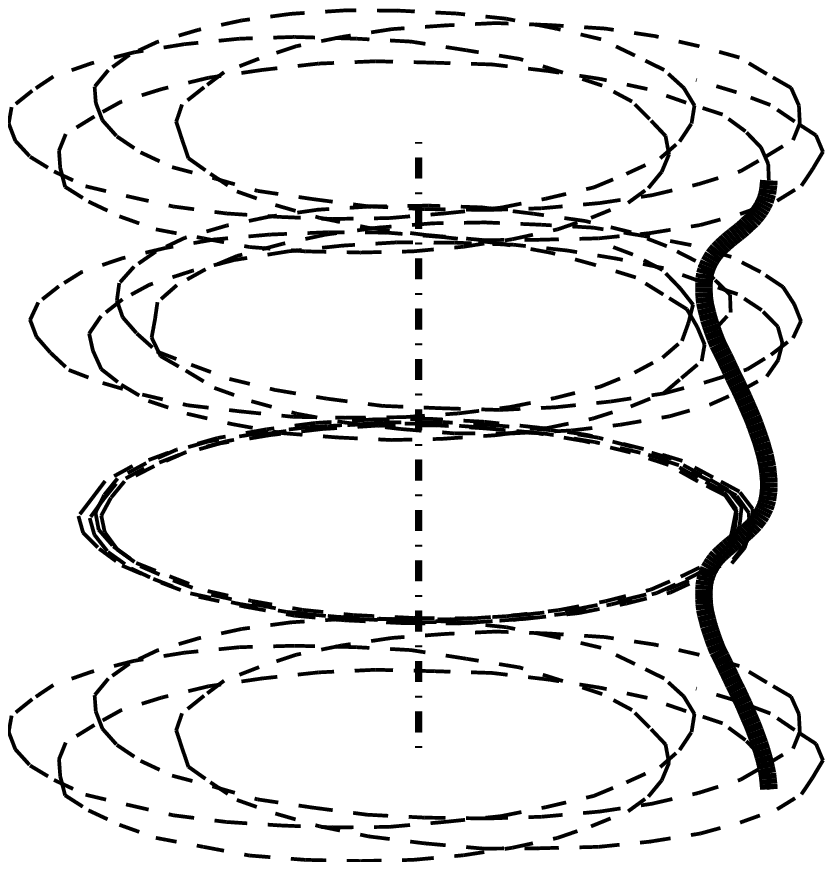}
\\
(c)& & (d)\\
\includegraphics[height=3.cm,width=3.cm]{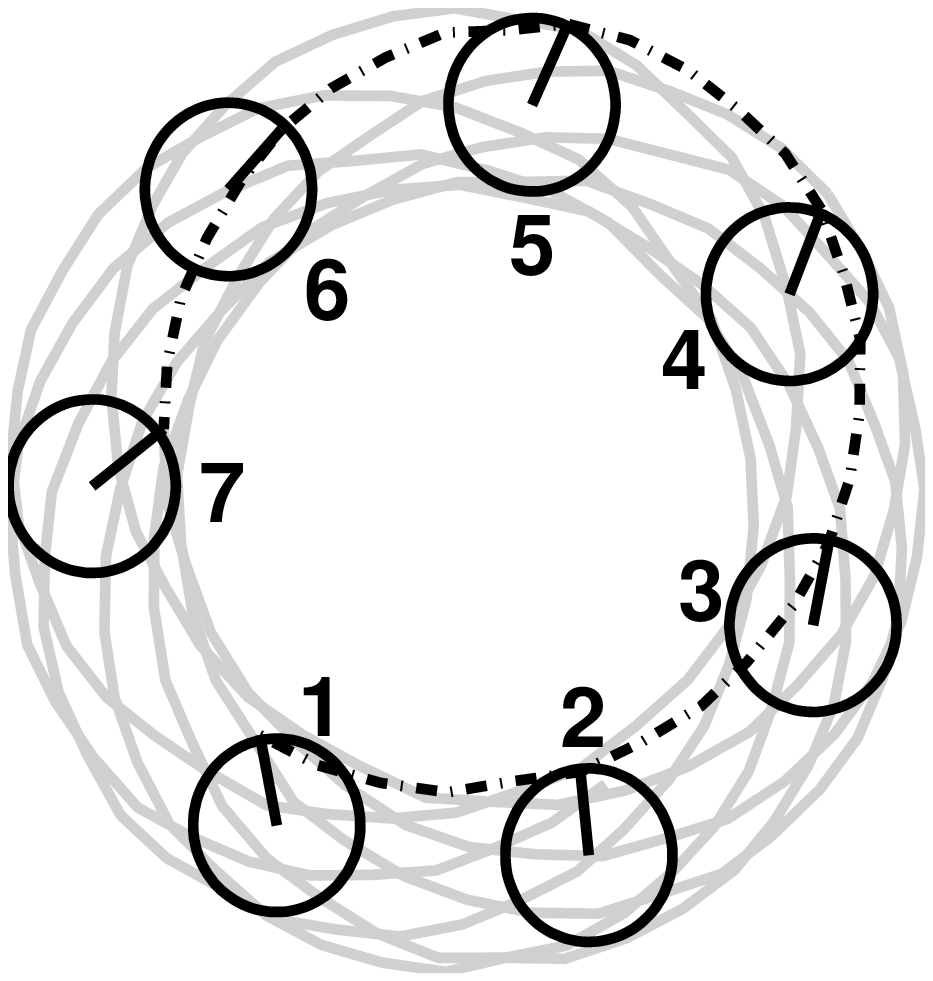}&
&
\includegraphics[height=3.cm,width=3.cm]{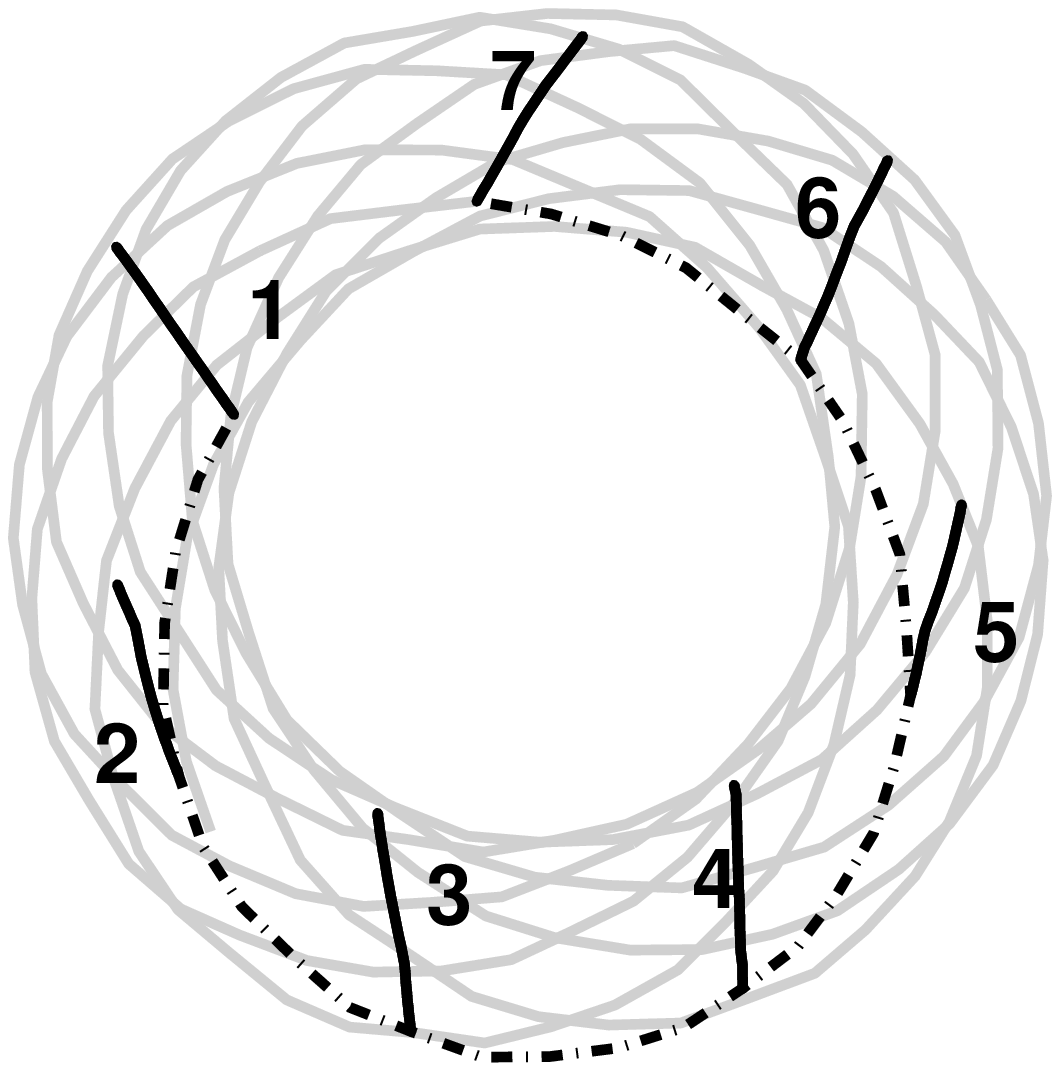}
\end{tabular}

\caption{(a): view of the restabilized state in the weakly non linear
regime. The parameters are~: $a=0.684,\ b=0.01\mbox{ and
}\epsilon=0.025$, the size of the simulation box is
$(128\times\, 128)\times \, 130$ with $dx=0.2$ 
and periodic
boundary conditions are used. The dashed lines are the trajectories of the tip 
of
the spiral in four regularly spaced horizontal planes. The bold line
is the instantaneous helical filament. The dash-dotted line is the
axis around which  the axis of the helical filament  rotates.
(c):  black bold circles, projections of the
instantaneous filament on an horizontal plane at different times. The bold
radius in each circle shows the instantaneous filament point at
height $z=0$.
This point trajectory is also shown for several
periods of rotation (bold dashed-dotted line: evolution between
circles 1 and 7; thin gray line: evolution for some time afterwards).
The pulsation of the axis of the filament is equal to $\omega_1=1.744$, the
pulsation of the meander in the rotating frame is equal to
$\omega_2=-2.159$. These values are to be compared with results of the linear
stability analysis~: $\omega_1=1.766$ and $\omega_2=\pm 2.194$. 
(b): view of the restabilized state with the same parameters and
same simulation box using no flux boundary conditions. The instantaneous
filament has a zig-zag  shape and the amplitude of meander varies with
$z$. (d) : black bold lines, projections in an horizontal plane of the
instantaneous filament at different times. The trajectory of the
spiral tip in a plane where the amplitude of meander is maximal   
is also shown for several
periods of rotation (bold dashed-dotted line: evolution between
filaments 1 and 7; thin gray line: evolution for some time
afterwards). Its pulsation is equal to
$\omega_1= 1.745$ whereas the pulsation of the instantaneous filament
in the rotating
frame is equal to $\omega_2=-2.159$. The mean distance between the instantaneous
filament points and the central (dash-dotted) axis is
$R_=0.4930$ in the (a) (periodic boundary condition) figure and
$R=0.4944$ in the (b) no-flux case. The core radius of the corresponding 2D
spiral is
 $R_0=0.4833$.
\label{filshape.fig}}
\end{center}
\end{figure}

Close to the instability threshold,
the
shape of the restabilized instantaneous filament
is closely approximated by an helix of circular cross section,
as shown in
Fig.~\ref{filshape.fig}, and its motion  can be
portrayed in a way similar to the epicycle description of meander. 
The axis of the instantaneous filament helix
rotates around a fixed vertical axis at
the frequency of the steady two dimensional spiral.
In the rotating  frame where these two axes are fixed, the helical instantaneous
filament itself rotates with a frequency close to
 the imaginary part of the meander linear eigenmode 
(i.e. the difference between these two frequencies is
of order $10^{-2}$ in a strong
meander regime and of order $10^{-3}$ in a weak meander regime).
Thus, for periodic boundary conditions the meander amplitude is found to
be independent of  height ($z$) while the phase of the epicycle motion
varies linearly with $z$.

We performed two different systematic studies in order
to better characterize the nature of the 3D meander
bifurcation.
In the first one, we kept the size of the simulation
box constant and varied the excitability using $a$. In the
second one, we kept $a$ constant and varied the size of the
simulation box, that is the wavenumber of the initial perturbation.

At the linear level,
the results of these direct numerical simulations are
in close agreement with the predictions of the linear
stability analysis, both for the
the instability threshold $a_c$
(with an accuracy of order $10^{-3}$) and for
the unstable wavenumber range $[k_-(a),\ k_+(a)]$.

\begin{figure}
\begin{center}

\begin{tabular}{ccc}
(a)& (b)
\\[.1cm]
\includegraphics[height=4.2cm,width=4.2cm]{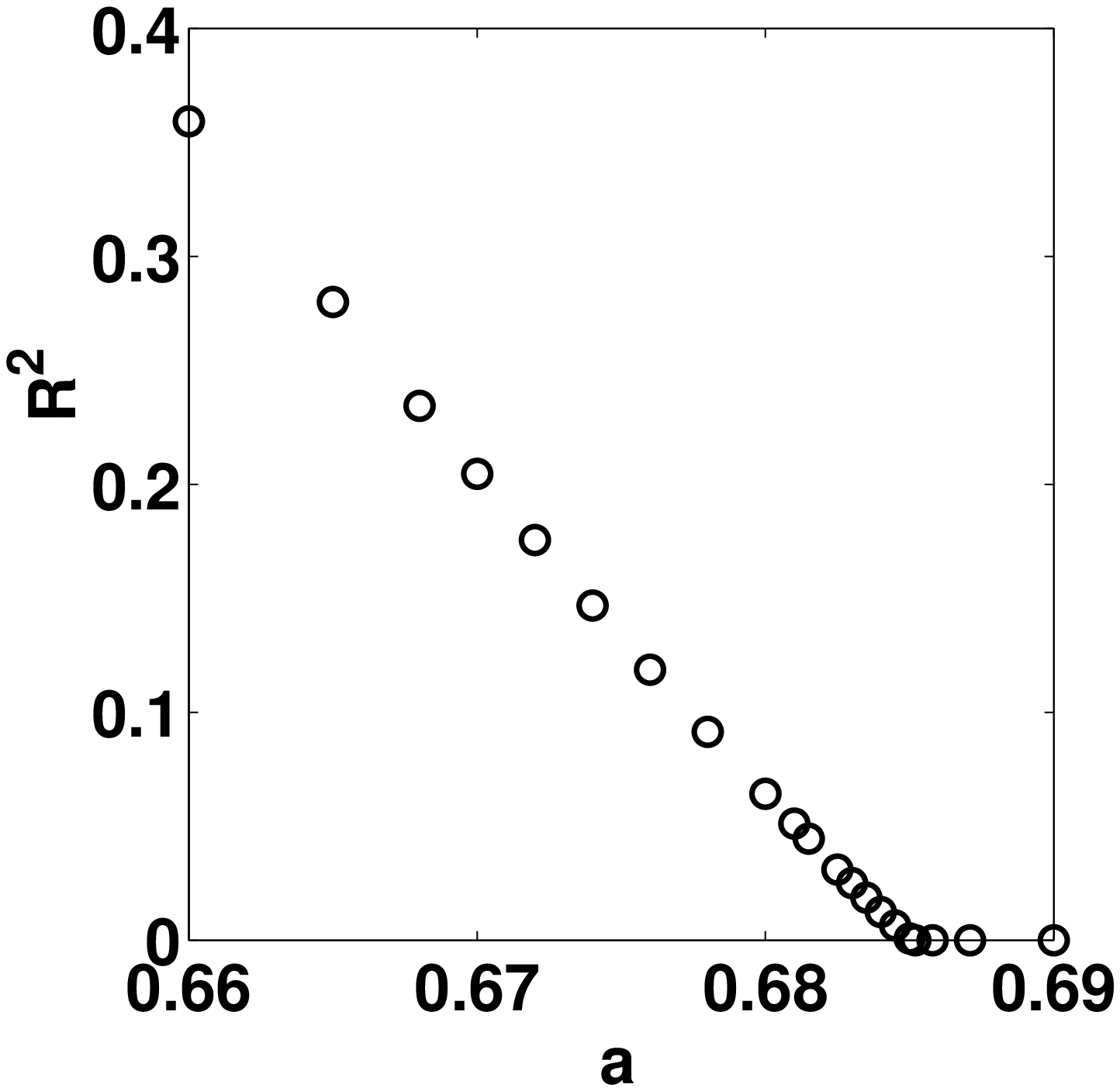}
&
\includegraphics[height=4.2cm,width=4.2cm]{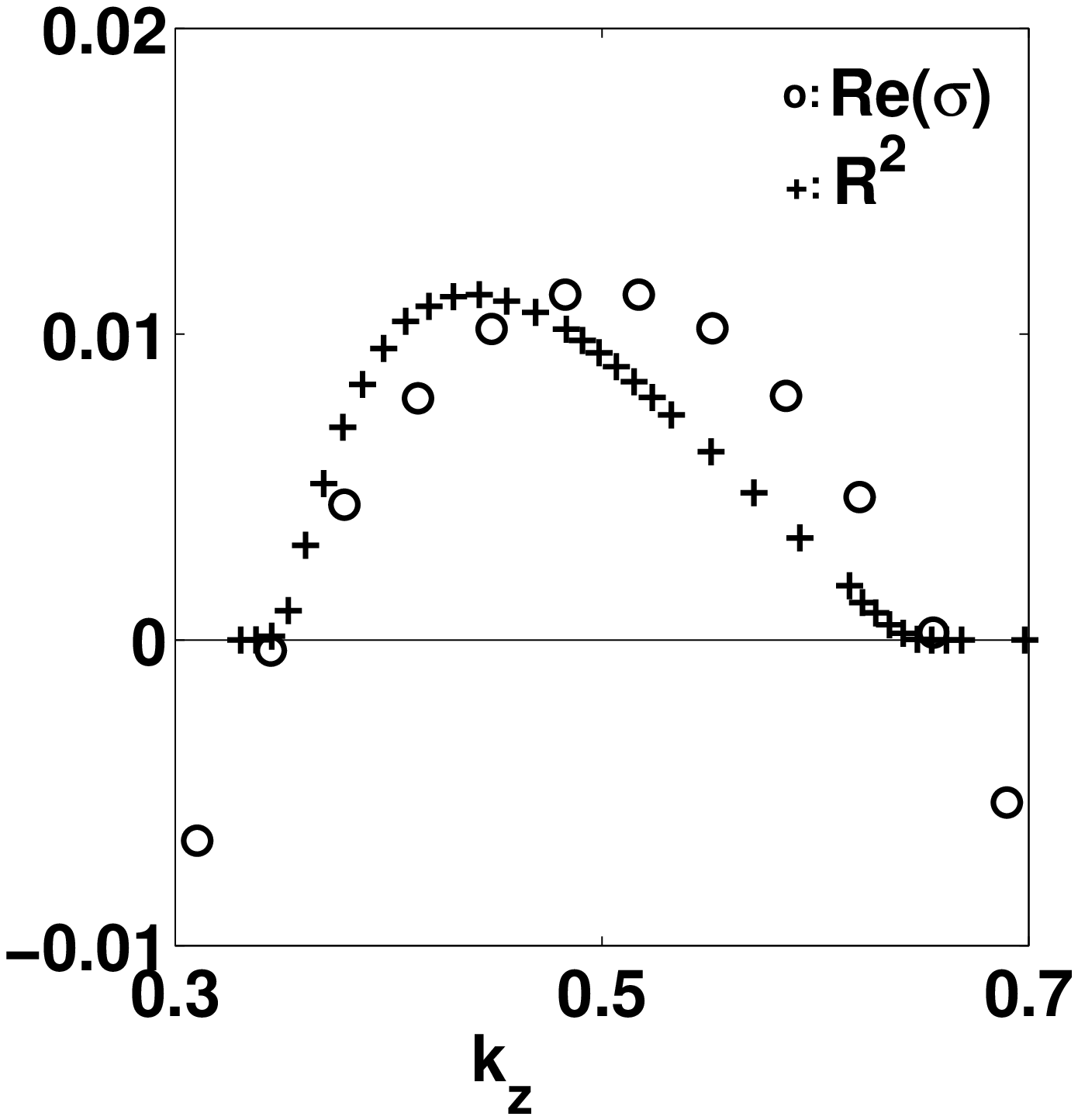}
\end{tabular}

\caption{\label{hopf.fig}(a)~: square of the amplitude  of the meandering 
restabilized
state as  a function of the parameter $a$. The parameters used are $\
b=0.01\mbox{ and }\epsilon=0.025$. The simulation box is $(128\times\,
128)\times \, 130$, with $dx=0.2$ and periodic boundary
conditions are used. (b)~: comparison of  the square of the
amplitude of the meandering restabilized state $R^2$ multiplied by 0.035 and
the growth rate of the meandering mode $Re(\sigma)$.
Same parameter regime as in (a) and $a$ is fixed to
$0.68$.}
\end{center} 
\end{figure}

The radius of the instantaneous filament helix can be taken as a measure
of the 3D meander instability amplitude.
As shown in Fig.~\ref{hopf.fig}, it is found to behave as the
square root of the distance to the threshold (either
$|a-a_c|$ or $|k_z-k_{\pm}(a)|$). Therefore as
reported previously \cite{armit}, the 3D
meander bifurcation  is  a supercritical Hopf bifurcation, as in 2D.

However, the $k_z$ dependence of the nonlinear term quickly becomes
important away from threshold. In Fig.~\ref{hopf.fig}, the 
square of the meander amplitude $R^2$
is compared to the growth rate of the unstable meander mode in the
whole band $[k_-(a),\ k_+(a)]$. A clear asymmetry of the $R^2$ curve is 
already seen,
with a slope at the $k_-(a)$-end about 3.4 times larger than the
$k_+(a)$-end slope. 

{\em No flux boundary conditions.}
For the box height $H$ 
used in our simulation (about  2 or 3 unstable wavelengths),
the boundary conditions chosen 
on the top and bottom boundary conditions have
a strong influence on  the restabilized state. For no-flux boundary
conditions, the spiral wave in each horizontal $xy-$plane has a meandering
motion. However, contrary to the case of periodic boundary conditions the
meander amplitude depends on $z$ and is well approximated by $|\cos(k_z
z)|$ (see Fig.~\ref{filshape.fig}). This implies in particular that in
in some horizontal planes the meander amplitude is zero and that the 
corresponding
spiral
tip performs a simple steady rotation. In the rotating frame where these special
tips are motionless,
the instantaneous  filament
takes  a planar  zig-zag shape that rotates around its vertical midline at a
pulsation close to  the imaginary part
of  the unstable meander  eigenmode corresponding to the
wavelength of the filament. Thus, for no-flux boundary condition, the meander
amplitude varies sinusoidally with height while the phase of the epicycle 
motion is independent of height.

This shape and motion are simply understood in a weakly nonlinear 
description where the restabilized state can be approximated as a
sum of the unperturbed
solution and the four unstable meander eigenmodes ($\sigma_m(k_z)$ and
$\sigma_m^*(k_z)$ at $\pm k_z$)
\begin{eqnarray}
 u&=&u_0(r,\psi)+[A {u}_1(r,\psi)\ {e}^{i(k_zz
+\omega_2 t)}+
B u_1 (r,\psi)\ {e}^{i(-k_zz r+\omega_2 t)}+ c.c.]\\
 v&=&v_0(r,\psi)+ [A {v}_1(r,\psi)\ {e}^{i(k_zz
+\omega_2t)}
+ B v_1(r,\psi)\ {e}^{i(-k_zz +\omega_2 t)}
+ c.c.]
\label{lsup}
\end{eqnarray}
with $u_1,v_1$  the eigenmode of
eigenvalue $\sigma_m(k_z)$ and $\omega_2=Im[\sigma_m(k_z)]$ 
 the meander frequency.
The no-flux boundary condition , $\partial_z u=0$, at $z=0$ and $z=H=2\pi/k_z$
enforces $A=B$ (i.e. for the height considered, the no-flux boundary
conditions stabilize the state with symmetric upward and downward propagating
deformation which
was observed to be unstable with periodic boundary conditions).
The instantaneous filament corresponding to the fields (\ref{lsup}) is 
easily determined by remembering that it is
the locus $u=u_{\mathrm{tip}}$, $v=v_{\mathrm{tip}}$. Its position 
is conveniently
parameterized as $x'_{\mathrm{tip}}(z)=x_0'+\delta x'(z,t), 
y'_{\mathrm{tip}}(z)=y_0'+\delta y'(z,t)$ using Cartesian coordinates 
in the rotating frame where the
 unperturbed filament is standing at $(x_0',y_0')$. For small $|A|$, one
obtains
\begin{eqnarray}
\partial_{x'}\! u_0\ \delta x' + \partial_{y'}\! 
u_0\ \delta y'  + \cos(k_z z) [2 A u_1 e^{i \omega_2
t} + c.c.]
&=& 0
\label{lfp1u}
\\ 
\partial_{x'}\! v_0\ \delta x' + \partial_{y'}\!  
v_0\ \delta y'  + \cos(k_z z) [2 A v_1 e^{i \omega_2
t} + c.c.]&=& 0
\label{lfp1v}
\end{eqnarray}
 where the field $u_1$, $v_1$, $u_0$, $v_0$ and their derivatives are
evaluated at the unperturbed filament position $(x_0',y_0')$.
Inversion of Eq.~(\ref{lfp1u},~\ref{lfp1v}) gives
\begin{eqnarray}
\delta x'= \cos(k_z z)[\alpha A e^{i \omega_2 t} +c.c.]\nonumber\\
\delta y'= \cos(k_z z)[\beta A e^{i \omega_2 t} +c.c.]
\label{fpnf}
\end{eqnarray}
where $\alpha$ and $\beta$ are complex constants which depend on $u_1,v_1$
and the derivatives of $u_0$ evaluated at the point $(x_0,y_0)$. This clearly
shows the planar zig-zag shape of the instantaneous filament since points in
different $xy$-planes simply differ by the real scale factor $\cos(k_z z)$.
As time evolves, Eq.~\ref{fpnf} also shows that the filament points follow 
scaled ellipses
in different $xy$-planes. Our simulations show that, as for 2D meander,
these ellipses are in fact almost circular \footnote{An explanation
can be provided by the proximity of the
$\omega_2=\omega_1$ point on the meander threshold line 
(as noted
in a particular limit in \cite{hk2}). The argument is that
i) the meander modes are close to the translation modes when $\omega_2$
is close to $\omega_1$ ii) the ellipse should reduce to
a circle for translations since the translated circular core is circular.
This can be explicitly seen from Eq.~(\ref{fpnf}). 
Namely,
Eq.~(\ref{fpnf}) gives $\delta x'+i\delta y'=\cos(k_z z)[A
(\alpha+i \beta) \exp(i\omega_2 t) +A^* (\alpha^*+
i \beta^*) \exp(-i\omega_2 t)]$. 
 The explicit inversion of 
Eq.~(\ref{lfp1u},~\ref{lfp1v}) shows that 
$\alpha+i \beta$ is proportional to
$(v_t u_1-u_t v_1 )$ and therefore vanishes when
 when $(u_1,v_1)$ tends toward
the translation eigenmode $(u_t,v_t)$. In this limit, it 
only remains the term proportional to
$\alpha^*+
i \beta^*$ and $\delta x'+i\delta y'$ follows a circle.}.

Finally, we briefly discuss the transition between the restabilized meander
regime seen on the "small core" side of the phase diagram
(Fig.~\ref{phasd.fig}) and the negative line tension dynamics which belongs
to its meander "large core" side. Since scroll waves do meander in this 
transition region,
the evolution of scroll waves as seen in direct numerical simulations
is not directly linked to the linear
spectrum of the steady scroll wave. For example, when the meander 
and translation bands are strongly hybridized, the translation bands are
only unstable for small values of $k_z$ (Fig.~\ref{lsmeantruns.fig}).
 Nonetheless,
direct simulations show 
that the  meandering scroll is unstable
and that its core grows as in the negative line
tension regime
in simulation boxes small enough to only contain unstable modes of the 
meander bands with
larger values of $k_z$. This happens 
on the whole small $a$ side of the dashed $\omega_1=\omega_2$
line of Fig.~\ref{phasd.fig}. This
line stands very close to the line where the external field
drift
of meandering spiral changes sign \cite{krins} 
and it is difficult to distinguish
the two in our 
simulations. So, the link between the spiral drift sign and the "negative line
tension" type of instability development continues to hold for meandering
scroll wave. 
\section{Influence of twist}
\label{sec:tw}
As noted in several previous studies \cite{win3d,pert,henze}, twist is
an important degree of freedom brought by the extension to 3D. It is well-known
from classic studies of elasticity \cite{love}(and everyday experience)
that straight rods and ribbons can be destabilized by twisting them beyond a
certain level. A somewhat similar instability was reported in ref.~\cite{henze}
in numerical simulations of excitable filaments. Beyond a threshold twist,
the rotation center line of an initially straight twisted filament was observed
to adopt an helical configuration. Observations of a similar 'sproing' 
\cite{henze} instability have since been made in the related context of
the complex Ginzburg-Landau equation vortex lines \cite{cgle}.
The characteristics of the excitable filament sproing instability have however
remained somewhat unclear. In the dynamical simulations of ref.~\cite{henze},
a single filament turn was imposed in a simulation box with periodic conditions.
and the box height was varied. A complicating feature of this procedure is that
both  twist and the available wavelength range are changed at the same time. 
On the theoretical side, the instability fails to be captured
by small twist approaches \cite{keen3d,bik3d}
since  
\cite{bik3d} the motion of the rotation center 
is not influenced by twist in this limit (see appendix \ref{keenav.sec}).

The present approach permits to relieve some of these 
problems since the twist $\tau_w$
can be varied from zero to large values and a whole
range of wavelengths can be examined in the linear stability
computations ($k_z$ is simply a parameter which can be given any chosen 
value independently of $\tau_w$).

\begin{figure}
\begin{center}
\includegraphics[height=4.cm]{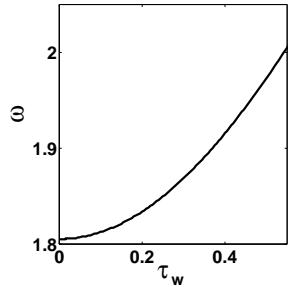}
\caption{Frequency $\omega_1$ of the scroll wave as a function of the twist 
for $a=0.8,\ b=0.01$, and $\epsilon=0.025$. The small $\tau_w$
behavior of the pulsation can be well approximated by
$\omega_1(\tau_w)=\omega_1(\tau_w=0)+0.7205\tau_w^2$ for low values
of $\tau_w$.
For higher values of $\tau_w$, a linear behavior
of $\omega_1$ as a function of $\tau_w$ is observed. 
The first-order perturbation result coefficient (\ref{omsmtw})
is $0.7203$ 
using the numerically determined rotation eigenmodes of $\mathcal{L}$ and of its
adjoint. 
\label{sttw.fig}}
\end{center}
\end{figure}

We restrict ourselves here mostly to parameter values 
for which a straight untwisted
filament is stable, that is on the large $a$ side of the (3D)
meander instability region. Fig.~\ref{sttw.fig} shows the frequency
and tip radius for a family of twisted scroll wave obtained by increasing
$\tau_w$ from $0$ to $\tau_w$ at one such parameter point ($a=0.8,b=0.01,
\epsilon=0.025$).
The frequency $\omega_1(\tau_w)$ increases quadratically at small twist
and almost linearly for larger twist values. The quadratic behavior
at small $\tau_w$ is simply obtained by applying first order perturbation theory
to Eq.~(\ref{2dst1},~\ref{2dst2}) which gives
\begin{equation}
\omega_1(\tau_w)=\omega_1(\tau_w=0)-\tau_w^2 \frac{\langle
\tilde{u}_{\phi}, \partial_{\phi\phi} u_0\rangle}
{\langle\tilde{u}_{\phi}, \partial_{\phi} u_0\rangle +
\langle\tilde{v}_{\phi}, \partial_{\phi} v_0\rangle} +O(\tau_w^4)
\label{omsmtw}
\end{equation}
The direct computation of the matrix element ratio on the r.h.s. of
Eq.~(\ref{omsmtw}) is in good agreement with a direct fit of
the $\omega_1(\tau_w)$ curve of Fig.~\ref{sttw.fig} (see caption). Analytic 
descriptions of the $\omega_1(\tau_w)$ curve
for larger twist values have recently been obtained in the free boundary
limit ($\epsilon\rightarrow 0$) both for small core \cite{mb} and large core
scroll waves \cite{hhk}.

\begin{figure}[h]
\begin{center}
\begin{tabular}{ccc}
(a)& &(b)
\\[.2cm]
\includegraphics[width=4.cm,height=4.cm]{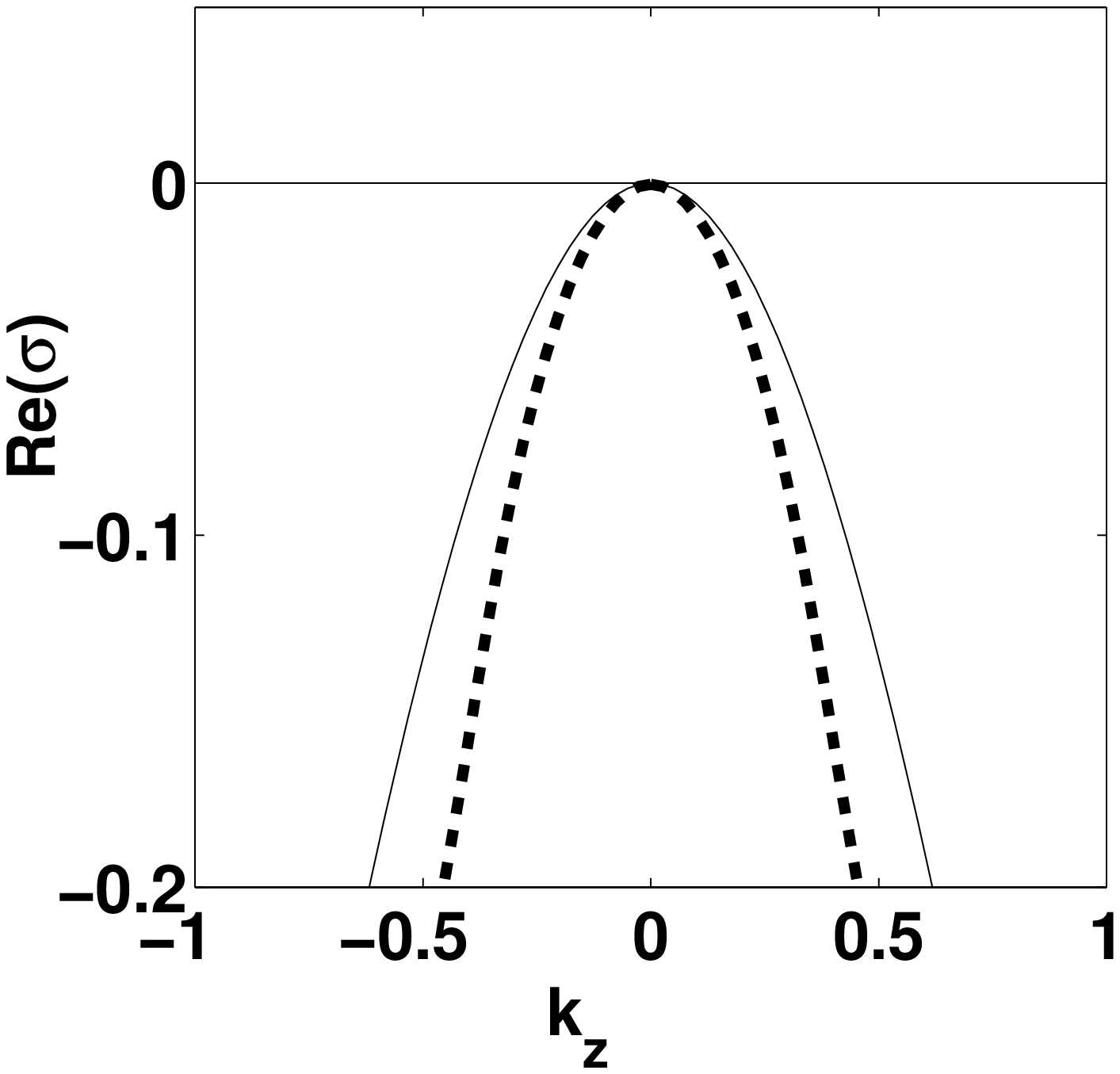}
& &
\includegraphics[width=4.cm,height=4.cm]{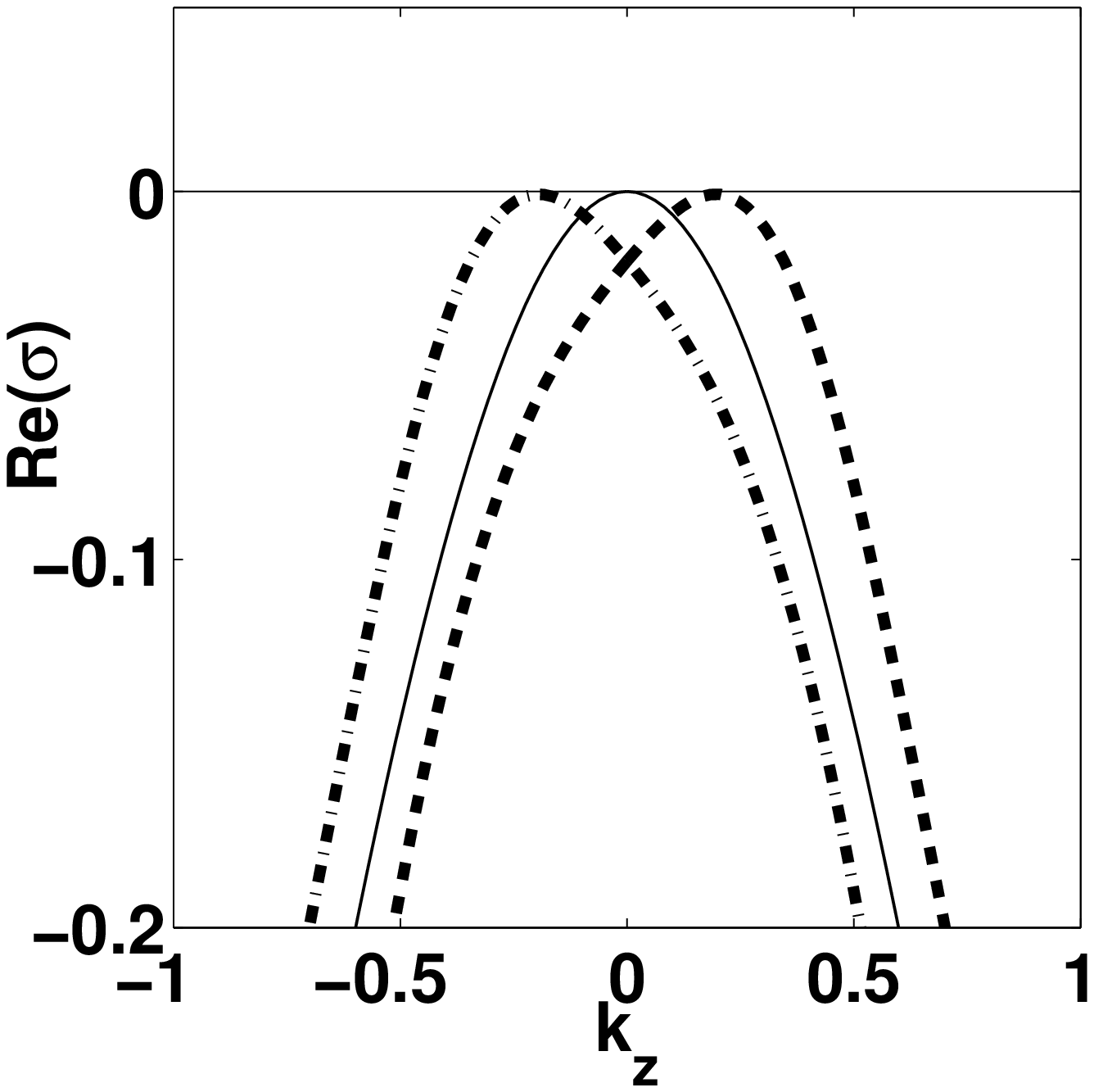}
\\
(c) & &(d)\\[.2cm]
\includegraphics[width=4.cm,height=4.cm]{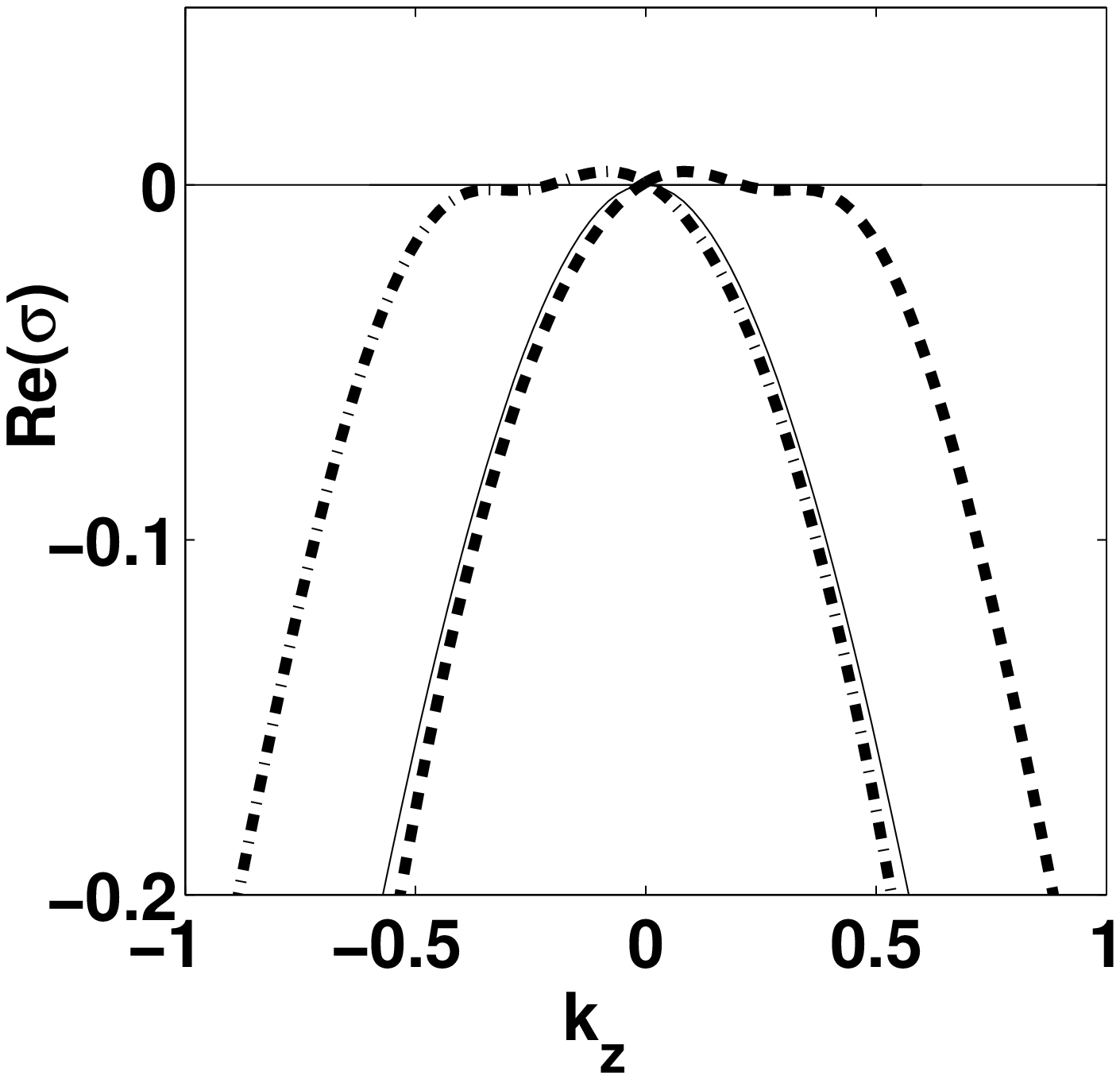}
& &
\includegraphics[width=4.cm,height=4.cm]{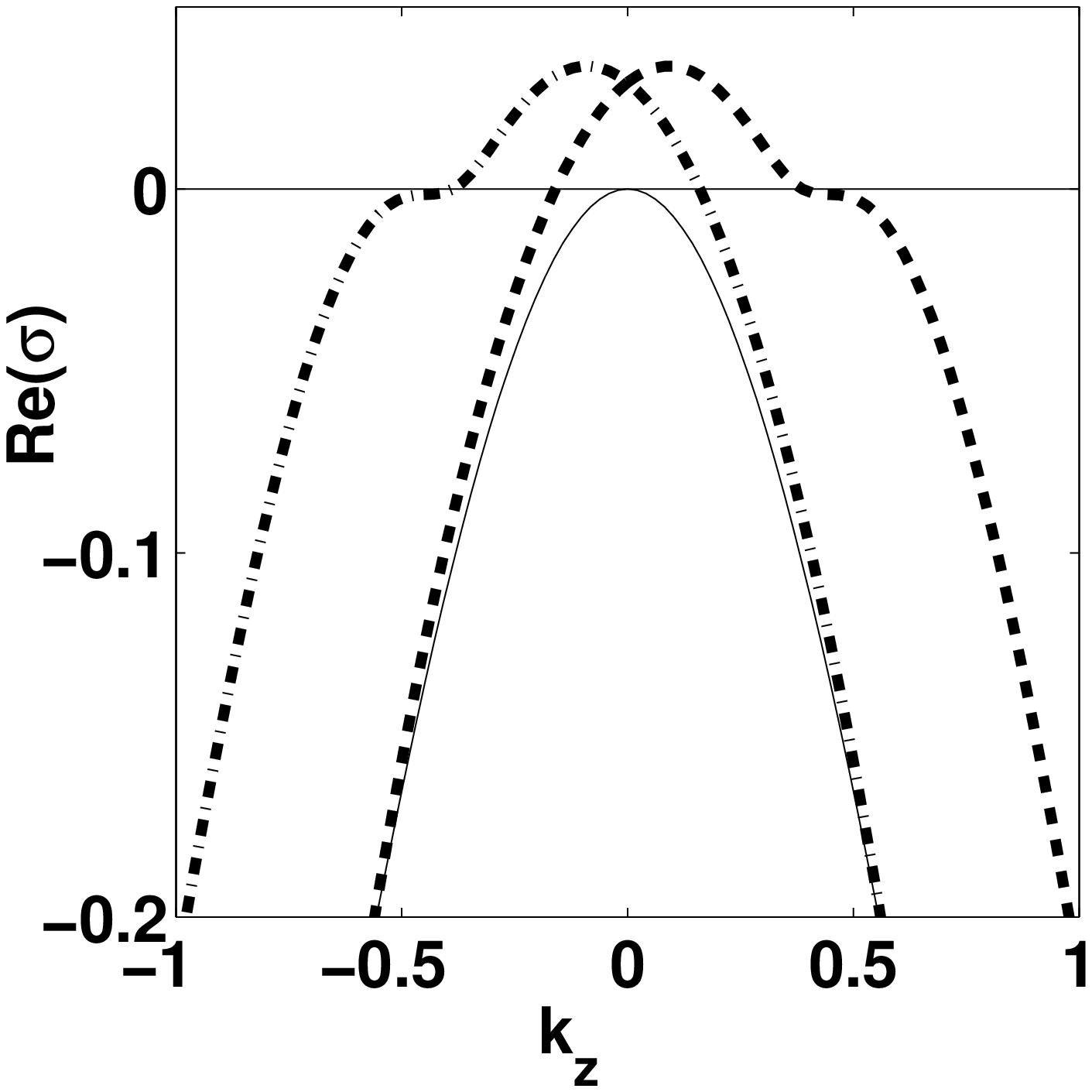}
\end{tabular}
\caption{Real parts of 
the rotation band (thin solid line) and the two translation  bands
(bold dashed and dash dotted lines) as a function of the wave number 
$k_z$ for the same parameter values
$a=0.8$, $b=0.01$ and $\epsilon=0.025$ and different values of twist~:
(a)$\tau_w=0.$,(b) $\tau_w=0.2$, (c) $\tau_w=0.35$  and (d)
$\tau_w=0.45$. The translation bands  have  maxima at
$k_z=\pm\tau_w$ with a zero growth rate. A secondary
maximum appears on the translation bands as the twist increases. At
a threshold value of the twist it becomes unstable at a non-zero
value of $k_z$.
\label{lstw.fig}}
\end{center}
\end{figure}

The determination of
a family of increasingly twisted steady scroll waves permits to determine
the evolution of the stability spectrum with $\tau_w$. The results of
such a computation are shown in Fig.~\ref{lstw.fig} for 
scroll waves of Fig.~\ref{sttw.fig}. As twist increases,
 the deformations of the translation
bands are particularly important. As expected from
general arguments (section \ref{linstab.sec}), 
for $\tau_w=0.2$ (Fig.~\ref{sttw.fig}b), 
the translation bands $\sigma_{t,1},\sigma_{t,2}$
are no longer even and related by complex 
conjugation. It only remains the lower symmetry 
$\sigma^*_{t,1}(k_z)=\sigma_{t,2}(-k_z)$. One can note also that
the
translation modes with $Re[\sigma(k_z)]=0$ stand at $k_z=\pm \tau_w$ and 
no longer
at $k_z=0$, in agreement
with the analytic expression given in section \ref{speceigen.sec} . When twist
 is further increased to $\tau_w\approx0.33$ (not shown) a second
maximum of $Re[\sigma(k_z)]$ appears near $k_z=0$. The value of  $Re[\sigma(k_z)]$ is
negative at first at this secondary maximum. However, it increases with
$\tau_w$ and it is slightly positive at $\tau_w=0.35$ (Fig.~\ref{sttw.fig}c).
The twisted scroll waves are then unstable for a finite range of wavevectors
near $k_z=0$. Increasing twist further, enlarges the range of unstable 
wavelengths and the instability growth rate, as shown on Fig.~\ref{sttw.fig}d.

Dynamical simulations reported in section \ref{spro.sec} show that this
twist-induced instability of the translation bands correspond to the
''sproing'' instability of ref.~\cite{henze}. Before describing these
results, it is worth explaining why the instability does not appear
around the translation modes  at $k_z=\pm \tau_w$ but 
a finite wavevector away from them. 
This is a direct consequence of 3D rotational invariance: a twisted scroll wave
the axis of which is tilted has the same frequency as one the axis of which is
vertical.
So, a small tilt perturbation should not change the translation mode
eigenvalues $\pm i\omega_1$ to linear order. Therefore, they
remain local extrema on the translation
bands, as it is observed in Fig.~(\ref{lstw.fig}). A direct mathematical
proof (based on the same reasoning) is offered in the next section.

\begin{figure}
\begin{center}
\begin{tabular}{ccc}
(a) & &(b)
\\[.1cm]
\includegraphics[width=4.cm,height=4.cm]{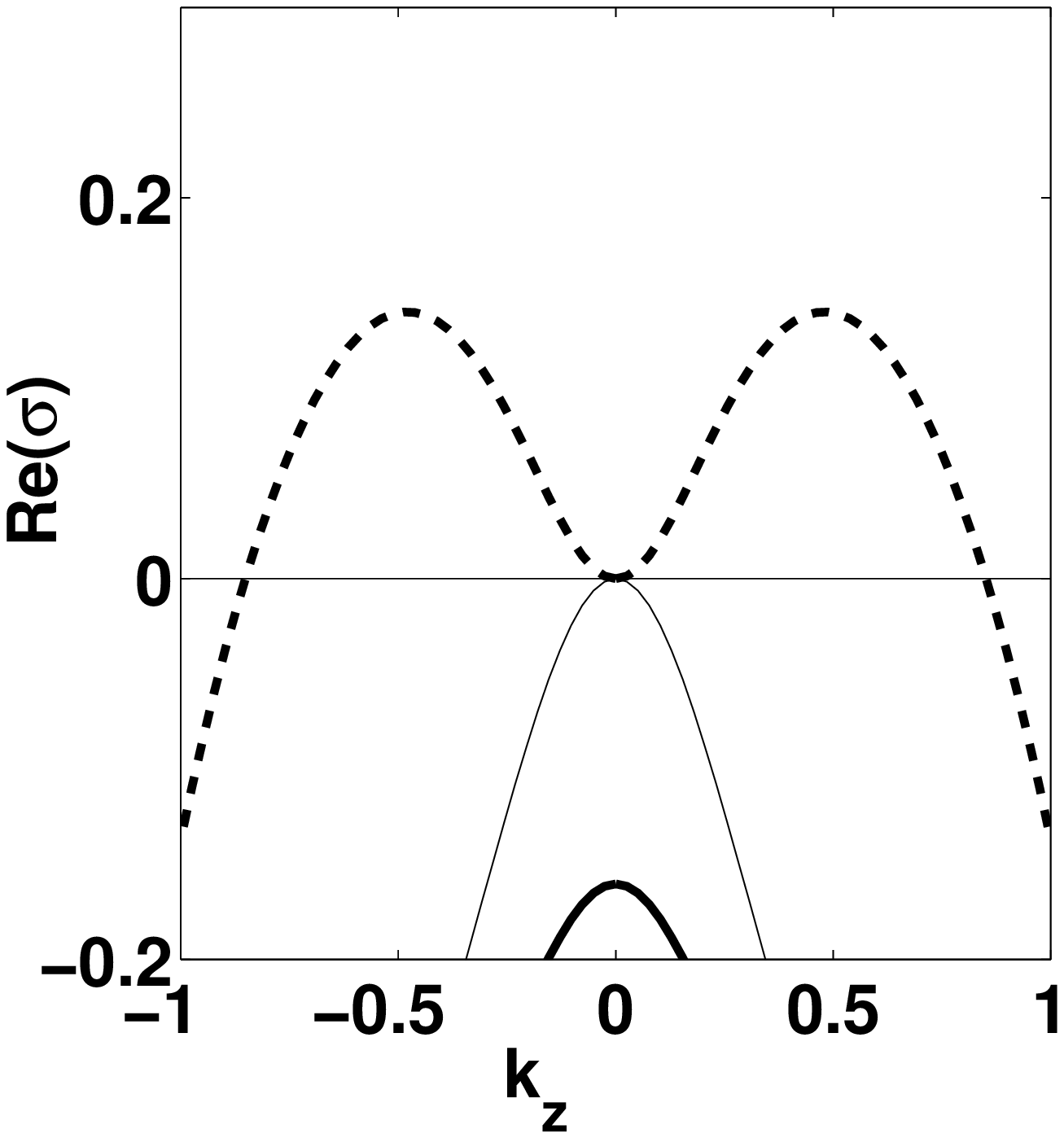}
& &
\includegraphics[width=4.cm,height=4.cm]{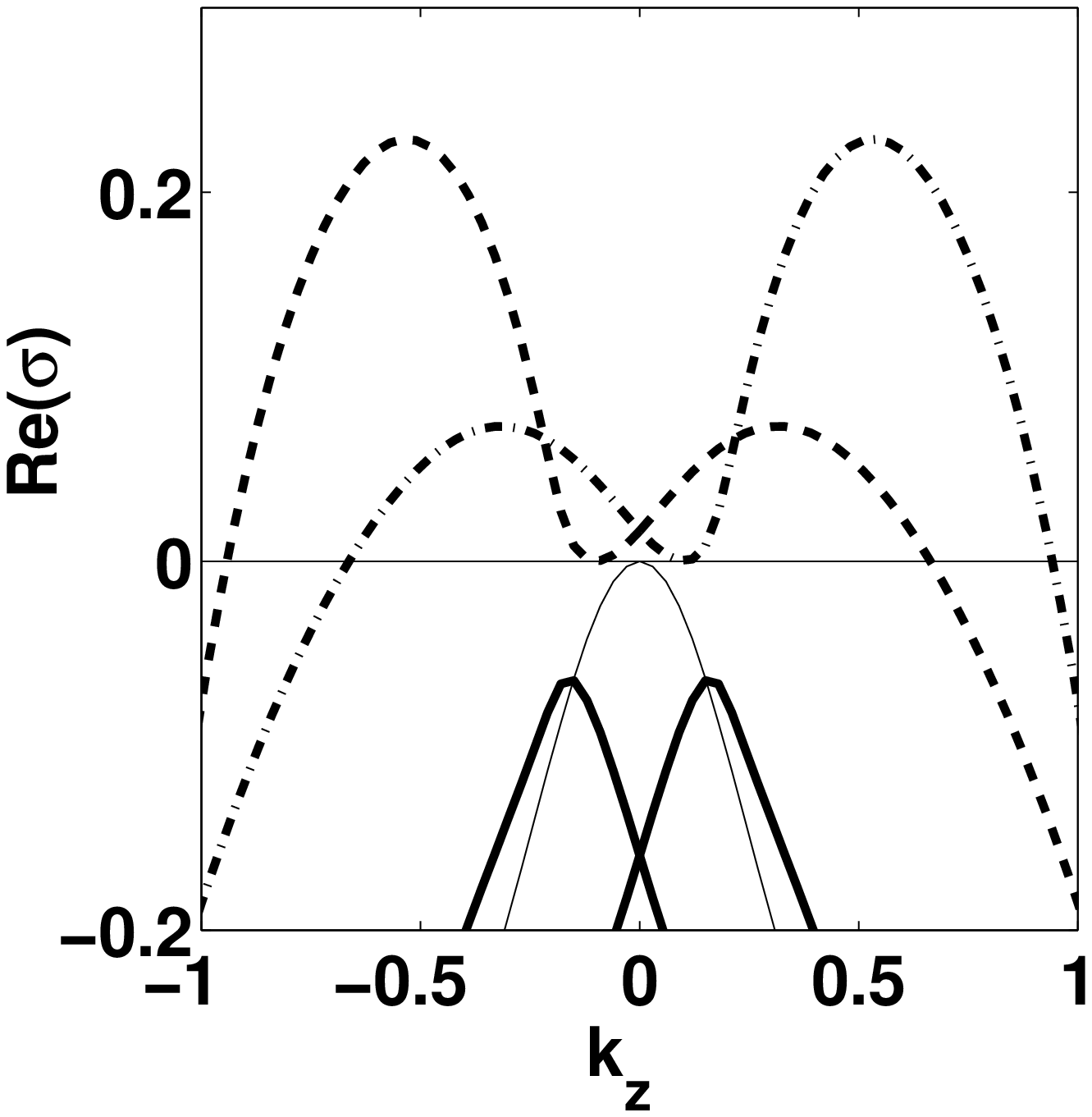}
\\
(c)& &(d)
\\[.1cm]
\includegraphics[width=4.cm,height=4.cm]{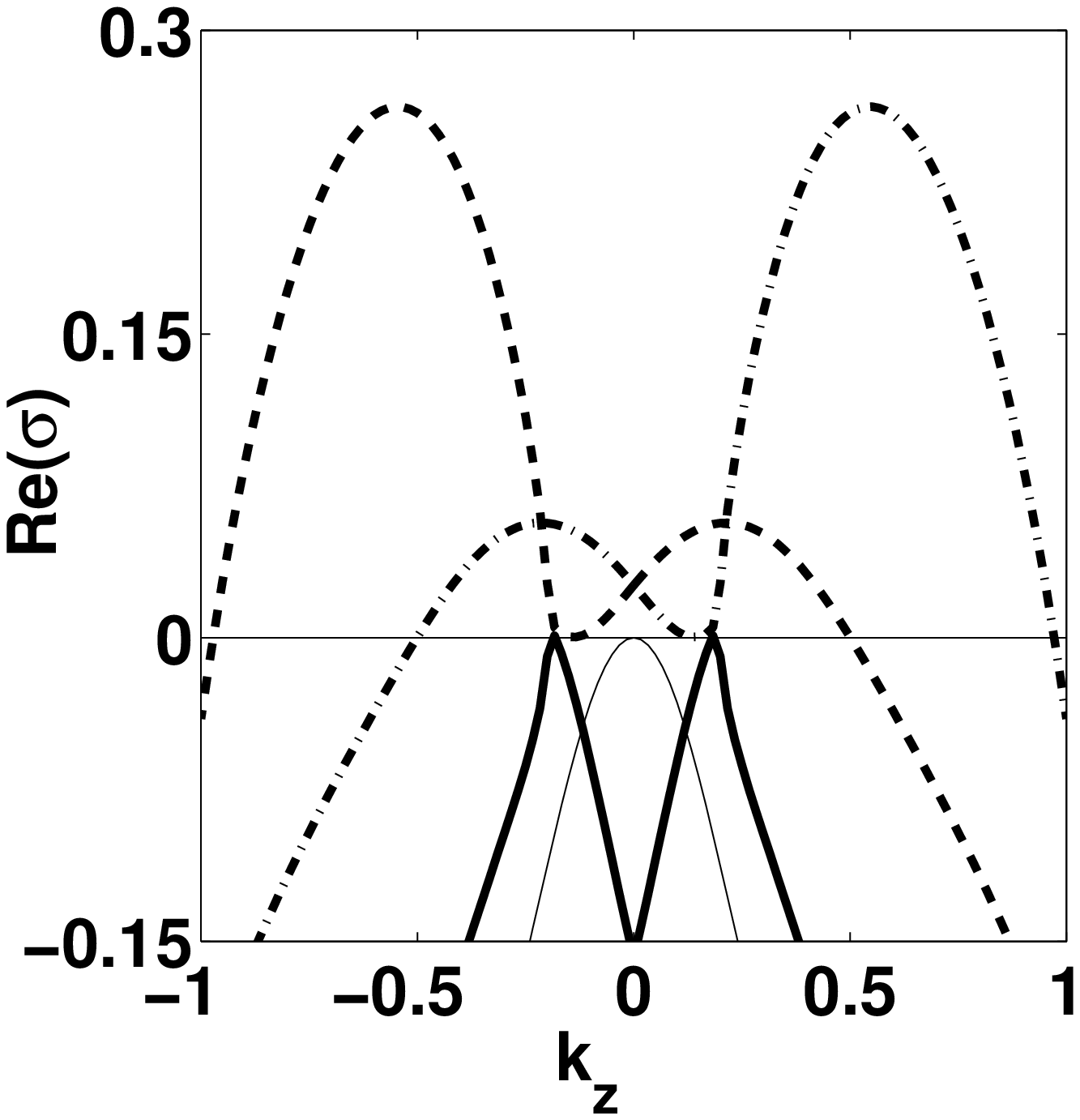}
& &
\includegraphics[width=4.cm,height=4.cm]{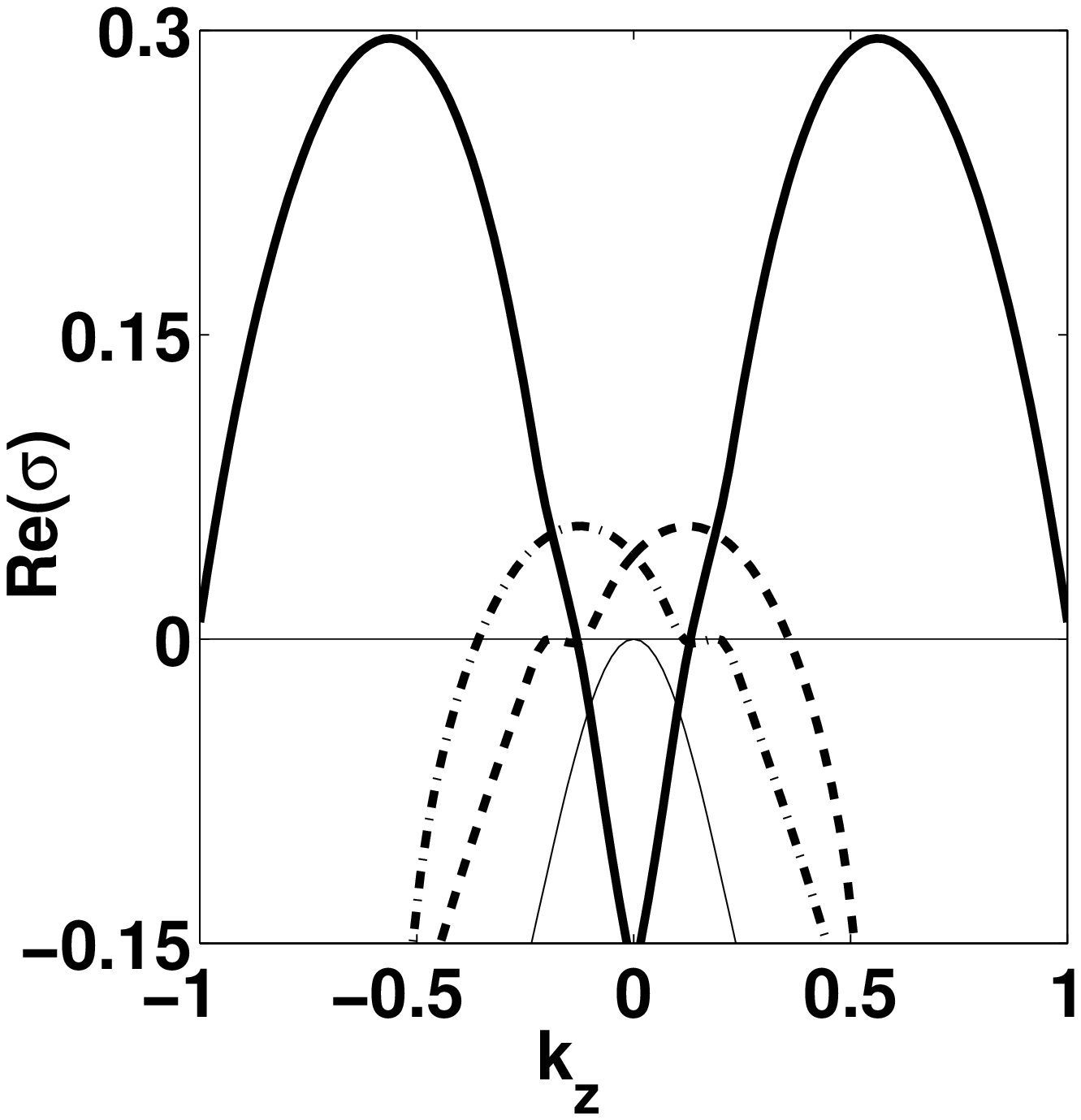}
\end{tabular}
\caption{Real parts of  
the rotation band (thin line) and the two translation bands
(dashed and dashed-dotted) as a function of the wave number $k_z$ for 
$a=0.44$, $b=0.01$ and $\epsilon=0.025$ and different values of twist~:
(a)$\tau_w=0.$, (b) $\tau_w=0.1$, (c) $\tau_w=0.14$ 
and (d) $\tau_w=0.19$ .
The translation bands have a minimum zero growth rate for
$k_z=\pm\tau_w$. 
The maximum
growth rate of the translation bands is increased by twist. 
The meander
modes also become less stable as twist is increased. 
The
change in the most unstable band for $\tau_w$ close to $0.14$ appears to be
due to
hybridization between the translation and meander bands like the one observed in
Fig.~\ref{lsmeantruns.fig}. The results of direct numerical
simulations 
show that twist does not qualitatively modify 
the development 
of the zero twist instability in this regime: a  "negative line tension" growth
of the filament
is observed.
\label{twgrayon}} 
\end{center}
\end{figure}

Finally, we find it interesting to show in
Fig.~\ref{twgrayon} the twist influence on the spectrum 
in the "negative line tension" parameter
regime of Fig.~\ref{lstruns.fig}, although the untwisted scroll wave is already
unstable in this case. Twist modifies the spectrum in
a way that is rather different from that seen in Fig.~\ref{lstw.fig}. It mainly
amplifies the instability of the large $k_z$ part of the spectrum .

\subsection{Helical destabilization and 3D rotational invariance}

In order to demonstrate that the translation modes eigenvalues
at $k_z=\pm \tau_w$ remain extrema on the translation bands, we
show that 
\begin{equation}
\left.\frac{d\sigma}{dk}\right|_{k_z=\pm \tau_w}=0
\end{equation}
We proceed in two steps. First, perturbation theory
is used
to compute the eigenvalues of modes close to the translation modes on 
the translation bands. For definiteness, we consider modes at 
$k_z=-\tau_w +\delta k$, close to the translation mode $(u_t,v_t)$ at
$k_z=-\tau_w$ with $\sigma=i\omega_1$. Eq. (\ref{eqlin1},~\ref{eqlin2}) 
can be written without approximation,
\begin{equation}
\sigma(k_z)  \left( \begin{array}{c} u_1\\v_1\end{array}\right) =
(-i\,\delta k^2+2 \delta k\,\tau_w+2 i\tau_w\delta k\,
\partial_{\phi}) \left( \begin{array}{c} u_1\\0\end{array}\right) +
{\mathcal{L}}_{k_z=-\tau_w}
        \left( \begin{array}{c} u_1\\v_1\end{array}\right)\label{pertkztau1}
\end{equation}
Seeking in perturbation, $u_1=u_t+\delta u_1, v_1=v_t+\delta v_1$, one
obtains to first order in $\delta k$,
\begin{equation}
\delta\sigma \left( \begin{array}{c} u_t\\v_t\end{array}\right)+
i\omega_1 \left( \begin{array}{c} \delta u_1\\\delta v_1\end{array}\right)=
2 \delta k\,\tau_w (1+i\partial_{\phi})\left( \begin{array}{c} u_t\\0\end{array}
\right)+
{\mathcal{L}}_{k_z=-\tau_w} \left( \begin{array}{c} \delta u_1\\\delta v_1
\end{array}\right)
\label{pertkztau}
\end{equation}
Multiplying by the left eigenvector
$(\tilde{u}_t,\tilde{v}_t)$ of
${\mathcal{L}}_{k_z=-\tau_w}$ for the eigenvalue $i\omega_1$ and taking
the scalar product gives,
\begin{equation}
\left.\frac{d\sigma}{dk}\right|_{k_z=\pm \tau_w}= 
\frac{2 \tau_w}{\langle\tilde{u}_t,u_t\rangle+
\langle\tilde{v}_t,v_t\rangle}\ 
\langle\tilde{u}_t,(1+i\partial_{\phi}) u_t\rangle
\end{equation}

Thus, the translation modes remain extrema on the translation
bands, if 
\begin{equation}
 \langle\tilde{u}_t,(1+i\partial_{\phi}) u_t\rangle=0.
\label{orth3d}  
\end{equation}
Eq.~(\ref{orth3d}) is a consequence of 3D rotational
invariance as we proceed to show. The perturbation corresponding to
inclining the scroll axis can be found by expressing the inclined
scroll in the vertical scroll referential, similarly to what
was done to determine translation modes (Eq.~(\ref{transspi},~\ref{tm})).

One obtains,
\begin{equation}
\left(\begin{array}{c} u_{inc}^{(1)}\\v_{inc}^{(1)}\end{array}\right)
=\exp[i(\omega_1 t+\tau_w z)]\left(\begin{array}{c} u_{inc}\\v_{inc}\end{array}
\right)
=\exp[i(\omega_1 t+\tau_w z)] \left[
z \exp(i\phi)(\partial_r+\frac{i}{r}\partial_{\phi})+\tau_w r \exp(i\phi)
\partial_{\phi} \right]\left( \begin{array}{c} u_0\\v_0\end{array}\right)
\label{uincdef}
\end{equation}
and the complex conjugate mode.
One can directly check that $u_{inc}^{(1)},v_{inc}^{(1)} $ obey the
linearized time dependent equations. Namely,
\begin{eqnarray}
(\partial_t +2\tau_w\partial^2_{\phi z} -\partial^2_{zz})u_{inc}^{(1)}&=&(
\omega_1
\partial_\phi+
\tau^2_w\partial^2_{\phi\phi}+\nabla^2_{2D})u_{inc}^{(1)}+[\partial_uf(u_0,v_0)
u_{inc}^{(1)}+\partial_vf(u_0,v_0)v_{inc}^{(1)}]/\epsilon\label{tdlin1}\\
\partial_t v_{inc}^{(1)}&=&\omega_1\partial_\phi v_{inc}^{(1)}+
\partial_ug(u_0,v_0)
u_{inc}^{(1)}+\partial_vg(u_0,v_0)v_{inc}^{(1)}
\label{tdlin2}
\end{eqnarray}
Eq.~\ref{tdlin2} shows that $(u_{inc},v_{inc})$ (Eq.~(\ref{uincdef}))obey
\begin{equation}
[{\mathcal{L}}_{k_z=-\tau_w}-i\omega_1]
        \left( \begin{array}{c} u_{inc}\\v_{inc}\end{array}\right)=
\left( \begin{array}{c} -i2\tau_w(1+i\partial_{\phi})u_t\\0\end{array}\right)
\label{linuinc}
\end{equation}
The inhomogeneous r.h.s. of Eq.~(\ref{linuinc})
comes from the fact that the
$\partial_z$ derivative terms in Eq.~(\ref{tdlin2}) 
act both on the exponential prefactor and on the intrinsic $z$ dependence of
 $u_{inc}$ (Eq.~(\ref{uincdef}), while in effect only their action
on the exponential term is taken into account in Eq.~(\ref{linuinc}) 
(through 
the $k_z$ dependence of ${\mathcal{L}}_{k_z}$).

Eq.~(\ref{linuinc}) directly gives the sought
orthogonality relation (\ref{orth3d}) after
multiplying both of its sides by the left eigenvector 
$(\tilde{u}_t,\tilde{u}_t)$ of
${\mathcal{L}}_{k_z=-\tau_w}$ with eigenvalue $i\omega_1$ and
taking the scalar product.

\subsection{The sproing bifurcation}
\label{spro.sec}
In order to study the nonlinear development of the twist-induced instability
shown in the parameter regime of Fig.~\ref{lstw.fig}c, d,
we performed direct numerical simulations of twisted scroll waves using
periodic boundary conditions at the top and bottom of the simulation
box.

Two kinds  of initial conditions were used.  The simplest one
consisted of two  dimensional spirals stacked along the vertical ($z$) axis. 
The twist was introduced by rotating them around this  vertical
axis. The main
disadvantage of this type of initial conditions 
was that they usually are
far from a stationary twisted scroll wave 
and from a restabilized wave when it existed. As a result, reaching the
asymptotic attracting
state could be very costly in computational time. 
In order to avoid this problem we  mainly used  initial conditions
constructed using results of previous direct simulations  
on a grid with the same values of the parameters and the  
same horizontal size but of a different vertical
extension,  interpolating linearly the values of the $u$ and $v$
field  on the new grid. 

For  definiteness, we describe
the result for the parameters of Fig.~\ref{lstw.fig}.

We first focus on the case
when  a single
turn of twist is initially imposed as in ref.~\cite{henze}. 
This case is special for the following reasons.
On one hand, the previous linear 
stability results (Fig.~\ref{lstw.fig}) show that all the potentially
unstable modes correspond to $|k_z|<\tau_w$. On the other hand, in a box
of height $H$, the only possible $k_z$ values
compatible with 
the imposed
top and bottom periodic boundary conditions are multiple of $2\pi/H$. Therefore,
for a single turn of twist $\tau_w=2\pi/H$, 
there is  a single potentially unstable mode in the simulation box and it
stands at $k_z=0$.

A series of dynamical simulations were performed in boxes of varying
heights $H$. The initial twist was correspondingly varied from
$\tau_w=0.3$ to $\tau_w=0.5$ .

For low values of twist, 
the twisted initial state
simply evolves toward
a straight twisted scroll wave. The instantaneous filament has an helical
shape that rotates at a fixed time independent frequency around its vertical
axis. In each horizontal plane, the wave is seen as a spiral steadily rotating
around the helix axis.

\begin{figure}
\begin{center}
\begin{tabular}{ccc}
(a)& &(b)
\\[.1cm]
\includegraphics[height=8.3cm,width=4.cm]{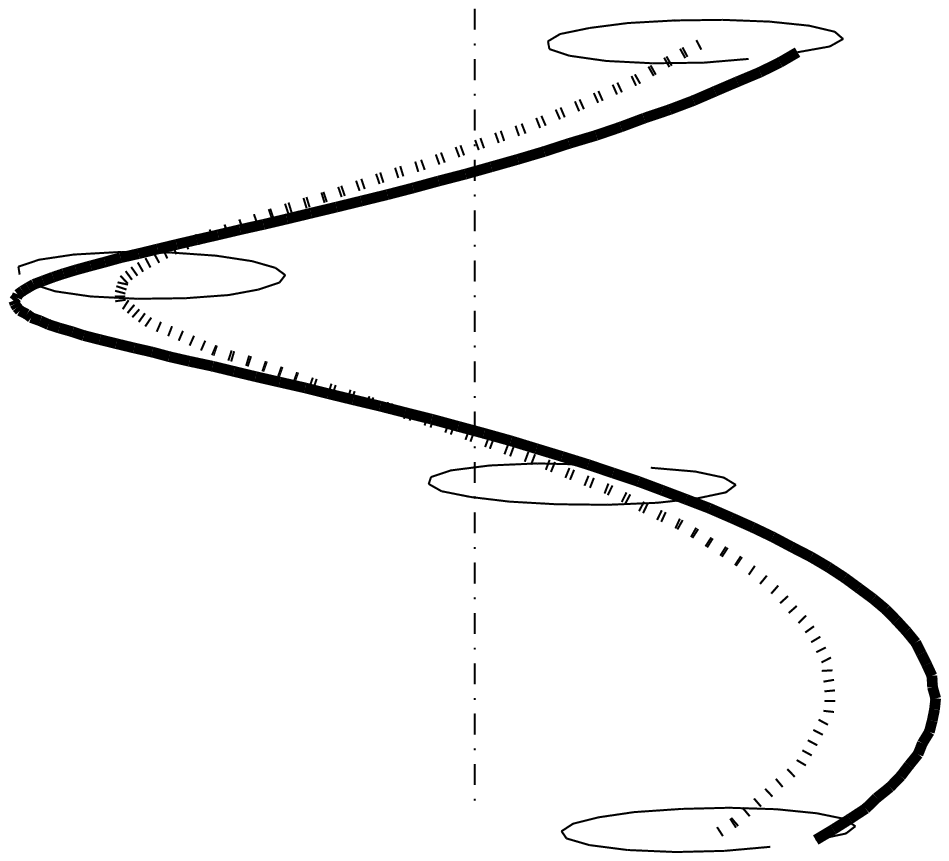}
& &
\begin{tabular}[b]{c}
\includegraphics[height=4.cm,width=4.cm]{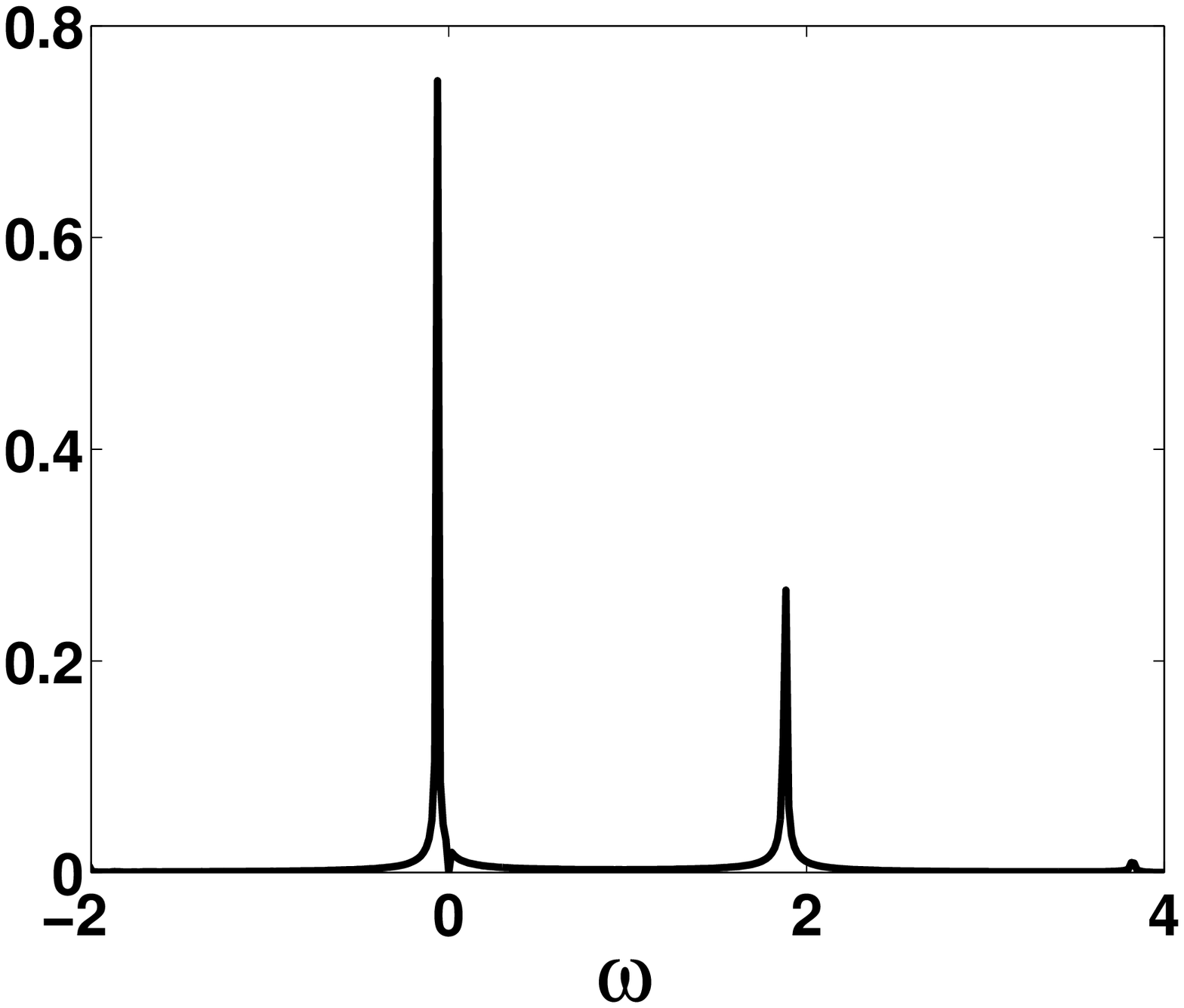}
\\
(c)
\\
\includegraphics[height=4.cm,width=4.cm]{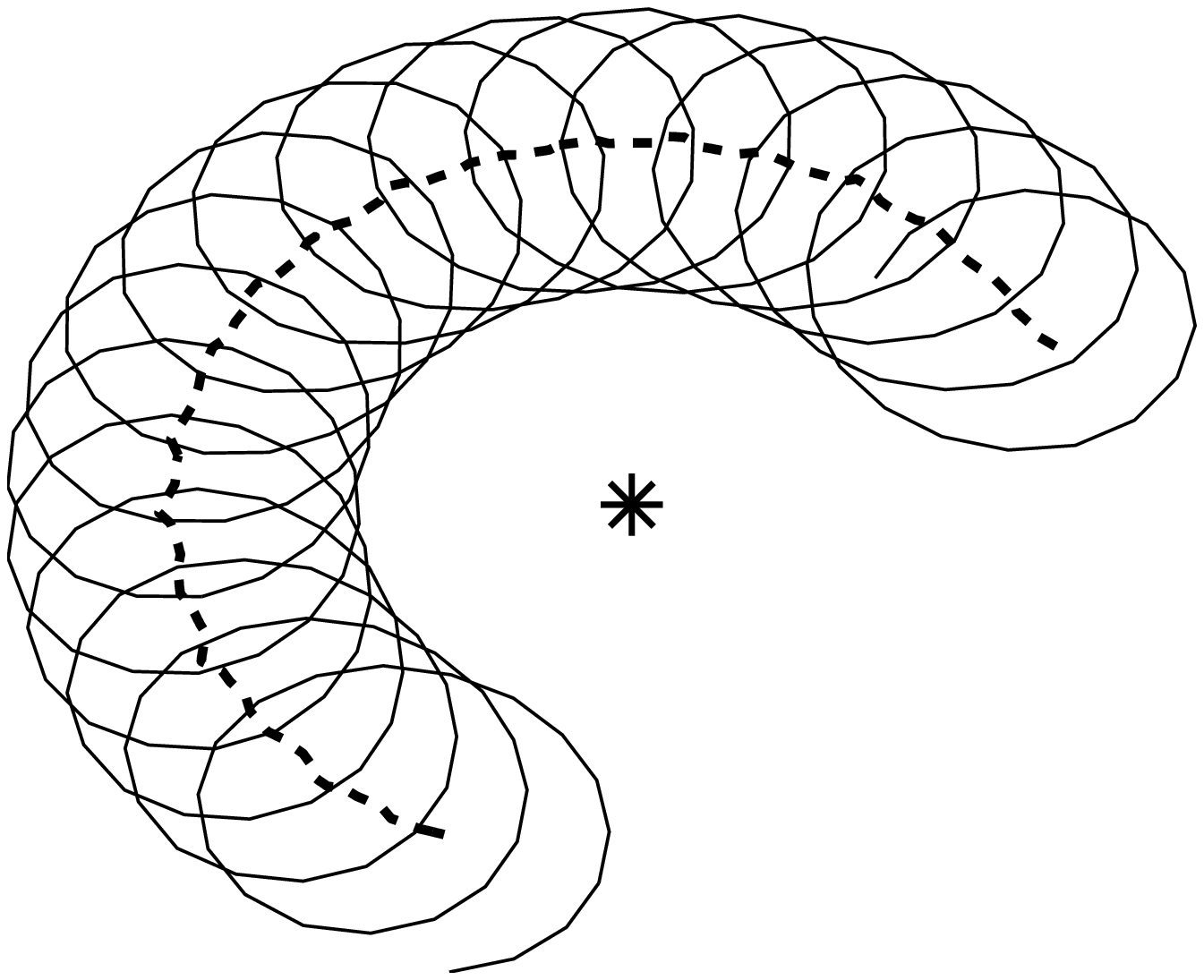}
\end{tabular}
\end{tabular}
\caption{\label{filamentsproing}(a): the bold solid line 
represents the  helical
instantaneous filament, the  dotted line the  mean
filament and  the thin solid lines the quasi-circular 
trajectories of the instantaneous filament in horizontal
planes of equally spaced $z$. 
Parameters values are $a=0.8$, $b=0.01$ and $\epsilon=0.025$,
the simulation box is $(128\times 128)\times 110$ with
$dx=0.15$, corresponding to $\tau_w=0.381$. (b): Modulus of the
Fourier transform of the spiral tip complex position ($x+iy$) in
an horizontal plane for the same parameter regime. The peak at
$\omega\approx-0.064$ corresponds to the slow movement of the mean
filament in the plane while the peak at $\omega=1.884$
corresponds to the rapid rotating motion of the spiral.  The Fourier
transform was performed using 2048 points with a time spacing of
$\delta t=0.1969$. 
(c): 
the
thin solid line is the trajectory of the spiral tip in an horizontal plane and
the bold dashed line is the trajectory of the mean filament in
that plane. The mean filament rotates clockwise while the fast rotation of the
spiral tip is counterclockwise. At a given time, the projection of the
mean filament and instantaneous filaments on the horizontal
plane are circles centered on the helices axis position marked by a star.   
}
\end{center}
\end{figure}

Beyond a threshold twist $\tau_c$, the twisted initial state evolves
toward a more complex state. The threshold twist  is estimated to
be $\tau_c=0.352$ from the direct numerical simulations. It closely 
corresponds to the value $0.350$ obtained from the linear stability 
analysis for
the instability of the $k_z=0$ modes (the instability
threshold twist at $k_z\ne 0$ is $\tau\simeq.345$).
The asymptotic scroll wave state is shown in Fig.~\ref{filamentsproing}
for $\tau_w=0.381$ and it is qualitatively similar for other values
$\tau_w>\tau_c$. The instantaneous filament takes an helical shape
at each time. The axis of this instantaneous helical shape is independent
of time but its other characteristics vary with time. The point of
the instantaneous filament in a given
horizontal plane (i.e. the spiral wave tip in that plane) closely follows an
epicycloidal motion (Fig.~\ref{filamentsproing}) and the instantaneous filament
global evolution can be accurately parameterized
as
\begin{equation}
\left\{ 
\begin{array}{l}
x=R_1 \cos(\omega_s t+\tau_w z+ \phi)+
R \cos(\omega_m t+\tau_w z)\\
y=R_1 \sin(\omega_s t+\tau_w z+ \phi)
+ R \sin(\omega_m t +\tau_w z)
\end{array}
\right.
\label{instfil}
\end{equation}

\begin{figure}
\begin{center}
\begin{tabular}{ccc}
(a)& &(b)
\\
\includegraphics[height=4.2cm,width=4.2cm]{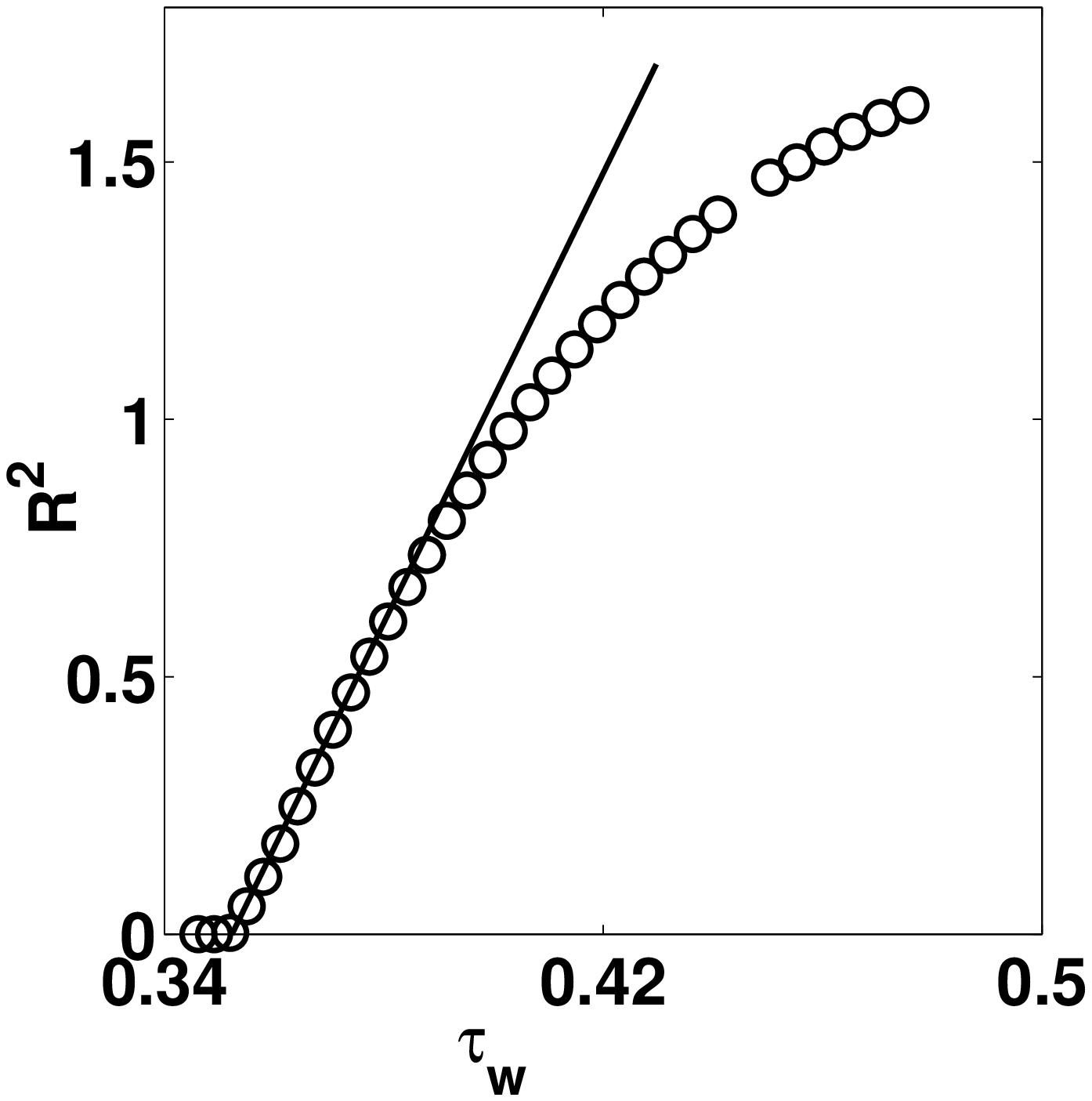}
& &
\includegraphics[height=4.2cm,width=4.2cm]{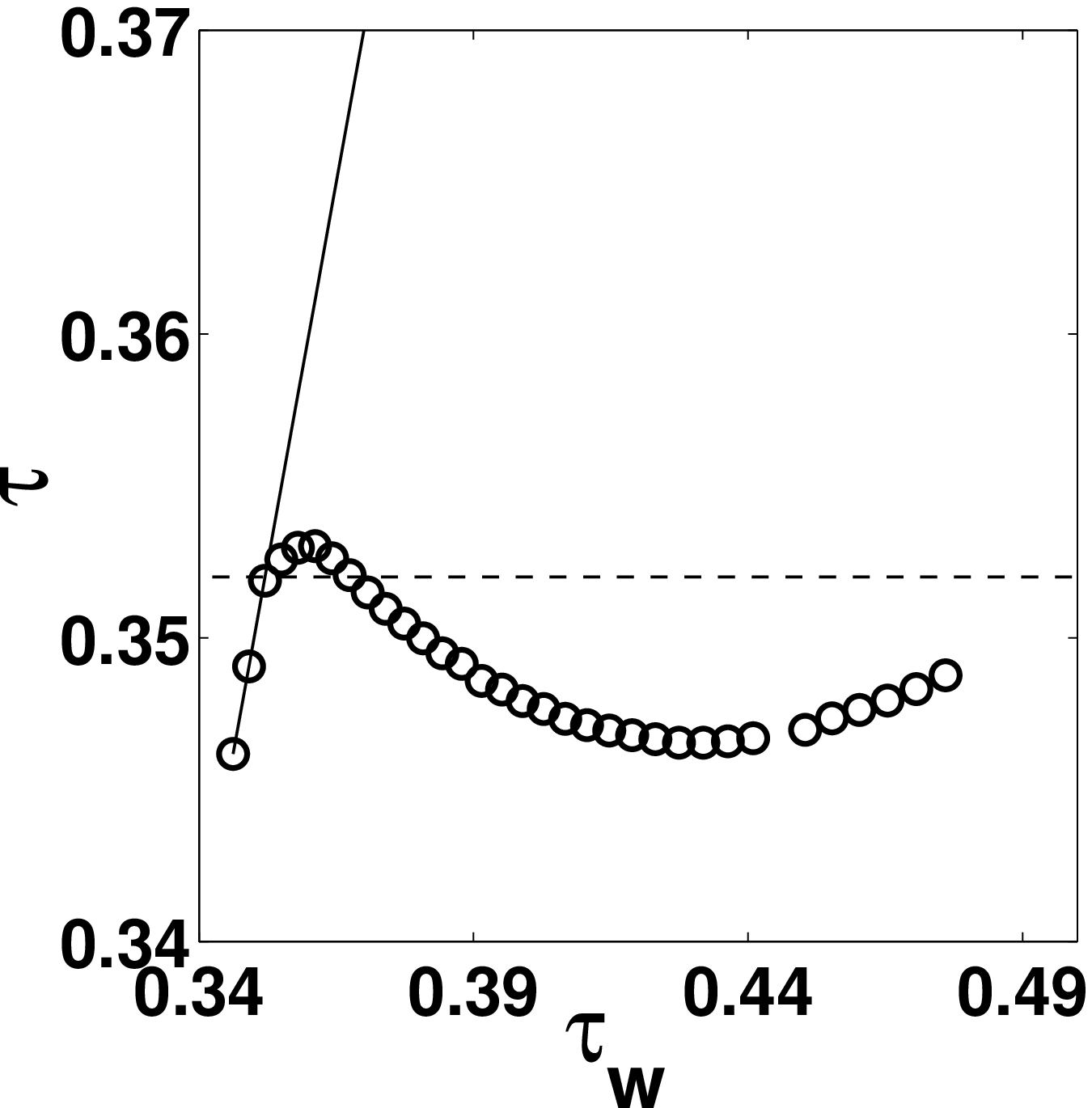}
\end{tabular}
\end{center}
\caption{\label{amplisproing}(a)($\circ$)
Radius ($R$) of the helical mean filament 
as a function of the twist $\tau_w$, for a stationary straight
twisted scroll wave and
linear interpolation of $R^2$ for small
values of $\tau_w-\tau_c$ (thin continuous line) where the
computed threshold twist is $\tau_c=0.352$.
 The parameters are $a=0.8\ b=0.01\
\epsilon=0.025$. 
(b): local twist of the restabilized
filament  as a function of
$\tau_w$. The dashed line is the line of equation $\tau=\tau_c$, the
continuous line is the line of equation $\tau=\tau_w$.
\label{ampspro.fig}}
\end{figure}

The pulsation $\omega_s$ and the radius $R_1$ are close to 
the pulsation and radius of the stationary
straight twisted scroll. The pulsation $\omega_m$ is found to be
small compared to $\omega_s$ (Table. \ref{compfrequences}). 
The radius $R$ is zero at the bifurcation
threshold. It increases and become comparable to $R_1$ as $\tau_w$ increases
past $\tau_c$ (Fig.~\ref{ampspro.fig}).
As $R \omega_m$ remains small compared to $R_1 \omega_s$,
the movement of the spiral tip in an horizontal plane 
can be described as a rapid rotation movement around a slowly moving 
"mean" point (Fig.~\ref{filamentsproing}). 
The bifurcation can thus be described as
in ref.\cite{henze} as a transition in the shape of
these slowly moving points, the "mean filament" , from a straight shape
to an helix of radius $R$ (with the same axis as the instantaneous
filament).

The amplitude of the sproing bifurcation can be measured by the radius $R$
of the helical mean filament\footnote{$R$ can be easily computed as 
$R=(R_{max}+R_{min})/2$ where
$R_{max}$ ($R_{min}$) is the maximum (minimum) distance of the 
instantaneous filament from its
axis in an horizontal plane 
($R_{min}$ should be counted
 negative value if the spiral tip trajectory 
 enlaces the axis of the helices during one rotation period).}.
The numerical simulations results (Fig.~\ref{ampspro.fig}) show
that $R$
behaves as $\sqrt{\tau_w-\tau_c}$ confirming the normal Hopf type of
the bifurcation. The motion of a filament point in an horizontal plane
is quasiperiodic with two frequencies (see Eq.~\ref{instfil} and 
Fig.~\ref{filamentsproing}) $\omega_m$ and $\omega_s$. As reported in
Table \ref{compfrequences}, the frequency $\omega_m$
closely agrees with the difference $\omega_1-\omega_{t,k_z=0}$ between
the stationary twisted scroll pulsation ($\omega_1$) and the imaginary part
of the unstable mode at $k_z=0$,
as expected from a Hopf bifurcation in a rotating frame. The other frequency
$\omega_s$ is equal to the twisted scroll pulsation $\omega_1$ at the 
bifurcation point but departs from it as one moves away from it. 

Some
insight
into the sproing bifurcation and the value
of $\omega_s$ can be gained by computing the local twist of the restabilized
scroll wave. The
mean filament can be taken as the central curve of a ribbon, one edge of which is the instantaneous filament of length. The
local twist of this ribbon is equal to (using Eq.~(\ref{white}) and 
(\ref{Wrhelix})
of appendix \ref{tw.app})
\begin{equation}
\tau=\frac{2\pi-Wr}{L}=\frac{\tau_w}{ 1+ (R\tau_w)^2}
\label{lotwist}
\end{equation}
where $L$ is the mean filament length and $\tau_w=2\pi/H$ is the twist 
imposed on the initial straight scroll wave.
Eq.~\ref{lotwist} shows that sproing decreases the twist. Moreover, when
the initial twist is increased the average helix radius $R$ also increases
so as to maintain the local twist $\tau$ approximately constant, as shown
in Fig.~\ref{ampspro.fig}b. The frequency $\omega_s$ appears to remain
close to the frequency
of twisted straight scroll wave with this value of the local twist, as
shown in Table \ref{compfrequences}.

\begin{table}
\begin{center}
\begin{tabular}{|c|c|c|c|c|c|c|}
\hline
$\tau_w$ & $\omega_{m}$ & $\omega_1(\tau_w) -Im(\sigma_t(k_z=0))$&
$\omega_1(\tau_w)$ & $\omega_s$&$\tau$&$\omega_1(\tau)$ \\ 
\hline
0.50     &  0.053        &  0.049 & 1.973  & 1.91 & 0.353 &  1.892 \\       
\hline
0.45     &  0.061         &  0.059 & 1.943  & 1.88 & 0.347& 1.889 \\
\hline
0.40     &  0.066         &  0.066 & 1.916  & 1.88 & 0.347& 1.889 \\  
\hline
0.355    &  0.063        &  0.063 & 1.891  & 1.88 & 0.349 & 1.890 \\
\hline
\end{tabular}
\caption{\label{compfrequences} For different values of the imposed
twist $\tau_w$, values of the pulsation of the 
restabilized mean filament
$\omega_m$, of the difference between the pulsation of the steady
scroll $\omega_1(\tau_w)$ 
and the imaginary part of the unstable translation mode for
$k_z=0$, the pulsation of the steady scroll  $\omega_1(\tau_w)$
, the pulsation of the
restabilized spiral around the mean filament
$\omega_s$ 
the local twist $\tau$ of the restabilized state
and the calculated corresponding pulsation $\omega_1(\tau)$. 
One notes that $\omega_s$ is close to $\omega_1(\tau)$ }
\end{center}
\end{table} 

\begin{figure}
\begin{center}
\begin{tabular}{ccc}
(a)& &(b)
\\
\includegraphics[height=4.cm,width=4.cm]{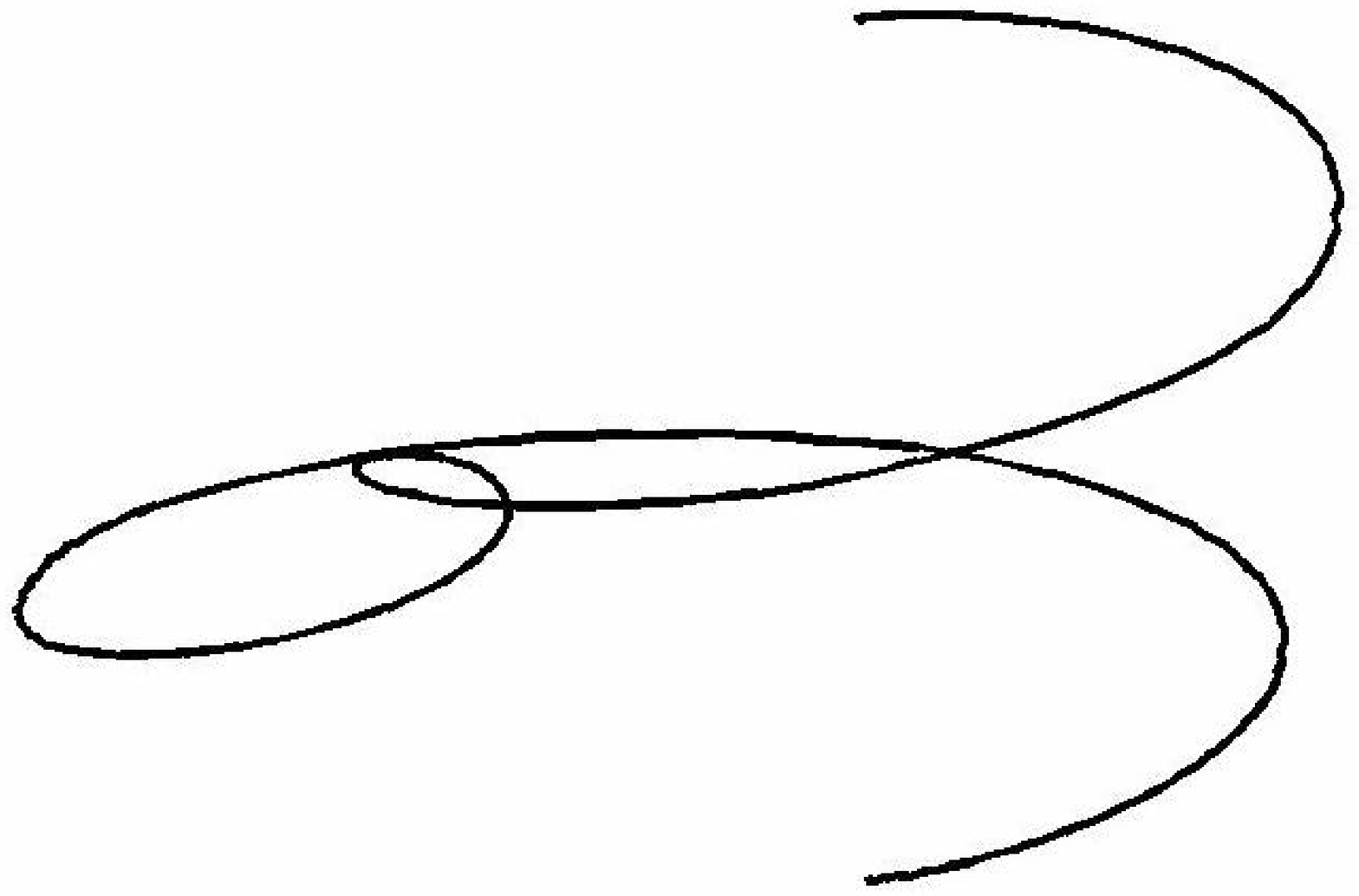}
& &
\includegraphics[height=4.cm,width=4.cm]{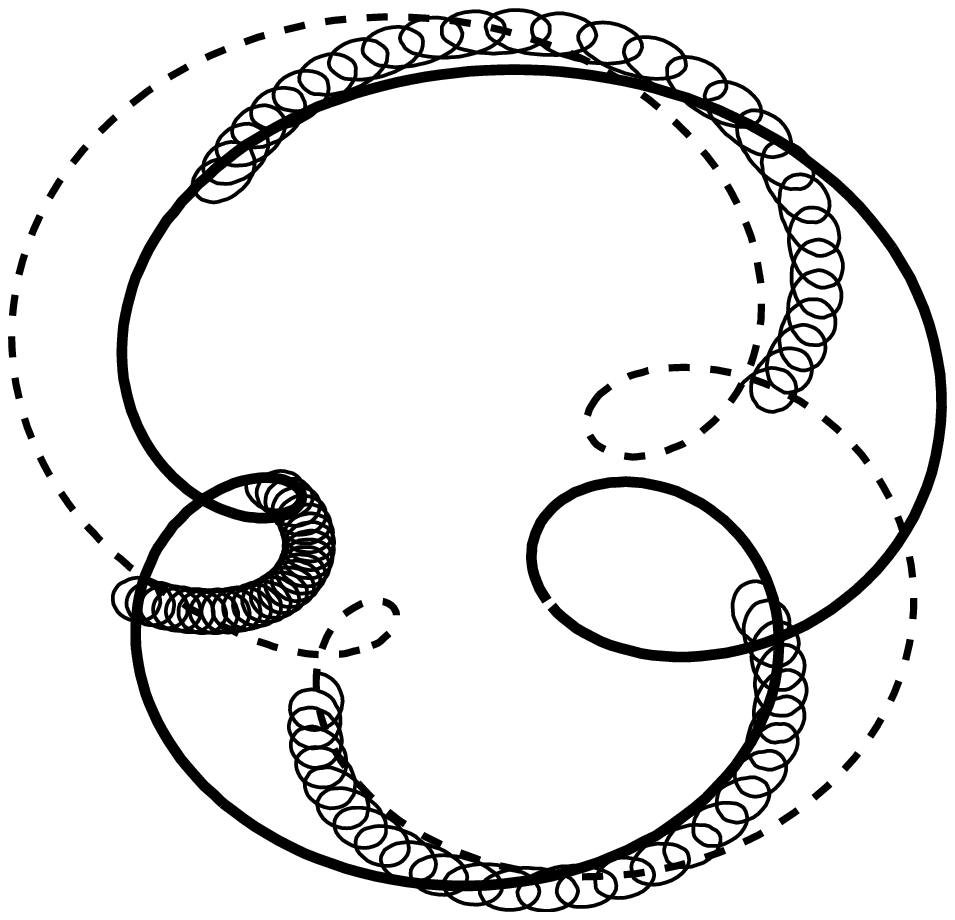}
\\
(c) & &(d)
\\[.1cm]
\includegraphics[height=4.cm,width=4.cm]{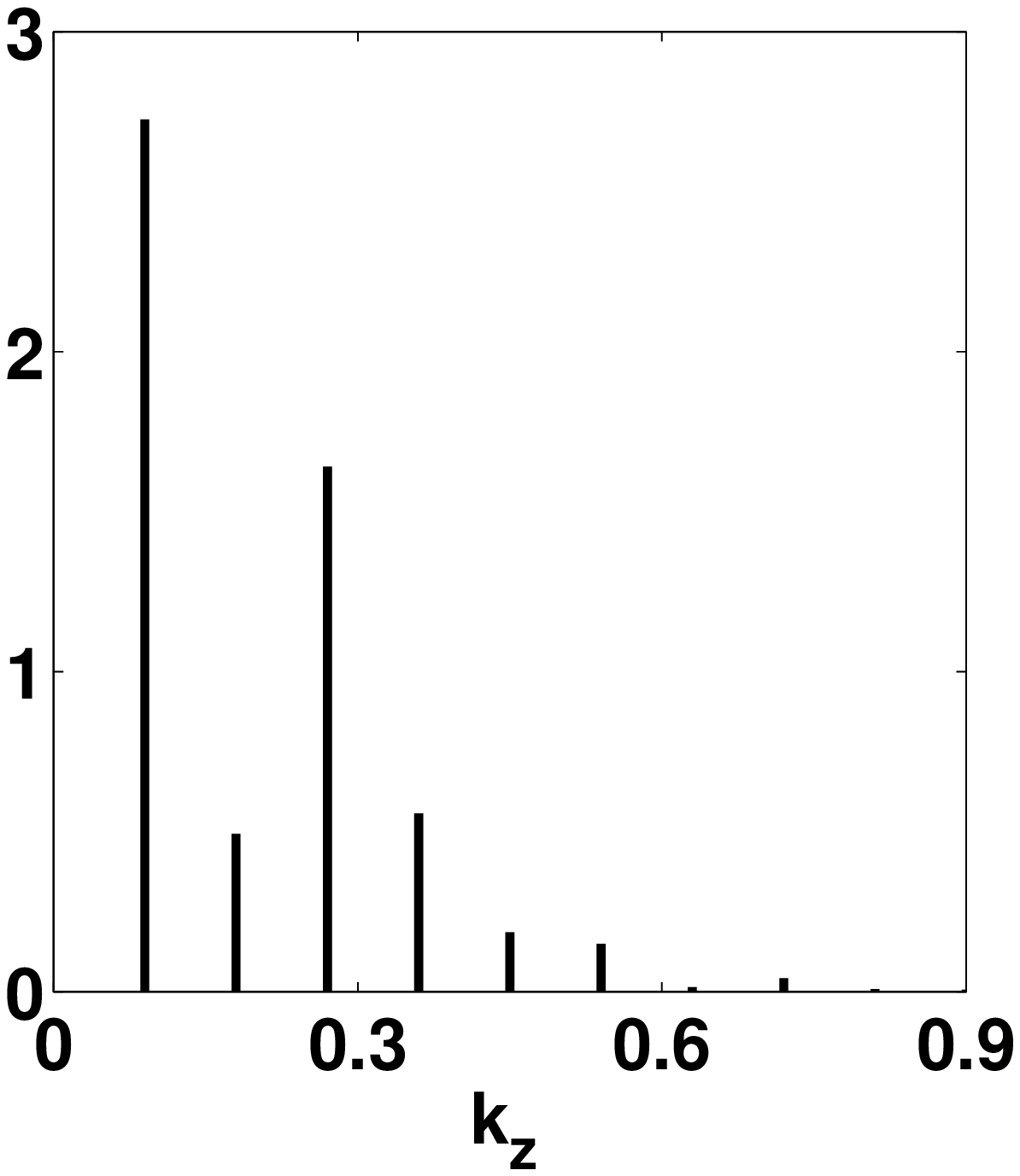}
& &
\includegraphics[height=4.cm,width=4.cm]{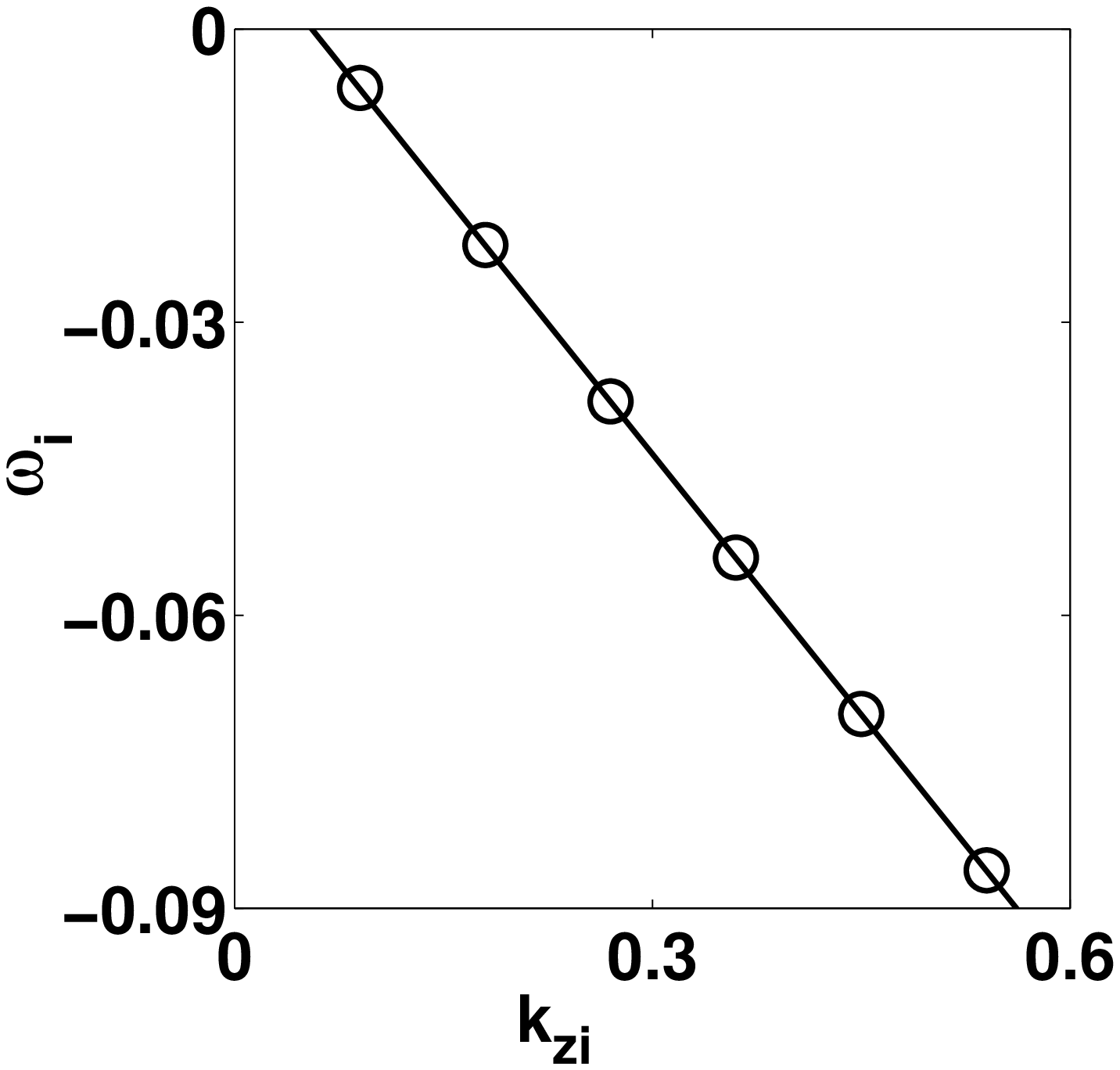}
\end{tabular}
\end{center}
\caption{\label{spro.fig} $a=0.8$, $b=0.1$, $\epsilon=0.025$, the simulation box size is ($128\times 128)\times 349)$, the space step is $dx=0.2$ and five turns of twist are imposed, which corresponds to $\tau_w=0.450$
(a): The bold solid line represents the restabilized mean
filament. The instantaneous filament and the quasi circular 
trajectories of instantaneous filaments in horizontal
planes are not shown. 
(b): The dashed and
solid bold lines represent a top view of the instantaneous filament
at  two given times. The three thin solid lines are typical
trajectories of the spiral tip in horizontal planes.
The motion of the spiral tip in each plane is the composition of a
fast counterclockwise and a slow clockwise rotations.
(c): Modulus of the space Fourier transform of the complex
mean filament position ($x(z)+i y(z)$). The amplitude of each mode 
is constant in time while its phase grows linearly in time with a
constant pulsation $\omega(k_z)$ shown in (d). For the Fourier modes
with an amplitude significantly different from zero, $\omega(k_z)$ is a
linear function of $k_z$.}
\end{figure} 

It is interesting to compare the above results with what happens when the 
initial condition contain $n$
turns of twist since the modes
with $k_z= j 2\pi/H, j=1,\cdots,n-1$ obey both $k_z<\tau_w$ and the periodic
boundary conditions. Analyzing moderate values of the imposed twist $\tau_w=
n 2\pi/H$ requires of course to extend the box height $H$  proportionally to
$n$ and restricted us to $n\le 5$.

The simplest interesting case occurs when a single unstable mode
with $k_z\ne 0$
can develop in the simulation box.
The linear stability results
shows that this can only happen close to the instability threshold
when the instability
growth rate is very low, otherwise the $k_z=0$ mode is also unstable.
This is achieved, for instance, for
$a=0.8,\ b=0.01 \mbox{ and } \epsilon=0.025$ 
and $n=5$ initial turns of twist in a box of height $H=613\times dx$ with
$ dx=0.15$.
The single
unstable mode $k_z=2\pi/H=0.069$ corresponds to a wavelength equal to the
box height $H$. With this parameter choice, a direct simulation
shows that 
the instability develops. The asymptotic state
is  similar to the 
previously described one for a single unstable mode at $k_z=0$.
The movement of the corresponding
instantaneous filament can be parameterized using polar coordinates
in each horizontal plane by:
\begin{equation}
x+iy=R_1 e^{i\omega_1 t+\tau_w z}+R_2 e^{+i\omega_2 t+(\tau_w-k_{z2}) z+\psi},
\end{equation}
where $R_1\simeq0.31$ and $\omega_1=1.85$ are  close to the radius 
and pulsation of
the straight twisted scroll  and where $R_2\simeq 0.1$ and $\omega_2=-0.047$
 are small
compared to $R_1$ and $\omega_1$. The wavenumber  $k_{z2}=0.069$ is equal to
the single unstable wavenumber that can develop in the simulation
box. The computation of $\omega_2$  shows that it is close of the
difference $\omega_1(\tau_w)-Im(\sigma_t(k_{z2})$. In contrast to the
single turn of twist case, the instantaneous filament shape 
is slightly different
from an 
helix since
its  radius varies along the vertical axis. The motion can nonetheless be 
interpreted in the same manner by considering that the scroll rotates uniformly around a
slowly moving helical mean filament of radius $R_2$ and pitch 
$2\pi/(\tau_w-k_{z2})$.

The case where several unstable
modes of the translation band can develop in the simulation
box, can only be studied in a box where several turns of twist are initially
imposed. 
 The asymptotic state reached in such a case after the instability
development, starting from an initially twisted straight scroll wave¸ is shown in Fig.~\ref{spro.fig}. It is significantly more 
complicated than in the single unstable mode case and the instantaneous 
filament shape is rather different from a simple helix.

In each horizontal plane, the spiral tip rotates uniformly
around a slowly moving point. The positions of these slowly moving
points can be numerically computed \footnote{This can be done by
considering the mean filament as the mean position of the
instantaneous filament over a rotation period, or by considering it as
the instantaneous center of rotation of the instantaneous filament in
a plane or by removing the high frequency peak in the 
Fourier spectrum of the instantaneous filament motion. These three methods
give very similar results.} and used to construct the position of a
mean filament. The motion  of the mean filament is found
to be well parameterized in each horizontal plane by
\begin{equation}
x+iy=\sum_{j=1...n} R_j e^{i (k_z^{(j)}z+\omega_j t+\phi_j})
\label{spro.param}
\end{equation}
where  $z$ denotes the vertical position of the plane and $k_{z}^{(j)}$
are the wavenumbers of the modes that have developed in the
simulation box. All the observed $k_{z}^{(j)}$ have the same sign as  $\tau_w$. As seen from Eq.~\ref{spro.param}, 
the corresponding modes have an amplitude $R_j$ 
which is constant in time and a phase $\omega_j t+\phi_j$
which evolves linearly in time. 
Moreover, the pulsations $\omega_j$ are linearly related to the $k_{z}^{(j)}$,
$\omega_j=\omega + k_{z}^{(j)}\,c$,
as shown in Fig.~\ref{spro.fig}.
Thus, the mean filament parameterization can be rewritten as 
\begin{equation}
x+iy= e^{i\omega t} F(z-ct)\ {\mathrm with}\  
F(z)=\sum_{j=1...n} R_j e^{i (k_z^{(j)}z}
\label{spro.param2}
\end{equation}

This explicitly shows that the mean filament deformation propagates as a 
nonlinear wave of constant shape in the vertical direction. It was
indeed directly
checked that the mean filament shape did not noticeably change at long times
in our simulation (i.e. for time intervals as long as $\Delta t=2000$).  
A direct computation of the twist
shows that it is almost
uniform and that its mean value is significantly
lower than the one obtained when  only $k_z=0$ mode can grow. In the case 
depicted in Fig.~\ref{spro.fig}, the mean local twist of the restabilized
state is $0.327$ whereas in the restabilized state when only the
$k_z=0$ mode can be destabilized, the mean local twist is equal to $0.347$.

The stability of the simple restabilized helices was tested in the parameter
regime were the more complex state of  Fig.~\ref{spro.fig} existed. 
To this end, a simulation was first performed with a single turn of twist in
a box height chosen such that the initial twist was equal to $\tau_w=0.45$.
It produced, as already described, a restabilized helix similar to the one
shown in Fig.~\ref{filamentsproing}. Five copies of this restabilized helix
were then vertically stacked. The resulting five turn helix
was used as the initial condition for a
simulation analogous to the one shown in Fig.~\ref{spro.fig}. It was observed
to evolve toward a complex state identical to the one shown
in Fig.~\ref{spro.fig}. This clearly shows that the simple helices are unstable
and strongly suggest that the complex state of  Fig.~\ref{spro.fig} is the
unique attractor in this parameter regime. We repeated this computation
with an initial twist of 0.357 and again a similar complex state was produced.

Systematically varying the initial twist would permit to see whether these
complex states 
directly appear at the instability threshold or 
arise from a secondary bifurcation of stable 
simple helical states. 
More generally,
further studying
the filament shapes in longer boxes with more unstable modes would be quite
instructive. Both tasks require more computer time and power than 
presently available to us and should be left for other studies.
 
\section{conclusion}
We have searched to gain and present a general view
of scroll wave linear instabilities and of their nonlinear developments for
a simple model of an isotropic excitable medium.
Different types of instabilities have
been shown to arise depending upon the band of modes to which they belong.
These different instabilities have been found to develop along different ways
and to give rise to distinctly different restabilized states that we have
endeavored to characterize.

The negative line tension type of instability has been found to occur
in the weakly excitable part of the phase diagram and to be strictly linked
to 2D spiral drift in an external field. In this respect, it seems worth to
try and better analyze the mechanisms of spiral drift change. 
The meandering instability is present in 3D 
in a larger parameter region than in 2D.  On the "small core"
side of the phase diagram, scroll wave meander in a regime 
where spirals are steadily rotating, as previously noted\cite{armit}. 
The long wavelength deformation of the meander band is
however not directly related to spiral drift. The introduction of twist
has been found to induce a deformation of the translation bands. 
Above a threshold twist, this gives rise to the
sproing instability which takes place a finite wavevector away from the 
translation mode. This has been shown to be a general consequence of 
3D rotational invariance and it explains that the sproing bifurcation is not
captured by small twist approaches. The bifurcated state arising from the growth
 of a single unstable mode
has been found to take a simple helical shape as previously described 
\cite{henze}. However, a more complex filament deformation has been observed
to result from the growth of several unstable modes. Simulations in larger
boxes appear needed to better analyze these states.

The present study appears worth extending along several other lines.
It certainly is important to investigate how the present results extend to more
realistic models of excitable media,
specially in the cardiac physiology context.
It will also be quite interesting to see how rotating  anisotropy 
\cite{pk,fk,wr,simber}.
or spatially
varying properties induce the different instabilities or interact with
them. 
Deeper insights
into the behavior of these complex waves in various situations may be gained
by developing and analyzing of reduced models reproducing the basic
phenomenology 
here uncovered. We hope to be able to report progress in this direction soon.
Advances in 3D visualization \cite{tombz,vis3d}  appear to render possible
detailed experimental
characterization of scroll wave instabilities and dynamics.
We hope that the present results will also be seen as a further motivation to 
undertaking this challenging task.
\appendix
\section{Numerical methods}

\subsection{Determination of the steady state}
\label{appss}

 Determining the stationary scroll waves consists in solving the nonlinear
eigenvalue problem (\ref{2dst2}). To this end, 
the equations are first discretized
on a polar grid of size $(N_r\times N_\phi)$. 
This provides a set  of $2 (N_r\times N_\phi)$ equations (the values

of the r.h.s of (\ref{2dst2}) on the grid points) with  $2
(N_r\times N_\phi)+1$ unknowns (the values of the field $u_0,\ v_0$ on
the points
of the grid and the pulsation $\omega$). This indetermination
comes from the rotational  invariance of the problem and can be taken
care of by setting the value of $u_o(N_r,N_\phi)$ to 0.5. One thus has
to find the zeros of $2 (N_r\times N_\phi)$ non linear equations  of 
$2 (N_r\times N_\phi)$ variables. 

 This can be done accurately using  Newton's method. The linear
operator involved in Newton's method is inverted using an
iterative technique (biconjugate gradient \cite{NRgradbiconj})
 and the starting point is
either the result of a direct numerical simulation interpolated on the
polar grid or the result of a previous computation with slightly
different parameters. We found that the Newton's method always
converged and that the convergence was exponential and allowed to
reach accuracies of order $10^{-8}$ in $L_2$ norm\footnote{We define
$\|u,v\|_2=\sum (u_{i,j}^2+v_{i,j}^2)$} in a few steps
(about 10 which take about an hour using a DEC alpha PWS 500 Workstation for
a grid of $80\times 160$ points) whereas direct numerical simulations 
allowed only an 
accuracy of order unity in $L_2$ norm.

\subsection{Computation of the linear stability spectrum}
\label{appls}
The eigenvectors of ${\mathcal{L}}_{k_z}$ are denoted by
$e_1, e_2,\cdots$ and indexed according to the real part of the corresponding
eigenvalues
$\sigma_1\ge\sigma_2\ge\cdots$. In order to accurately compute the
eigenvalues of of ${\mathcal{L}}_{k_z}$ of
largest real parts, we have used an 
iterative method proposed and analyzed in \cite{Gold87}. 
It is briefly described in this appendix 
and some details of our implementation are
provided.

The idea of the algorithm is to diagonalize a projection of
${\mathcal{L}}_{k_z}$ in a subspace (approximately) spanned by the eigenvectors
$e_1,\cdots,e_m$
corresponding to a given number $m$ of eigenvalues of largest real parts.

The algorithm proceeds in three main steps.

The first step consists in creating an appropriate vector $x_1$ for generating
the diagonalization subspace. A suitable choice is to take 
$x_1=\exp({\cal L}_{k_z}t)x_0$ for a generic vector $x_0$. This
suppresses
the components of $x_1$ on the eigenvectors on high index
[note that these eigenvectors correspond to eigenvalues of large modulus].
 The end 
consequence is that truncation at level $m$ produces an error of
order $\exp[t(Re(\sigma_{m+1}-\sigma_i)]$ on the representation of
the $i$-th eigenvector (for $i<m$)\cite{Gold87}.  
In practice,  multiplication by the exponential  is approximately
achieved
by
computing by iterations
$x_1=(1+dt{\mathcal{L}}_{k_z})^{t_0/dt} x_0$, for
an arbitrary vector $x_0$,  a sufficiently large
integer $t_0/dt$ and a sufficiently small
$dt$ to prevent  the time stepping scheme from being unstable (this
means that $dt$ is lower than $1/\mbox{max}(\|\lambda_i\|)$, here
$dt\simeq 10^{-5}$).

The second step consists in generating an appropriate subspace $E$ for
diagonalization. This is taken to be  the space $E$ spanned
by $(G({\mathcal{L}}_{k_z}))^n x_1,\ n=0...m-1$
where $G$ is an 
polynomial, the choice of which is discussed below.
Explicitly,
the computation
of an orthonormal base of $E$ and of the matrix $A=a_{i,j}$ of the projection
of 
$(G({\mathcal{L}}))_m$ on $E$ proceeds recursively as follows. Let $x_1,\ ...,\ x_n$ be the n first element
of the base of $E$ and $ G({\mathcal{L}}_{k_z})x_n=y_{n}$. Then, at the next 
step
we compute:
\begin{equation}
x_{n+1}=\frac{y_{n}-\sum_{i=1...n}(y_{n}.x_i)x_i}{\| y_{n}-\sum_{i=1...n}(y_{n}.x_i)x_i\|_2}.
\end{equation}
The construction ends at step $m$.
This building of an
orthonormalized base of $E$ presents the advantage of decreasing 
the contribution of the first eigenmodes in the
elements of higher order of the base. Otherwise, this contribution would be
dominant and would result in a lower accuracy in the computation.
In our case, the scalar product is defined by
a discretized
version of the standard scalar product (\ref{scaprod}):
\begin{equation}
(x_i,y_{j})=\sum_{i_r} \sum_{i_\theta} x_i(i_r,i_\theta)
y_{j}(i_r,i_\theta )\,  i_r\, dr d\theta dr
\end{equation}

The third step consists in the diagonalization of an appropriate truncation 
of
$(G({\mathcal{L}}_{k_z}))$. Since
$E$ is not invariant
under the application of $(G({\mathcal{L}}_{k_z}))$, the orthogonal
projection $G_m$ 
of $G({\mathcal{L}}_{k_z})$  onto $E$ is considered.
Its matrix elements are:
\begin{equation}
g_{i,j}=(x_i,y_{j}),\ (i,j)\in \{1...m\}\times \{1...m\} 
\end{equation} 
The obtained matrix is diagonalized and both its   
eigenvectors  and its 
eigenvalues are computed. 
It is checked at the end that the leading eigenvectors of $G_m$
are good approximations of the leading eigenvectors of
${\mathcal{L}}_{k_z}$.

 The choice of the polynomial $G$ is of some
importance. Indeed,  the simplest and computationally most efficient
choice would be to take $G(X)=X$. This  would result in the
amplification of the contribution of the 
eigenmodes of ${\mathcal{L}}_{k_z}$ of large index [i.e corresponding
to eigenvalues of large modulus] and, ${\mathcal{L}}_{k_z}$
being ill-conditioned,  would  prevent the success of the method.
We have used
$G(X)=(1+dtX)^{t_1/dt}$, with $t_1$ chosen large enough to make 
$G({\mathcal{L}}_{k_z})$ differ significantly from the identity (a typical
value was $t_1=0.5$). Despite the
increase in computational cost, this choice significantly increases
the accuracy of the method by decreasing the contribution of 
the eigenmodes of  negative eigenvalues.
Using $m=50$, the 10 most unstable
eigenmodes and eigenvalues are obtained with a good accuracy: $\|(\sigma
-{\mathcal{L}}_{k_z})(u_1,\ v_1)\|_2<10^{-6}$.

\subsection{Direct numerical simulations}
\label{appdns}
Direct numerical simulations of three dimensional
excitable media were performed using a forward Euler explicit
timestepping scheme . The diffusion operator was evaluated using finite
differences and a 19 points formula \cite{dow97},

\begin{eqnarray}
6 dx^2 \delta u_{i,j,k}&=& -24u_{i,j,k}\nonumber\\ 
&+& 2(u_{i+1,j,k}+ u_{i-1,j,k}+ u_{i,j+1,k}+ u_{i,j-1,k}+ u_{i,j,k+1}+
u_{i,j,k-1})\nonumber\\
&+&u_{i+1,j+1,k}+ u_{i+1,j-1,k}+ u_{i+1,j,k+1}+
u_{i+1,j,k-1}\nonumber\\ 
&+&u_{i-1,j+1,k}+ u_{i-1,j-1,k}+ u_{i-1,j,k+1}+
u_{i-1,j,k-1}\nonumber\\ 
&+&u_{i,j+1,k+1}+ u_{i,j+1,k-1}+ u_{i,j-1,k+1}+ u_{i,j-1,k-1}
\end{eqnarray}
 This method has two main advantages compared with the classical 7
points formula. First its stability limit  allows greater time steps
and therefore the computing time in spite of the additional
operations involved in the Laplacian evaluation is found to be
1.3 times smaller. Second, the  error made when evaluating the
diffusion operator is isotropic at the dominant order(order $dx^2$), 
whereas with the 7 points formula it depends of the grid orientation.  

No-flux 
boundary conditions
$\vec{n}.\vec{\nabla}u=0$ are imposed on the vertical sides of the box and
either no-flux or periodic boundary conditions are chosen on the
top and bottom 
boundaries.
 
The position of the instantaneous filament in the horizontal planes
is computed as the intersection of a
$u=0.5$ and a $v=0.75\, (0.5 a-b)$ isosurfaces.
 
\section{Spiral drift in an external field}
\label{drift.app}
In the presence of a small external field $\boldmath{E}$,
the spiral rotation center
drifts at a constant velocity proportional to the field 
magnitude but at an
angle with the field direction \cite{ag,stein,bel} as given by 
Eq.~(\ref{2ddrift}) of the main text.
The drift coefficients $\alpha_{\parallel}$ and
$\alpha_{\perp}$ have been computed in the free boundary limit 
($\epsilon \ll 1$) both for small core \cite{mak} and large core spirals
\cite{hk2}. We derive here a general formula (Eq.~(\ref{driftcoef}) below)
for the drift coefficient
$\alpha=\alpha_{\parallel}+i \alpha_{\perp}$ as a matrix element between
the translation eigenvector $(u_t,v_t)$ of $\cal{L}$ and the corresponding
right eigenvector $(\tilde{u}_t,\tilde{v}_t)$ of  
$\cal{L}$.

With an added external field $\boldmath{E}$, Eq.~(\ref{eq1},~\ref{eq2}) read
\begin{eqnarray}
\partial_t u&=& \nabla^2u+f(u,v)/\epsilon -\boldmath{E}.\nabla u,
\label{eqE1}
\\
\partial_t v&=& g(u,v).
\label{eqE2}
\end{eqnarray}
We choose a coordinate system with the $x$-axis along the field direction
and corresponding polar coordinates $(r,\theta)$.
We analyze the motion of a counterclockwise rotating spiral which is stationary,for
$\boldmath{E}=0$,
in the rotating referential $(r,\phi)$ with $\phi=\theta-\omega_1 t$. 
Since $\partial_x=\cos(\theta)\partial_r
-\sin(\theta)/r \partial_{\theta}$, the field term $ 
\boldmath{E}.\nabla u$ can be written in the rotating referential
\begin{eqnarray}
-E\partial_x u&=&-E[\cos(\phi+\omega_1 t)\partial_r u-\sin(\phi+\omega_1 t)/r
\partial_{\phi}u],\nonumber\\
&=& -E/2 [\exp(i\omega_1 t) \exp(i\phi)(\partial_r u+i\partial_{\phi} u/r)
+ c.c.].
\end{eqnarray}

As will be seen below, secular terms appear 
when the $E$ term on the r.h.s. of Eq.~(\ref{eqE1}))
is treated in perturbation.
Their origin is, of course, the
induced spiral drift. Anticipating this phenomenon, we suppose that the spiral 
rotates steadily in the frame with coordinates $(r,\theta)$
which drifts with respect to the lab frame with coordinates $(x,y)$.
That is,
\begin{eqnarray}
x&=&x_0(t)+ r \cos(\phi+\omega_1 t),\\
y&=&y_0(t)+ r \sin(\phi+\omega_1 t),
\end{eqnarray}
and
\begin{equation}
\partial_t|_{r,\phi}=\partial_t|_{x,y}+\omega_1\partial_{\phi}+\dot{x}_0\partial_x+
\dot{y}_0\partial_y .
\label{driftcoor1}
\end{equation}
The supplementary terms proportional to $\dot{x}_0$ and $\dot{y}_0$
should be chosen to cancel
the unwanted secular terms. This determines the spiral drift.
Explicitly, we rewrite Eq.~(\ref{driftcoor1}) using the rotating coordinates
as
\begin{equation}
\partial_t|_{x,y}=\partial_t|_{r,\phi}-\omega_1\partial_{\phi}-1/2[(\dot{x}_0-i
\dot{y}_0)\exp(i\omega_1 t)\exp(i\phi)(\partial_r +i\partial_{\phi} /r)+c.c.].
\label{driftcoor2}
\end{equation}
Substitution of this formula in Eq.~(\ref{eqE1},\ref{eqE2}) and linearization
under the form $u=u_0(r,\phi)+ u_p \exp(i\omega_1 t)+c.c.,\ 
v=v_0(r,\phi)+ v_p \exp(i\omega_1 t)+c.c.$ gives,
\begin{equation}
(i\omega_1 -{\mathcal{L}})\left( \begin{array}{c} u_p\\v_p\end{array}\right)
-1/2 (\dot{x}_0-i\dot{y}_0)\left( \begin{array}{c} u_t\\v_t\end{array}\right)
=-E/2 \left( \begin{array}{c} u_t\\0\end{array}\right)
\end{equation}
Multiplying by the left eigenvector $\mathcal{L}$ of eigenvalue $i\omega_1$
would show the need for secular terms if we had not introduced
the drift terms. Here, however it simply determines the drift as
\begin{equation}
\dot{x}_0-i\dot{y}_0=E\,\frac{\langle\tilde{u}_t,u_t\rangle}{
\langle\tilde{u}_t,u_t\rangle+
\langle\tilde{v}_t,v_t\rangle} 
\label{fdrift}
\end{equation}
Equivalently, this gives the sought formula for the drift coefficients
\begin{equation}
\alpha_{\parallel}-i\alpha_{\perp}=\frac{\langle\tilde{u}_t,u_t\rangle}{
\langle\tilde{u}_t,u_t\rangle+
\langle\tilde{v}_t,v_t\rangle}
\label{driftcoef}
\end{equation}

\begin{table}
\begin{tabular}{|c|c|c|}
\hline
 a &  r.h.s of Eq.~(\ref{driftcoef} )& $\alpha_{\parallel}-i\alpha_{\perp}$ \\
\hline
0.44 & -2.05  -0.78i& -1.97-0.84i\\
\hline
0.62 & 3.5 - 0.47i& 3.4 -0.42i\\
\hline
0.7  &1.59 - 0.80i & 1.62 -0.83i\\
\end{tabular}
\caption{\label{table.drift} For  $b=0.01$, 
$\epsilon=0.025$ and several values of $a$, the drift
coefficients given by Eq.~(\ref{driftcoef}) computed using the 
eigenmodes of  ${\mathcal{L}}_{k_z}$ and of its
adjoint 
and  the
drift coefficients $\alpha_\parallel-i\,
\alpha_\perp$ measured in direct numerical simulations.}
\end{table}

\section{Twist and writhe of a ribbon}
\label{tw.app}

We recall the definition of
quantities  associated to
closed ribbons and some useful mathematical properties 
for analyzing the ``sproing''
bifurcation. In particular we give a mathematical definition of the 
twist and its value
in the case of a uniformly
twisted ribbon with an helical central curve.

The local twist of a ribbon is classically defined as (\cite{love})
\begin{equation}
 \tau= \left(\frac{d}{ds}\left(\vec{p}\right)\wedge\vec{p}\right).\vec{t}
\label{deftwistlocal}
\end{equation}
where $\vec{t}$ is the unit tangent vector to the mean curve of the
ribbon,  $s$ is
the curvilinear coordinate along the mean curve  and $\vec{p}$, the
unit vector perpendicular to that curve  that directs the line
intersecting one of the edges of the ribbon. 
The twist measures  the spatial rotation rate of the
edges of the ribbon  around the mean curve. 

For a closed ribbon, the linking number $L_k$ is the integer
which measures the entanglement of the two edges of the
ribbon. $L_k$ is a topological invariant which is constant under a continuous
deformation of the ribbon (as long as it does not intersect itself).

The linking number is related to the integral of twist by a
formula \cite{white} (which has been popularized
by its use in a molecular biology context)
\begin{equation}
L_k=W_r+\frac{1}{2\pi}\int_s \tau ds\label{white}
\end{equation}
The writhing number $W_r$ depends only of the mean curve $\vec{r}(s)$of the ribbon
and is equal to:
\begin{equation}
 W_r=\frac{1}{4\pi}\int ds\int ds'
\frac{
\partial_s\vec{r}(s)\wedge \partial_s\vec{r}(s').[\vec{r}(s)-\vec{r}(s')]
}
{
\|\vec{r}-\vec{r'}\|^3
}
\end{equation}

The tangent vector to $r(s)$ traces a closed curve on the unit sphere as $r(s)$
goes around the ribbon. The writhing number is also
equal, up to an even integer, 
to the area enclosed on the unit sphere by this closed curve
divided by $2\pi$\cite{PNASfuller}. 

An example of interest is 
the writhing number of a single turn of an helix of radius $R$ and pitch $H$
(linking the
two free ends by a non self intersecting planar curve to obtain a
closed curve). The tangent vectors curve encloses a 
spherical cap on the unit sphere of normalized area equal
to $(1-\cos \theta)$ where
$\theta$ is the angle made by the tangent vector with the vertical
axis. This gives the writhing number (up to an even integer 
which is seen to be zero by using the continuity 
of $Wr$ and the fact that the writhing
number of a straight line is equal to zero),
\begin{equation}
Wr=1- \frac{1}{\sqrt{1+(2\pi R/H)^2}}\label{Wrhelix}
\end{equation}
\section{Link with averaged equation}
\label{keenav.sec}
In ref.~\cite{keen3d}, equations were derived for the motion of the mean 
scroll filament and the evolution of twist for a weakly twisted and weakly
curved scroll wave. It was subsequently noted \cite{bik3d} that many 
coefficients in ref.~\cite{keen3d} original equations were identically zero
and that only four non-trivial ones remained to be determined. This approach
has recently been extended to take into account fiber rotation anisotropy
\cite{simber}.

In this
appendix, we find it
of some interest to explicitly relate the averaged equations of 
ref.~\cite{keen3d,bik3d} to the 
computations that are performed in the main part of the present paper. 
One coefficient in the equations of ref.~\cite{keen3d,bik3d} is given by
the quadratic scroll rotation frequency dependence at small twist 
(Eq.~(\ref{omsmtw})). Unsurprisingly, the three other ones are given by the
curvature of the rotation and translation bands around the corresponding 
symmetry eigenvalues. This explicitly confirms that the sproing instability
cannot be captured in the limit considered to derive the averaged equations 
since the instability takes place a finite
distance away from the translation symmetry eigenvalues on the translation
bands.

We provide a simple-minded derivation of the equations ref.~\cite{keen3d,bik3d}
using a Cartesian frame instead of the more sophisticated intrinsic
mean filament coordinates used in ref.~\cite{keen3d}. This limits us
to consider a weakly inclined filament but we can proceed very similarly
to the derivation of spiral drift in appendix \ref{drift.app} and the extension
to non-isotropic or non-homogeneous medium is straightforward (but not 
considered here).

We denote the fixed Cartesian coordinates by $(x,y,z)$ and by $(r,\phi,z)$ 
the cylindrical coordinates
of a frame rotating at the 2D spiral frequency and centered on the
mean filament position $(x_0(t,z),y_0(t,z))$,
\begin{eqnarray}
x&=&x_0(t,z)+ r \cos[\phi+\omega_1 t +\psi(t,z)]\\
y&=&y_0(t,z)+ r \sin[\phi+\omega_1 t+\psi(t,z)]
\end{eqnarray}
The corresponding relations between the time and vertical ($z$) derivatives 
in the two 
referentials are thus,
\begin{eqnarray}
\partial_t|_{r,\phi}&=&\partial_t|_{x,y}+\omega_1 \partial_{\phi}+
\partial_t\psi\,\partial_{\phi}+
\partial_t{x}_0\,\partial_x+
\partial_t{y}_0\,\partial_y\\
\partial_z|_{r,\phi}&=&\partial_z|_{x,y}+\partial_z\psi\,\partial_{\phi}
+\partial_z{x}_0\,\partial_x+
\partial_z{y}_0\,\partial_y
\end{eqnarray}
These relations can be rewritten
 using
$\partial_x+i\partial_y=\exp[i\phi+i\omega_1 t+i\psi(z,t)]
[\partial_r+i\partial_{\phi}/r]$ and introducing $w_0=x_0-i y_0$,
\begin{eqnarray}
\partial_t|_{x,y}&=&\partial_t|_{r,\phi} -\omega_1 \partial_{\phi}
-\partial_t\psi\,\partial_{\phi}-
\frac{1}{2}\left[\partial_t{w}_0\,e^{i\phi+i\omega
_1 t+i \psi}(\partial_r+i\partial_{\phi}/r)+c.c.\right]\label{t}\\
\partial_z|_{x,y}&=& \partial_z|_{r,\phi}- \partial_z\psi\,\partial_{\phi}-
\frac{1}{2}\left[\partial_z w_0\,e^{i\phi+i\omega
_1 t+i\psi}\,(\partial_r+i\partial_{\phi}/r)+c.c.\right]\\
\partial^2_{zz}|_{x,y}&=& \partial^2_{zz}|_{r,\phi}
+(\partial_z\psi)^2\partial^2_{\phi\phi}-2\partial_z\psi
\partial^2_{\phi z}
-\left[\partial_z{w}_0
\,e^{i\phi+i\omega
_1 t+i\psi}(\partial^2_{rz}+i\partial^2_{\phi z}/r)+c.c.\right]
\nonumber\\
&+&\frac{1}{4}\left[(\partial_z{w}_0)^2\,  e^{2(i\omega
_1 t+i\psi)}\left(e^{i\phi}(\partial_r+i\partial_{\phi}/r)\right)^2+c.c.\right]
\nonumber\\
&+&\frac{1}{2} |\partial_z{w}_0|^2 \nabla^2_{2D}-\frac{1}{2}
\left[\partial^2_{zz}{w}_0
e^{i\phi+i\omega
_1 t+i\psi}\,(\partial_{r}+i\partial_{\phi}/r)+c.c.\right]
\nonumber\\
&-&\partial^2_{zz}\psi\,\partial_{\phi}+\partial_z\psi\, [\partial_zw_0
 e^{i\phi+i\omega_1 t+i\psi}\,(\partial^2_{r\phi}+i\partial^2_{\phi\phi})+c.c.]
\label{zz}
\end{eqnarray}
With Eq.~(\ref{t}, \ref{zz}) , the governing reaction-diffusion
equations (\ref{eq1},\ref{eq2}) become
\begin{eqnarray}
\left(\partial_t -\omega_1 \partial_{\phi}-\nabla^2_{2D}\right)u-
f(u,v)/\epsilon
&=&
\partial_t\psi\,\partial_{\phi}u+
\frac{1}{2}\left[\partial_t{w}_0\,e^{i\phi+i\omega
_1 t+i\psi}(\partial_ru+i\partial_{\phi}u/r)+c.c.\right]
+
\partial^2_{zz}u
\nonumber\\
&+& (\partial_z\psi)^2\partial^2_{\phi\phi}u-2\partial_z\psi
\partial^2_{\phi z}u-\left[\partial_z{w}_0
\,e^{i\phi+i\omega
_1 t +i\psi}(\partial^2_{rz}u+i\partial^2_{\phi z}u/r)+c.c.\right]
\nonumber\\
&+&\frac{1}{4}\left[(\partial_z{w}_0)^2\,  e^{2(i\omega
_1 t+i\psi)}\left(e^{i\phi}(\partial_r+i\partial_{\phi}/r)\right)^2u+c.c.\right]
\nonumber\\
&+&\frac{1}{2} |\partial_z{w}_0|^2 \nabla^2_{2D}u-\frac{1}{2}
\left[\partial^2_{zz}{w}_0
e^{i\phi+i\omega
_1 t+i\psi}\,(\partial_{r}u+i\partial_{\phi}u/r)+c.c.\right]
\nonumber\\
&-&\partial^2_{zz}\psi\,\partial_{\phi}u+\partial_z\psi\, [\partial_zw_0
 e^{i\phi+i\omega_1 t+i\psi}\,(\partial^2_{r\phi}u+i\partial^2_{\phi\phi}u/r)
+c.c.]
\label{tdk1}\\
\left(\partial_t -\omega_1 \partial_{\phi}\right)v-g(u,v)&=&
\partial_t\psi\,\partial_{\phi}v+
\frac{1}{2}\left[\partial_t{w}_0\,e^{i\phi+i\omega
_1 t+i\psi}(\partial_rv+i\partial_{\phi}v/r)+c.c.\right]
\label{tdk2}
\end{eqnarray}
For a weakly curved and weakly twisted scroll
wave, the r.h.s. of Eq.~(\ref{tdk1},\ref{tdk2}) can be treated in perturbation
starting from the 2D spiral fields $(u_0,v_0)$. At first order, the
inhomogeneous r.h.s are a superposition of time independent terms and of
terms oscillating at frequencies $\omega_1$ and $2 \omega_1$. One can
therefore seek $(u,v)$ in perturbation as
\begin{equation}
\left( \begin{array}{c} u\\v\end{array}\right)=
\left( \begin{array}{c}  u_0\\v_0\end{array}\right)+\left[
\left( \begin{array}{c}  u_1^{(2\omega_1)}\\ v_1^{(2\omega_1)}
\end{array}\right)
e^{2i\omega_1 t+2i\psi} +c.c.\right]+
\left[\left( \begin{array}{c}  u_1^{(\omega_1)}\\ v_1^{(\omega_1)}
\end{array}\right)
e^{i\omega_1 t+i\psi} +c.c.\right]+
\left( \begin{array}{c}  u_1^{(0)}\\v_1^{(0)}\end{array}\right)
\label{pertk}
\end{equation}
where $( u_1^{(2\omega_1)},v_1^{(2\omega_1)},
(u_1^{(\omega_1)},v_1^{(\omega_1)}$
and $(u_1^{(0)},v_1^{(0)})$ are time-independent
functions characterizing the three
different first order perturbative corrections.
 
The $2 \omega_1$-functions obey
\begin{equation}
(i2\omega_1-{\cal L})
\left( \begin{array}{c}  u_1^{(2\omega_1)}\\ v_1^{(2\omega_1)}
\end{array}\right)
  =
\frac{1}{4}(\partial_z{w}_0)^2\,
\left( \begin{array}{c}\left(e^{i\phi}(\partial_r+
i\partial_{\phi}/r)\right)^2u\\0
\end{array}\right)
\end{equation}
 The operator $(i2\omega_1-{\cal L})$ is invertible and
the $2 \omega_1$ functions $( u_1^{(2\omega_1)},v_1^{(2\omega_1)})$
can be determined. They simply describe the inclined circular
core which is viewed as elliptical in the chosen $xy$ coordinates 
(these terms are absent
in the filament coordinates used in ref.~\cite{keen3d} where in effect
$\partial_zw_0$ is zero).
 
On the contrary, the $\omega_1$ and constant functions arise from resonant
forcing and can only be determined when solvability conditions are verified.
 
The $\omega_1$ functions obey,
\begin{equation}
(i\omega_1-{\cal L})
\left( \begin{array}{c}  u_1^{(\omega_1)}\\ v_1^{(\omega_1)}
\end{array}\right)
  =\frac{1}{2}\partial_t{w}_0\
\left( \begin{array}{c}u_t\\v_t\end{array}\right)
-\frac{1}{2}
\partial^2_{zz}{w}_0 \left( \begin{array}{c}u_t\\0\end{array}\right)
+\partial_z\psi\, \partial_zw_0
\left( \begin{array}{c}e^{i\phi} (\partial^2_{r\phi}u_0+i
\partial^2_{\phi\phi}u_0/r)\\0\end{array}\right)
\label{eqo1}
\end{equation}
Since $(u_t,v_t)$ is an eigenvector of $\cal{L}$ with eigenvalue $i\omega_1$,
Eq.~(\ref{eqo1}) is solvable only if its r.h.s has no component on this
eigenvector. More explicitly, one obtains by multiplying Eq.~(\ref{eqo1})
by the associated left eigenvector,
\begin{equation}
\partial_t{w}_0 - \partial^2_{zz}{w}_0
\frac{\langle \tilde{u}_t,u_t\rangle}{\langle \tilde{u}_t,u_t\rangle+
\langle \tilde{v}_t,v_t\rangle}+\partial_z\psi\, \partial_zw_0
\frac{\langle \tilde{u}_t,e^{i\phi} (\partial^2_{r\phi}u_0+i
\partial^2_{\phi\phi}u_0/r)\rangle}{\langle \tilde{u}_t,u_t\rangle+
\langle \tilde{v}_t,v_t\rangle}=0
\label{kee1}
\end{equation}
The last term on the l.h.s should vanish by rotational invariance since
simply inclining a twisted scroll cannot induce its drift. This explicitly
follows from Eq.~(\ref{orth3d}) since
\begin{equation}
\langle \tilde{u}_t,e^{i\phi} (\partial^2_{r\phi}u_0+i
\partial^2_{\phi\phi}u_0/r)\rangle=\langle \tilde{u}_t,(1+i\partial_{\phi})u_t
\rangle=0.
\end{equation}
 
Eq.~(\ref{kee1}) [without the last term]
 is the equation of motion for the mean filament obtained in
ref.\cite{keen3d,bik3d}. In the limit considered, the mean filament motion
is independent of the scroll twist and only depends on the filament curvature
($\partial^2_{zz}{w}_0$) with a coefficient which gives both
the small $k_z^2$
dependence of the translation bands (Eq.~(\ref{fop})) and spiral drift
in an external field (Eq.~(\ref{driftcoef})).
 
The time independent component $(u_1^{(0)},v_1^{(0)})$ of the first
0 correction (\ref{pertk}) obey
\begin{equation}
-{\cal L}
\left( \begin{array}{c}  u_1^{(0)}\\ v_1^{(0)}
\end{array}\right)
  =\partial_t\psi\,
\left( \begin{array}{c}\partial_{\phi}u_0\\\partial_{\phi}v_0\end{array}\right)
-\partial^2_{zz}\psi
\left(\begin{array}{c}\partial_{\phi}u_0\\0\end{array}\right)
+(\partial_z\psi)^2
\left( \begin{array}{c}\partial^2_{\phi\phi}u_0\\0\end{array}\right)
+\frac{1}{2} |\partial_z{w}_0|^2
\left( \begin{array}{c}\nabla^2_{2D}u_0\\0\end{array}\right)
\label{eqo0}
\end{equation}
Again since the rotation mode is an eigenvector with eigenvalue zero of
${\cal L}$, this equation can be solved only if
\begin{equation}
 \partial_t\psi - \partial^2_{zz}{\psi}
\frac{\langle \tilde{u}_{\phi},u_{\phi}\rangle}{\langle \tilde{u}_{\phi},
u_{\phi}\rangle+
\langle \tilde{v}_{\phi},v_{\phi}\rangle}+(\partial_z\psi)^2
\frac{\langle \tilde{u}_{\phi},\partial^2_{\phi\phi}u_0
\rangle}{\langle \tilde{u}_{\phi},u_{\phi}\rangle+
\langle \tilde{v}_{\phi},v_{\phi}\rangle}+\frac{1}{2} |\partial_z{w}_0|^2
\frac{\langle \tilde{u}_{\phi}, \nabla^2_{2D}u_0
\rangle}{\langle \tilde{u}_{\phi},u_{\phi}\rangle+
\langle \tilde{v}_{\phi},v_{\phi}\rangle}
=0
\label{kee2}
\end{equation}
Rotational invariance again implies that the last term on the l.h.s.
of Eq.~(\ref{kee2}) vanishes (since
simply inclining a scroll wave cannot change
its rotation frequency) as explicitly shown below (Eq.~(\ref{dil}).
 The remaining
Eq.~(\ref{kee2}) is the
equation obtained in ref.~\cite{keen3d,bik3d} for the scroll twist dynamics.
The coefficient of $(\partial_z\psi)^2$ simply describes the twist dependence
of steady scroll rotation
frequency (Eq.~(\ref{omsmtw})) while the coefficient of $\partial^2_{zz}{\psi}$ governs the
small $k_z^2$ dependence of the rotation band 
[Eq.~(\ref{fop}) with the subscript $t$ replaced by $\phi$].
 
The computations reported in the present paper permit 
to explicitly evaluate the four real coefficients which appear in 
Eq.~(\ref{kee1},~\ref{kee2}). For
example, for $a=0.8, b=0.01, \epsilon=0.025$, the reported results give
\begin{equation}
\frac{\langle\tilde{u}_t,u_t\rangle}{\langle\tilde{u}_t,u_t\rangle+
\langle\tilde{v}_t,v_t\rangle}=0.8842-0.662,\,
\frac{\langle \tilde{u}_{\phi},u_{\phi}\rangle}{\langle \tilde{u}_{\phi}
,u_{\phi}\rangle+
\langle \tilde{v}_{\phi},v_{\phi}\rangle}=0.578,\,
\frac{\langle \tilde{u}_{\phi},\partial^2_{\phi\phi}u_0
\rangle}{\langle \tilde{u}_{\phi},u_{\phi}\rangle+
\langle \tilde{v}_{\phi},v_{\phi}\rangle}=-0.7203.
\end{equation}

More generally, the two coefficients in Eq.~(\ref{kee2}) do not change
sign as one traverses the different regions of the phase diagram: a small twist
increases the scroll rotation frequency and  a small modulation in the
$z$-direction increases the stability of the rotation band
modes. The complex coefficient of Eq.~(\ref{kee1}) is directly linked  to
the scroll wave
line tension
stability/instability and its real part change sign as 2D spiral drift.

Finally, it may be worth comparing the above derivation to that of
\cite{keen3d}. Since the l.h.s of Eq.~(\ref{pertk}) is a superposition
of terms with different time dependences, this is also the case
of its solution and many coefficients formally introduced
in ref.~\cite{keen3d}
do not even appear in our derivation, as previously noted in \cite{bik3d} (in
a slightly different formulation). On the other side, we have chosen a simple
parameterization for which rotational invariance is not manifest, in contrast
to ref.~\cite{bik3d}. This forces us to
explicitly show that coefficients which do not
appear in ref.~\cite{bik3d} vanish.

We conclude by showing that indeed,
\begin{equation}
\langle \tilde{u}_{\phi}, \nabla^2_{2D}u_0
\rangle=0.
\label{dil}
\end{equation}
This is a simple consequence of 
the transformation property of the reaction-diffusion
equations (\ref{2dst1},~\ref{2dst2}) under dilation. 
Namely, $(u_0(r(1+\eta),\phi),
v_0(r(1+\eta),\phi)$ is a solution of (\ref{2dst1},~\ref{2dst2}) with
$\nabla^2_{2D}$ replaced by $1/(1+\eta)^2 \nabla^2_{2D}$. Expanding for
$\eta\ll 1$ gives the infinitesimal version of this transformation
\begin{equation}
{\cal L}\left(\begin{array}{c}r\partial_{r}u_0\\
r\partial_{r}v_0\end{array}\right)=2
\left(\begin{array}{c}\nabla^2_{2D}u_0\\0\end{array}\right)
\label{dilsym}
\end{equation}
which can also be directly checked by differentiating 
Eq.~(\ref{2dst1},~\ref{2dst2}) with respect to $r$. The multiplication of
Eq.~(\ref{dilsym}) on both sides by $(\tilde{u}_{\phi},
\tilde{v}_{\phi})$, the left eigenvector of
${\cal L}$
of eigenvalue zero, gives the desired identity (\ref{dil}).

\end{document}